\documentclass[useAMS,usenatbib]{mn2e}

\DeclareSymbolFont{cmletters}{OML}{cmm}{m}{it}
\DeclareMathSymbol{v}{\mathalpha}{cmletters}{"76}

\usepackage{amsmath}
\usepackage{amssymb}
\usepackage{epsfig}
\usepackage{graphicx}
\usepackage{ifthen}
\usepackage{latexsym}
\usepackage{rotating}
\usepackage{subfigure}
\usepackage{times,epsf}
\usepackage{txfonts}
\usepackage{varioref}
\usepackage{verbatim}
\usepackage{array}
\usepackage{color}
\usepackage[dvipsnames]{xcolor}
\usepackage[T1]{fontenc}

\newcolumntype{M}{>{$\vcenter\bgroup\hbox\bgroup}c<{\egroup\egroup$}}

\newcommand{\be}{\begin{equation}}
\newcommand{\ee}{\end{equation}}
\newcommand{\bea}{\begin{eqnarray}}
\newcommand{\eea}{\end{eqnarray}}

\newcommand{\Em}{{\cal M}}

\newcommand{\sgra}{Sgr~A$^*$ }

\newcommand\apj{Astrophysical Journal}
\newcommand\apjl{Astrophysical Journal Letters}

\newcommand\aap{Astronomy \& Astrophysics}

\newcommand\nat{Nature}

\newcommand{\nar}{{NAR}}
\newcommand\mnras{Monthly Notices of the Royal Astronomical Society}

\newcommand\pasj{Publications of the Astronomical Society of Japan}

\title[G2 impact on \sgra]{
Radio light curves during the passage of cloud G2 near \sgra }
\author[A. S\k{a}dowski, L. Sironi, D. Abarca, X. Guo, F. {\"O}zel, R. Narayan]
     {\parbox{\textwidth}{Aleksander S\k{a}dowski$^1$,  Lorenzo
         Sironi$^{1,2}$,   David
         Abarca$^{3}$, Xinyi Guo$^{1}$, Feryal
         {\"O}zel$^{1,4}$ and
 Ramesh Narayan$^{1}$\thanks{E-mail:
           asadowski@cfa.harvard.edu (AS); lsironi@cfa.harvard.edu
           (LS); 
david.abarca@college.harvard.edu (DA); xinyi.guo@cfa.harvard.edu (XG); 
fozel@email.arizona.edu (FO); rnarayan@cfa.harvard.edu (RN)}}\vspace{0.4cm}\\
        $^1$ Harvard-Smithsonian Center for Astrophysics, 60 Garden
        St., Cambridge, MA 02138, USA\\
 $^2$ NASA Einstein Postdoctoral Fellow\\
 $^3$ Harvard College, Massachusetts Hall, Cambridge, MA 02138, USA\\
 $^4$ University of Arizona, 933 N. Cherry Avenue, Tucson, AZ 85721,
 USA}

\begin{document}

\maketitle

\label{firstpage}

\begin{abstract}
We calculate radio light curves produced by the bow shock that is
likely to form in front of the G2 cloud when it penetrates the
accretion disk of Sgr~A$^*$. The shock acceleration of the
radio-emitting electrons is captured self-consistently by means of
first-principles particle-in-cell simulations. We show that the
radio luminosity is expected to reach maximum in early 2013, roughly a
month after the bow shock crosses the orbit pericenter. We estimate
the peak radio flux at $1.4\,\rm GHz$ to be $1.4 - 22\,\rm Jy$
depending on the assumed orbit orientation and parameters. We show
that the most promising frequencies for radio observations are in the
$0.1<\nu<1\,\rm GHz$ range, for which the bow shock emission will be
much stronger than the intrinsic radio flux for all the models
considered.
\end{abstract}


\section{Introduction}
\label{s.introduction}

The center of our Galaxy is known to host a black hole with a moderate
mass of $M = 4.3 \times 10^6 M_\odot$ \citep[e.g.,][]{genzel+10}.
This low-luminosity black hole, which is identified with the radio
source Sgr~A$^*$, accretes at a very low accretion rate (see, e.g.,
Quataert et al.\ 1999), has not been observed to produce relativistic
outflows, and only sporadically shows flares in X-rays (Baganoff et
al.\ 2001). The quiet and unimpressive life of \sgra is slated to
change in the late summer of 2013 when a dense cloud of gas will
approach it along a highly eccentric orbit
\citep{gillessen+12a,gillessen+12b}, with periapsis well inside the
Bondi radius of the central black hole, $R_{\rm B} \approx 0.04\rm pc
\approx 2\times 10^5R_{\rm G}$, where the gravitational radius is
defined as $R_{\rm G} = GM/c^2=2\times 10^{-7}\rm pc$.

\cite{gillessen+12a} discovered the gas cloud, called G2, during their
extended observational campaign of stars in the vicinity of the
Galactic Center. They estimated the cloud's mass to be $\lesssim
3\;M_{\oplus}$ \citep{gillessen+12a} and determined the orbital
parameters of its center of mass. A semi-major axis of $a=666\pm39 \rm
mas$, eccentricity of $e=0.9664\pm0.0026$, and epoch of periastron of
$t_{0,\rm CM}=2013.69\pm 0.04$, yield a periapsis distance that is
only $4400\pm600 R_{\rm G}$ from the black hole
\citep{gillessen+12b}.  The interaction of the cloud with the
accretion flow presents a unique opportunity to probe the local
conditions of gas at radii $R\sim10^{3.5}R_{\rm G}$. This corresponds
to a few percent of the Bondi radius of \sgra
\citep{yuan+03} and is a largely unexplored region of the accretion
flow around the black hole.

Using theoretical models of the accretion flow around Sgr~A$^*$,
\cite{narayan+12a} recently predicted that a bow shock will form ahead
of G2 for a few months around the time of closest approach. This shock
will likely accelerate particles and produce detectable levels of
synchrotron radiation. The expected parameters of the shock correspond
to a highly unusual regime for particle acceleration, one that has
hardly been studied until now. Although the shock velocity is
non-relativistic, at a temperature of $T \sim 10^{8.5}$\,K the
upstream electrons are nearly relativistic. Therefore, the electrons
are more easily accelerated to high Lorentz factors and the injection
problem often encountered in shock acceleration is mitigated.  In
addition, the unshocked gas will be already magnetized, leading to
conditions that could be particularly ideal for seeing synchrotron
radiation from particles accelerated by G2's bow shock.

In this paper we extend the work by \cite{narayan+12a} in several
ways. We explore in more detail the propagation of G2 through the
accretion flow around \sgra using a general relativistic
magnetohydrodynamic (GRMHD) numerical simulation of the pre-impact
accretion flow (Narayan et al. 2012b). We consider various
orientations of the orbit of G2 with respect to the accretion flow and
utilize the simulation to determine the physical parameters of the
pre-shock gas.  We perform first-principles particle-in-cell
calculations of particle acceleration in the shock for the range of
physical parameters sampled along the orbit. We consider the net
effect of particle acceleration along the cloud's trajectory using two
limiting approximations: one in which the moving cloud collects and
holds on to all the relativistic electrons accelerated by the bow
shock (the plowing case) and one in which electrons accelerated in the
shock are left behind and radiate in the unshocked magnetic field (the
local case). We calculate the spectrum of accelerated electrons along
G2's trajectory and the corresponding synchrotron radio light curves
for these two possibilities.

The structure of the paper is as follows. In Section~\ref{s.disk} we
discuss the adopted model for the accretion flow around Sgr~A$^*$. In
Section~\ref{s.orientation} we discuss the parametrization of the
cloud orbit with respect to the disk equatorial plane. In
Section~\ref{s.results} we show results of our study. We begin 
in Section~\ref{s.physical} by calculating profiles of physical 
parameters of the accretion flow along the cloud orbit. In
Section~\ref{s.acceleration} we discuss the physics of electron
acceleration and present the particle-in-cell simulations that we
performed for relevant gas and shock parameters in order to provide
estimates of the acceleration efficiency and the shape of the spectrum
of non-thermal electrons. In Section~\ref{s.lightcurves} we present
the predicted radio light curves and the spectrum of the radio
emission from the bow shock and discuss their dependence on the orbit
orientation. In Section~\ref{s.discussion}, we comment on the
uncertainties and optimal observing strategies of the G2
impact. Finally, we summarize our work in Section~\ref{s.summary}.


\section{Model of Sgr A$^*$ accretion disk}
\label{s.disk}

Accretion on to Sgr A$^*$ is believed to occur via an
advection-dominated accretion flow
\citep[ADAF:][]{narayanyi94,narayanyi95,narayan+95,yuan+03,narayanmcclintock08}. 
This mode of accretion is present whenever the mass accretion rate is
low, roughly $<1\%$ of Eddington; hence, it operates in the vast
majority of galactic nuclei in the Universe. In the case of Sgr A$^*$,
observations in radio, infrared and X-rays, coupled with theoretical
models, have provided partial information on the parameters of the
accretion flow.  \cite{yuan+03} used the number density of electrons
at the Bondi radius $n_{\rm e}(R_{\rm B})\approx130\, \rm cm^{-3}$
(Baganoff et al. 2003), along with a model of the radio and
submillimeter emission from thermal electrons in the inner accretion
disk, to estimate the mass accretion rate on to the black hole. They
found $\dot M_{\rm BH}\approx 4\times10^{-8}\, \rm M_\odot\,{\rm
yr^{-1}}$.  Other spectral models, based on radiative transfer
calculations using post-processed numerical simulations of global
accretion disks, led to somewhat lower estimates, $\dot M_{\rm
BH}\approx 2\times10^{-9}\, \rm M_\odot\,{\rm yr^{-1}}$
\citep{moscibrodzka+09}, while measurements of Faraday rotation in the
magnetized accreting gas gave values somewhere in between the two
estimates: $\dot M_{\rm BH}\approx 10^{-8}\, \rm M_\odot\,{\rm
yr^{-1}}$ \citep{marrone07}.

Although constraining, the results above do not specify the entire
structure of the accretion flow. In particular, while the models are
fairly reliable near the black hole and near the Bondi radius, they
are poorly constrained at the intermediate radii probed by G2's
orbit. To fill in this gap, studies rely on numerical calculations.
GRMHD simulations provide a proper treatment of the innermost regions
as well as of the flow on larger scales.\footnote{If the innermost
regions are not of interest and the important timescales are short, a
non-relativistic approach, such as the treatments of \cite{pen+03} and
Chan et al.\ (2009) are sufficient.}. Recently, a number of groups have
studied radiatively inefficient accretion flows using such numerical
models
\citep[e.g.,][]{devilliersetal03,gammieetal03,anninosetal05,delzannaetal07,narayan+12b,mtb12,tm12a}.
For the work reported here, we make use of one of the ADAF models
around a non-spinning black hole (spin parameter $a_*=0$) described in
\cite{narayan+12b}. The simulated flow corresponds to a Magnetically
Arrested Disk (MAD, Narayan et al. 2003), i.e., a system in which the
poloidal magnetic field is strong enough to partially ``arrest'' the
accreting gas.  The MAD state is believed to be a reasonable
approximation of most ADAFs in the Universe fed by gas with initially
non-zero poloidal magnetic flux. The most important feature of the
particular MAD simulation under consideration is that it was run
for an unprecedentedly long time. The results reported in
\cite{narayan+12b} corresponded to a duration of $10^5GM/c^3$, but
since then the simulation has been run further, up to a time of
$2\times10^5GM/c^3$. As a result, the accretion flow has achieved
steady state out to a large radius $\sim200R_{\rm G}$. This is
important for the present work, as we explain below.  The fact that
the simulation assumes zero black hole spin, whereas the black hole in
\sgra probably has a moderate non-zero spin \citep{broderick+11}, is
not important since spin affects only the innermost regions of the
accretion flow, whereas G2 interacts with gas that is much farther
out.

At periapsis, G2 is estimated to be located at a radius of $R=4400
R_{\rm G}$ from the black hole \citep{gillessen+12b}. Therefore, in order 
to predict synchrotron emission from G2's bow shock, we need to know the
properties of the ambient accreting gas at this radius. However, no
global GRMHD simulation can hope to obtain direct estimates at such
large radii because of the prohibitive computational times required. 
Because of this, we extrapolate the physical quantities we obtained 
from smaller radii. Such an extrapolation is acceptable, provided that
the simulation has reached steady state out to a large enough radius 
where the flow shows self-similar behavior (at small radii, the 
influence of the black hole causes large deviations from self-similarity.) 
This is what makes the MAD simulation described above particularly good 
for the present application. Having reached steady state out to radii $R\sim
200 R_{\rm G}$, the flow properties in this simulation have almost
certainly entered the self-similar regime (see Narayan et al. 2012b).
Hence, we may safely extrapolate out to the radii of interest to us.
In the present work, we choose $R=150 R_{\rm G}$ as the anchor radius
from which we carry out the extrapolation.

\begin{figure*}
  \centering
\subfigure{\includegraphics[width=.463\textwidth]{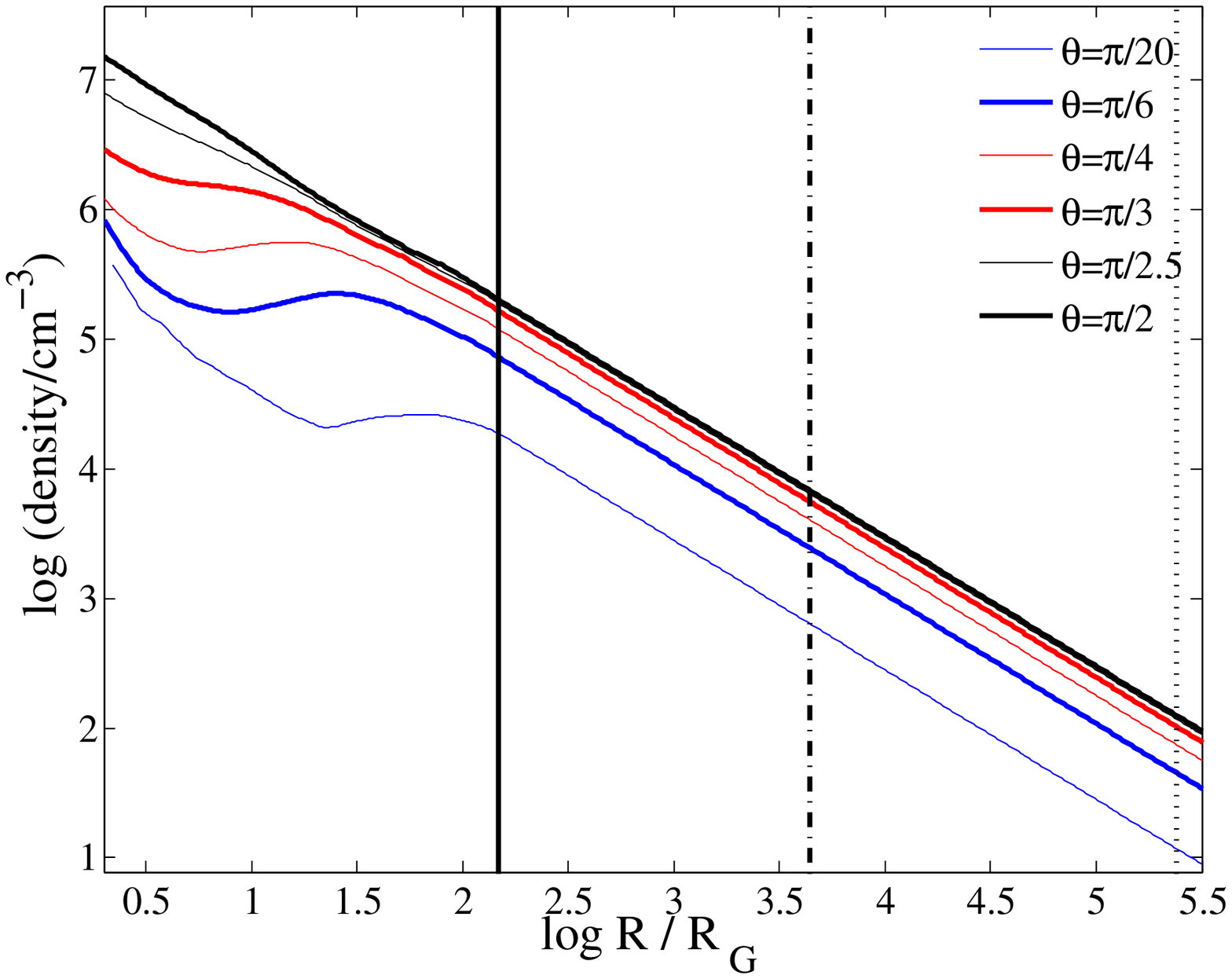}}
\subfigure{\includegraphics[width=.463\textwidth]{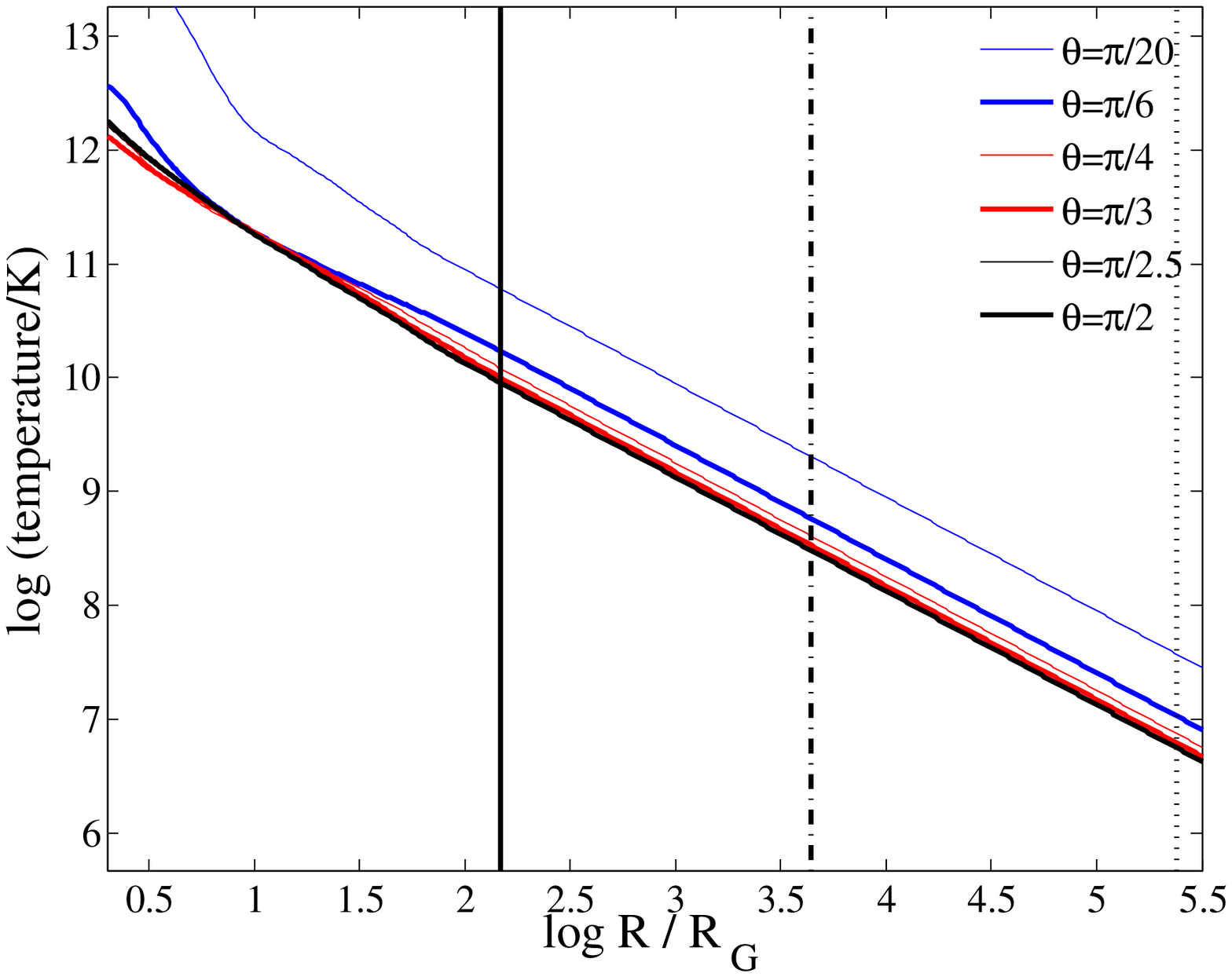}}
\subfigure{\includegraphics[width=.463\textwidth]{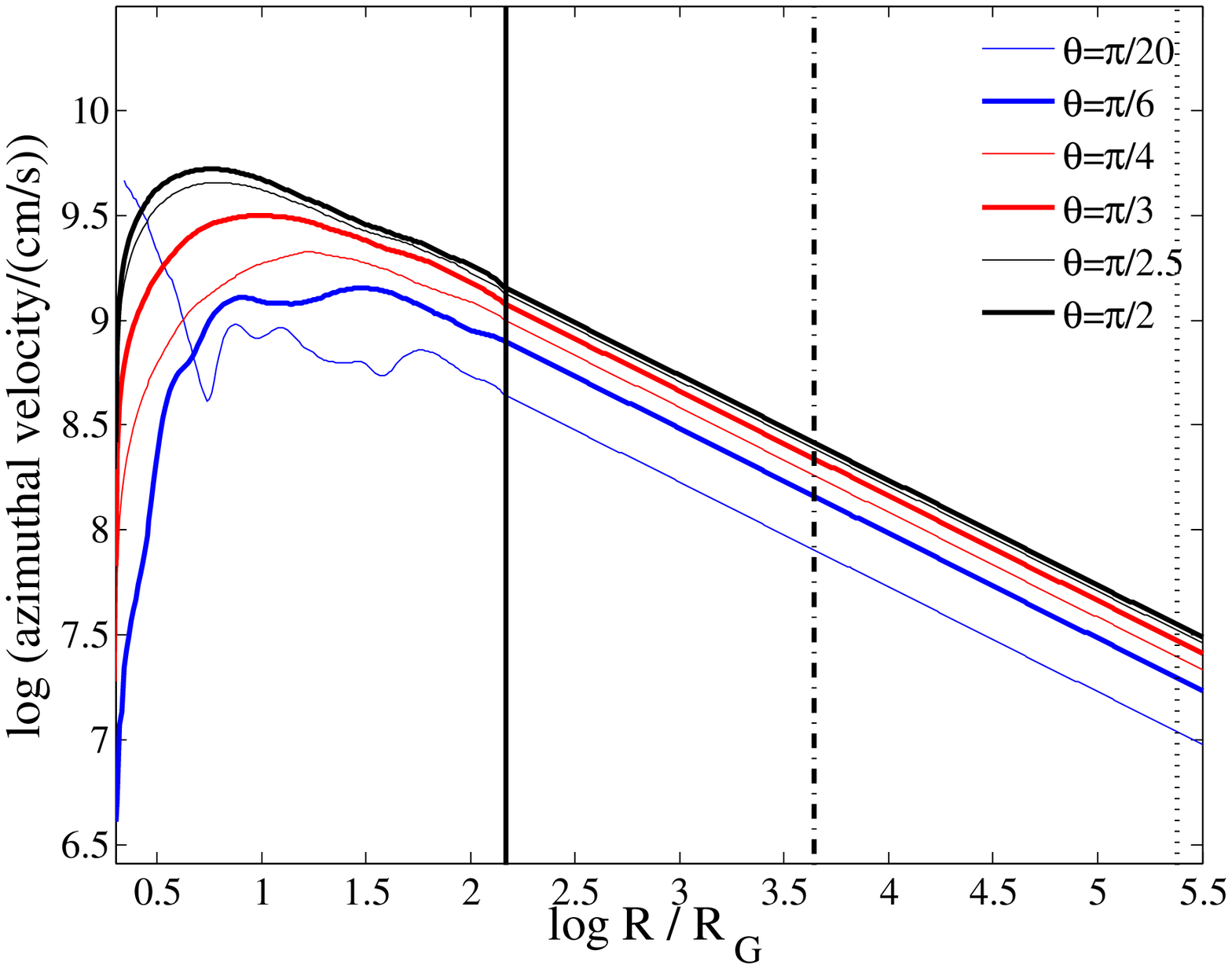}}
\subfigure{\includegraphics[width=.475\textwidth]{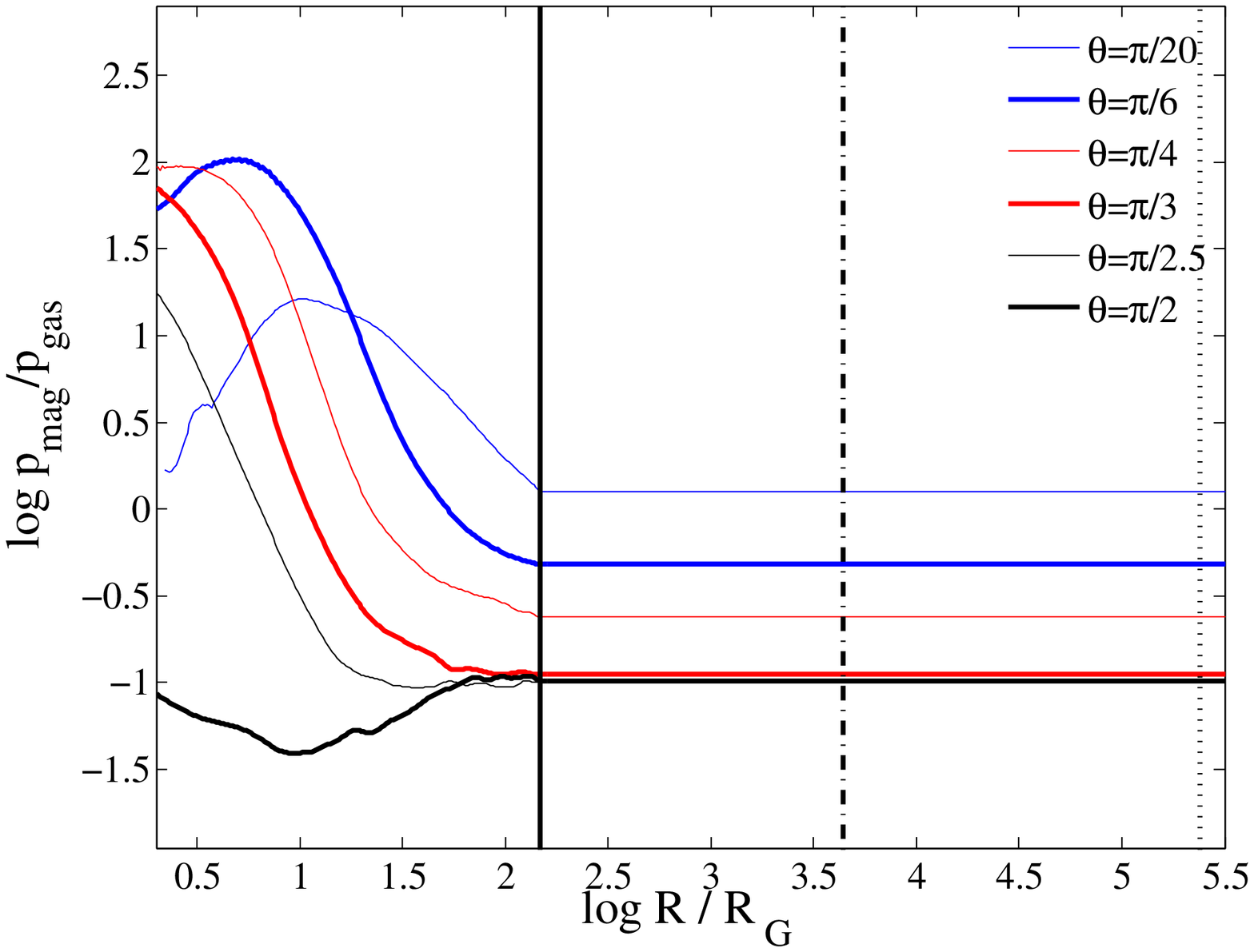}}
\caption{Numerical disk model and its extrapolation. Panels show
  logarithms of density (top left), gas temperature (top right),
  azimuthal velocity (bottom left), and magnetic to gas pressure ratio
  (bottom right panel).  Vertical lines show the
  radius of extrapolation ($R=150 R_{\rm G}$, solid), the periapsis of
  the cloud orbit ($R=4400 R_{\rm G}$, dot-dashed) and the Bondi
  radius ($R=2\times 10^5 R_{\rm G}$, dotted). $\theta$ is the polar
  angle and $\theta=\pi/2$ corresponds to the equatorial plane. }
  \label{f.extr}
\end{figure*}

We present in Figure~\ref{f.extr} the radial profiles of density
(top-left), temperature (top-right), azimuthal velocity (bottom-left)
and magnetic to gas pressure ratio (bottom-right). We plot the MAD
simulation results for $R \leq 150R_{\rm G}$ (left of the vertical
solid line) and show power-law extrapolations at larger radii, $R\geq
150 R_{\rm G}$. We indicate the radius corresponding to G2's
periapsis with the vertical dash-dotted line. The various colors and
line thicknesses correspond to different values of the polar angle
$\theta$.

We adjusted the density normalization to match the measured density at
the Bondi radius, $n_{\rm e}(R_{\rm B})=130\,\rm cm^{-3}$ (Baganoff et
al. 2003), using a power-law extrapolation $R^{-1}$.  The
corresponding mass accretion rate at the black hole is $\dot M_{\rm
BH}=5\times10^{-8} \rm M_\odot/yr$, which is consistent with the
estimates given above. The slope we chose for the density
extrapolation differs from the canonical slope for an ADAF disk
$R^{-3/2}$. However, as emphasized by Yuan et al. (2012a, b), an
$R^{-1}$ profile provides a better description of nearly all global
ADAF simulations to date\footnote{\cite{pen+03} found a $R^{-0.72}$
scaling for the density in a set of large-scale non-relativistic MHD
simulations. However, subsequent work by \cite{pang+11} concluded that
an $R^{-1}$ scaling is a better description of non-convective
magnetized ADAFs.}  This is true of our MAD simulation as well.
Especially in the vicinity of the equatorial plane, we see that the
simulation results at smaller radii match seamlessly to an $R^{-1}$
extrapolated power-law at larger radii. As expected, the largest
densities are found at the equatorial plane ($\theta=\pi/2$) and the
smallest near the polar axis ($\theta=0$).

The profiles of gas temperature show equally good convergence. These
again show an $R^{-1}$ dependence, indicating that the gas temperature
scales proportional to the virial temperature.  At a given radius, the
hottest gas is above the disk surface, near the polar axis. The
simulation does not distinguish between the ion and electron
temperatures, whereas detailed ADAF models generally allow for a two
temperature gas with $T_i \neq T_e$ (Narayan \& Yi 1995).
In most models, the plasma becomes two-temperature only at radii
well below $10^3R_G$ (e.g., Fig.~2 in \citealt{narayan+95}). At
larger radii, Coulomb collisions between protons and electrons are
sufficiently frequent to enforce a single temperature. For the model
of the accretion flow employed in the present work, the Coulomb
equilibration time between protons and electrons at pericenter of G2
is $\sim10^6\,\rm s$, which is shorter than the orbital time by a
factor of $\sim50$ (and of course, shorter than the radial accretion
time by an even larger factor).

The azimuthal velocity scales with radius according to the expected
Keplerian dependence $R^{-1/2}$. At a given radius, the largest
azimuthal velocity is seen at the equatorial plane. The further from
the disk plane, the slower is the rotation. We assume the radial
and vertical (in $\theta$) velocities to be zero.

The magnetic to gas pressure ratio shows the worst convergence among
the four quantities shown. Only inside the bulk of the disk does it
settle down to a relatively radius-independent value $\sim0.1$, which
is a typical value for most MHD accretion disks (but see
\citealt{gaburov+12} for a discussion of the impact of the initial
magnetic field configuration on the saturated value of the pressure
ratio). Off the mid-plane, the gas is strongly magnetized and the
pressure ratio does not settle down to a constant value even at
$R=150R_{\rm G}$. Thus, extrapolation under the assumption of a
constant ratio is not well justified. However, this will have
relatively little impact on the light curves when compared to other
uncertainties (see Section~\ref{s.discussion}).

For convenience, in Appendix~\ref{ap.1} we give analytic formulae
which reasonably approximate the disk structure for $R>150 R_{\rm G}$.

\section{Orbit orientation}
\label{s.orientation}

The orbital parameters of the center of mass of G2 are well
constrained \citep{gillessen+12b}. In this work we assume that the
shock follows precisely this orbit, neglecting possible hydrodynamical
interactions between the cloud and disk gas which might affect the
cloud front. Modeling these hydrodynamic effects will require MHD
simulations that include both the moving cloud and the orbiting
accretion flow, which we will present in future work.

The cloud orbital plane is inclined with respect to the observer's
line-of-sight and is roughly aligned with the well-known ring of stars
around \sgra (Bartko et al. 2009; Levin \& Beloborodov 2003).
The orientation of the accretion disk around \sgra is, however, poorly
constrained.  Therefore, in this paper, we allow the full range of
possible relative orientations between the cloud orbital plane and the
disk plane.  Given the uncertainties in our model (see
Section~\ref{s.discussion}), it is unlikely, but not impossible, that
the observed lightcurve will put constraints on the disk orientation.

\begin{figure}
  \centering
\hspace{-.5cm}
\includegraphics[width=.94\columnwidth]{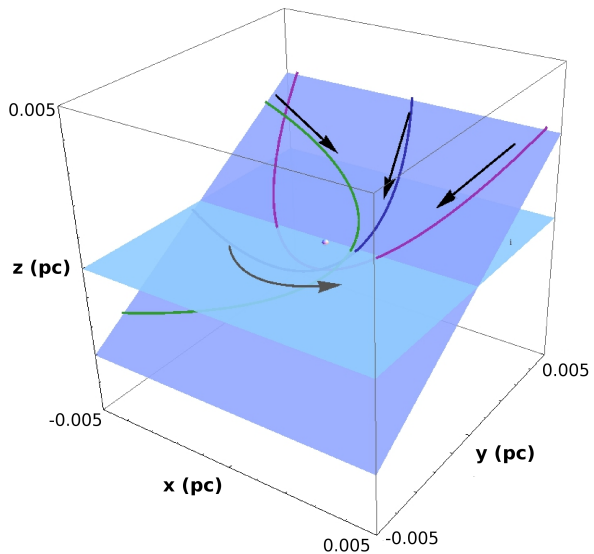}\\
\caption{Three-dimensional visualisation of the orbit orientation. The
  orbital plane is inclined to the disk plane by an angle $i$. The
  argument of periapsis for the orbits shown is $\omega=0$, $\pi/4$ and
  $\pi/2$ for green, blue and purple orbits, respectively. The arrows
  denote the cloud direction of motion and the disk rotation.}  
\label{f.trajectories}
\end{figure}

Because of the assumed axisymmetry of the accretion flow, the
following two angles completely specify the geometry: the inclination
angle, $i$, and the argument of periapsis,
$\omega$. Figure~\ref{f.trajectories} shows a three-dimensional
visualisation of three sample orbits whose orbital parameters
correspond to those of G2. The inclination $i$ corresponds to the
angle between the orbital plane of G2 and the equatorial plane
of the accretion disk. Cloud orbits that counter-rotate with respect
to the disk correspond to $i<\pi/2$, while co-rotating orbits
correspond to $\pi/2<i<\pi$. The argument of periapsis $\omega$ is the
angle between the line connecting the black hole and orbit pericenter
and the line of nodes where the two planes cross. The orbits shown in
the figure correspond to $\omega=0$ (green), $\pi/4$ (blue) and
$\pi/2$ (purple line).

\begin{figure}
  \centering
\begin{tabular}{MM}
\begin{sideways}time (yr)\end{sideways}&\hspace{-.5cm}
\includegraphics[width=.95\columnwidth]{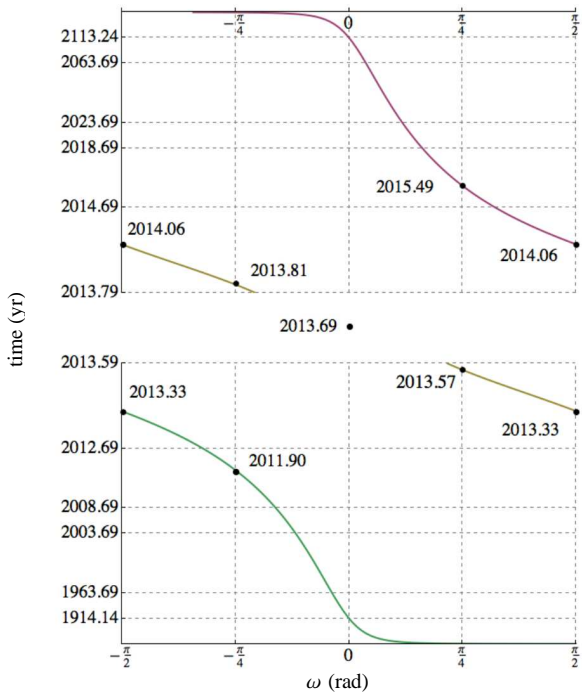}\vspace{-.25cm}\\
&\hspace{-.15cm}$\omega\,\, \rm(rad)$\\
\end{tabular}
\caption{Crossing time of the disk equatorial plane (in years) 
by the cloud center of mass as a function of the argument of periapsis
$\omega$. The vertical axis is non-uniform and follows the logarithm
of $t-t_{0,\rm CM}$, where $t_{0,\rm CM}$ is the epoch of pericenter
for the center of mass. Points and labels denote crossing times for
$\omega=-\pi/2$, $-\pi/4$, $0$, $\pi/4$, and $\pi/2$.}
\label{f.crossings}
\end{figure}

The cloud crosses the disk equatorial plane twice during each full
orbit.  The locations and times of these crossings depend on $\omega$,
but not $i$. Figure~\ref{f.crossings} shows as a function of
$\omega$ the epochs when the two crossings occur for the cloud center
of mass (CM). For $\omega=0$ (green orbit in
Fig.~\ref{f.trajectories}), one of the crossings takes place exactly
at periastron ($t=2013.69$), while the other occurs half an orbital
period, i.e., $\sim 100 \,\rm yr$ later. For $\omega=\pi/4$, the CM
crossings take place at $t=2013.57$ and $t=2015.49$\footnote{Note
that, for $\omega=-\pi/4$, the crossing times will be reflected (in
time) with respect to the epoch of periastron. Later, we describe
results corresponding to $\omega=\pi/4$. Disk properties along the
orbit for $\omega=\pi/4$ can be easily converted to $\omega=-\pi/4$ by
reflection. However, the radio light curves presented in
Section~\ref{s.lightcurves} are not related so simply.}. For
$\omega=\pi/2$, the disk equatorial plane crossings are symmetric with
respect to periastron and occur at $t=2013.33$ and $2014.06$.

The particular choices of $i$ and $\omega$ that we considered in this
work are listed in Table~\ref{t.models} where the crossing times are given
relative to the epoch of pericenter $t_0$.

\begin{table}
\caption{Orbit orientation parameters}
\label{t.models}
\centering\begin{tabular}{@{}ccc}
\hline
\hline
Inclination ($i$)&    \\
\hline
$\pi/3$ & counter-rotating \\
$\pi/4$ & counter-rotating \\
$\pi/6$ & counter-rotating\\
$\pi-\pi/3$ & co-rotating \\
$\pi-\pi/4$ & co-rotating  \\
$\pi-\pi/6$ & co-rotating  \\
\hline
\hline
\end{tabular}\vspace{.5cm}
\begin{tabular}{@{}ccc}
\hline
\hline
 Argument& Disk equatorial plane crossings\\
of periapsis ($\omega$) & relative to the epoch of pericenter $t_0$\\
\hline
$0$ & single crossing at $t=t_0$\\
$\pi/4$ & two asymmetric crossings at $t=t_0-0.12 \rm yr$ and $t_0+1.8 \rm yr$\\
 $\pi/2$ & symmetric crossings at $t=t_0-0.37 \rm yr$ and $t_0+0.37 \rm yr$\\
\hline
\hline
\end{tabular}
 \end{table}

\section{Results}
\label{s.results}

\subsection{Accretion gas properties along the orbit}
\label{s.physical}

The radio synchrotron emission that we expect as a function of time as
G2 orbits around \sgra depends on the properties of the ambient
accretion disk at various points along the orbit. Key quantities are
the density, temperature and magnetic field strength in the ambient
gas, and the Mach number of the cloud relative to the gas. Using
the extrapolated profiles discussed in Section~\ref{s.disk}, it is
straightforward to compute all these quantities along the orbit as a
function of time for a given choice of the cloud orbital parameters,
$i$ and $\omega$.

\begin{figure*}
  \centering
\subfigure{\includegraphics[width=.33\textwidth]{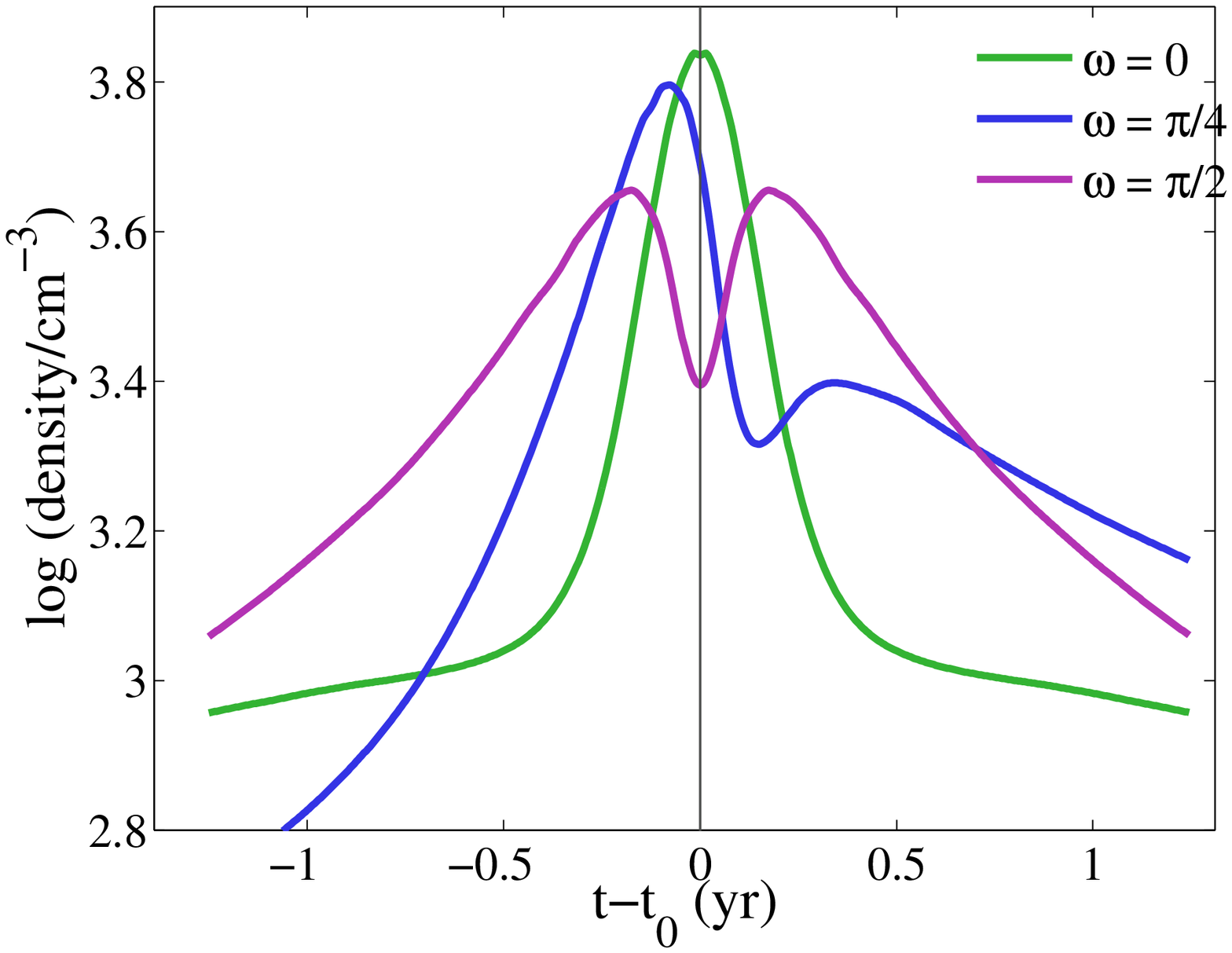}}
\subfigure{\includegraphics[width=.33\textwidth]{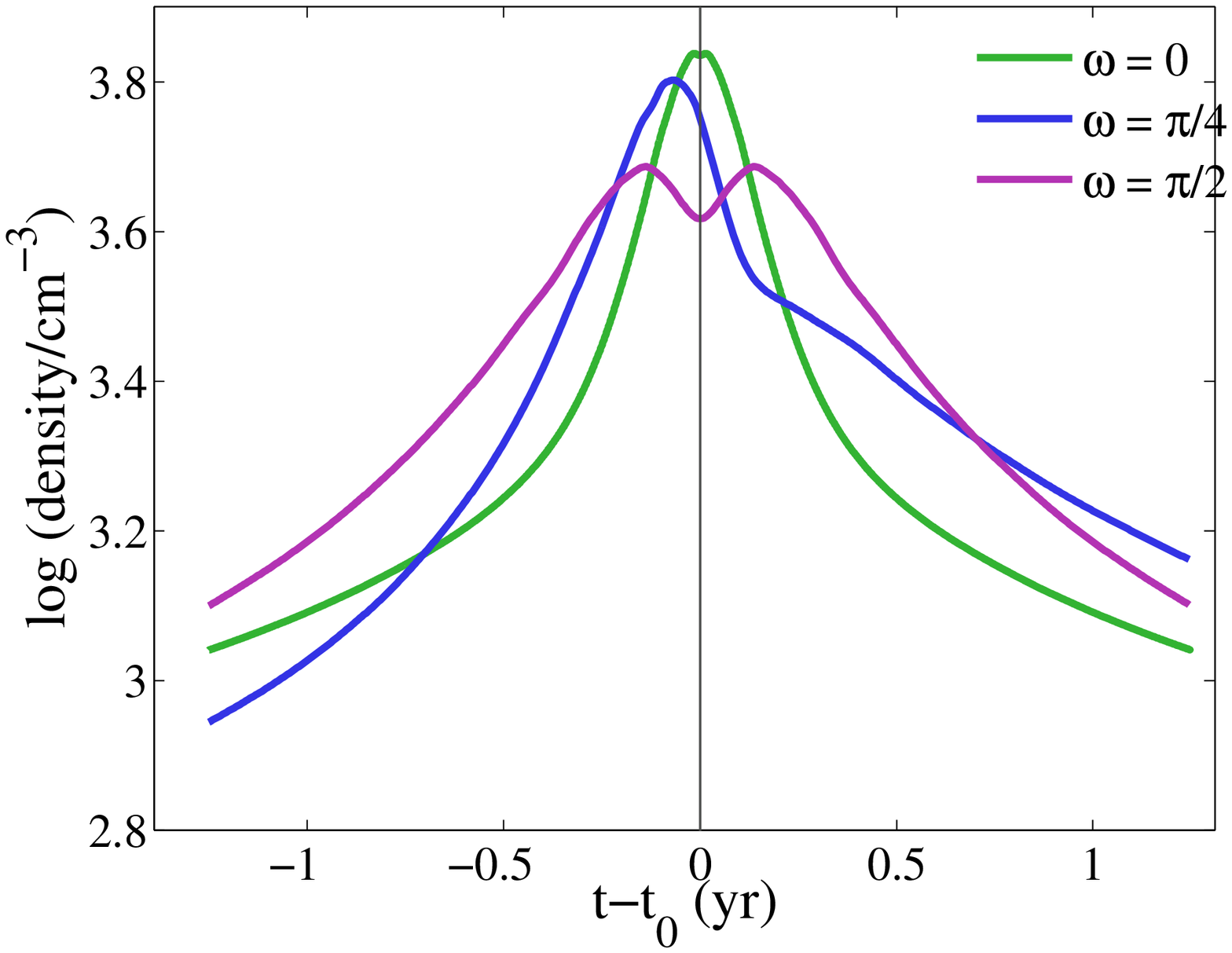}}
\subfigure{\includegraphics[width=.33\textwidth]{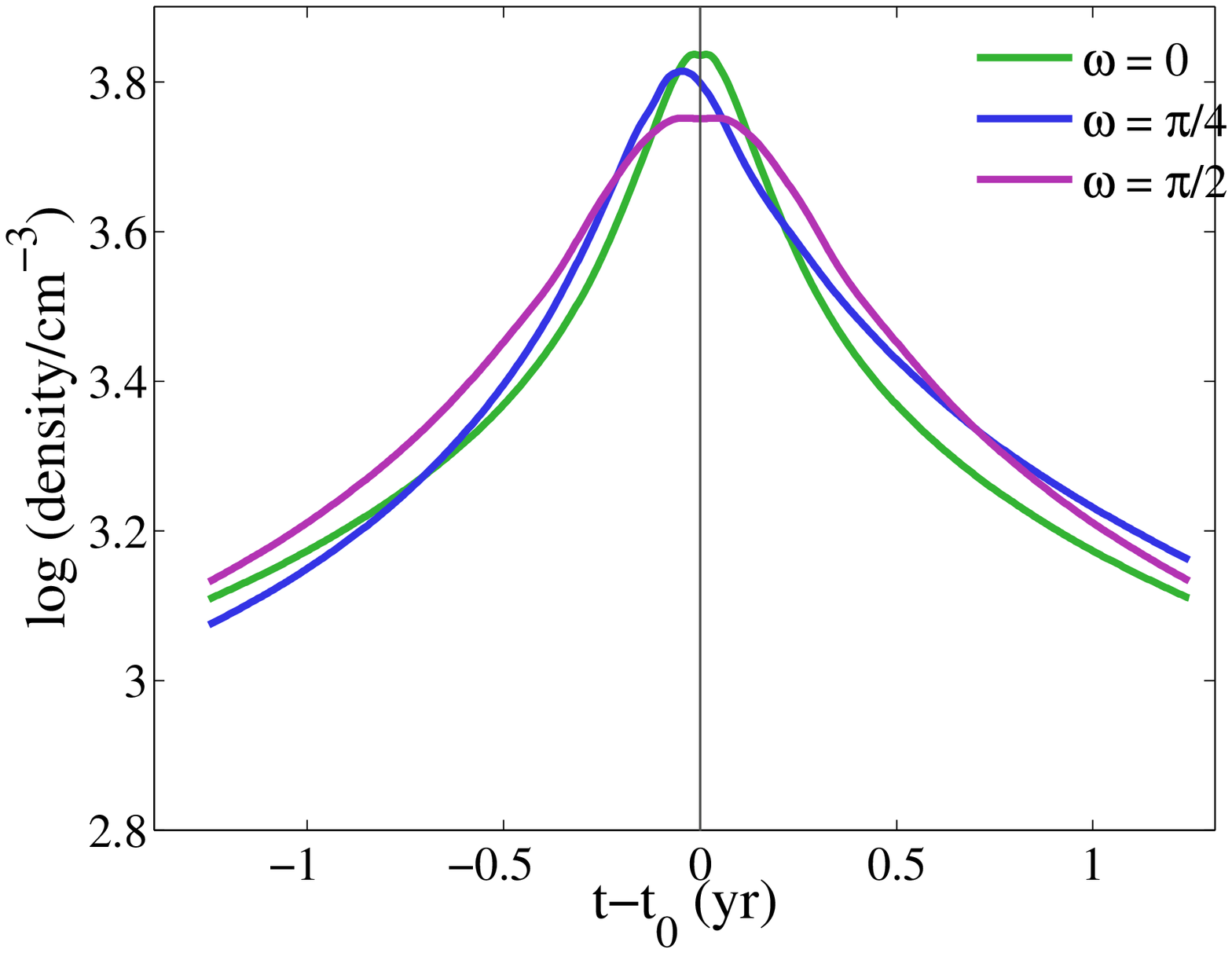}}\\\vspace{-.4cm}
\subfigure{\includegraphics[width=.33\textwidth]{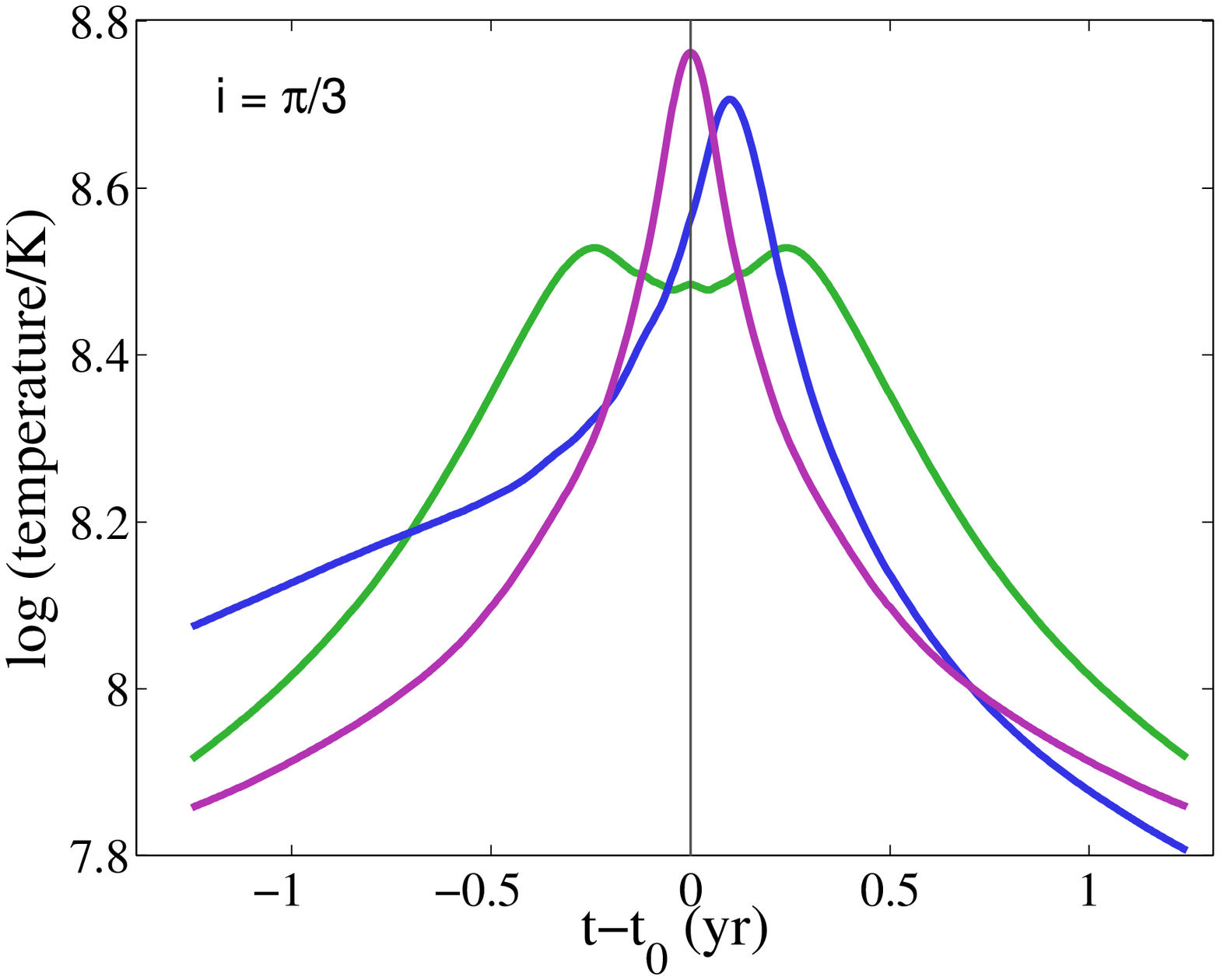}}
\subfigure{\includegraphics[width=.33\textwidth]{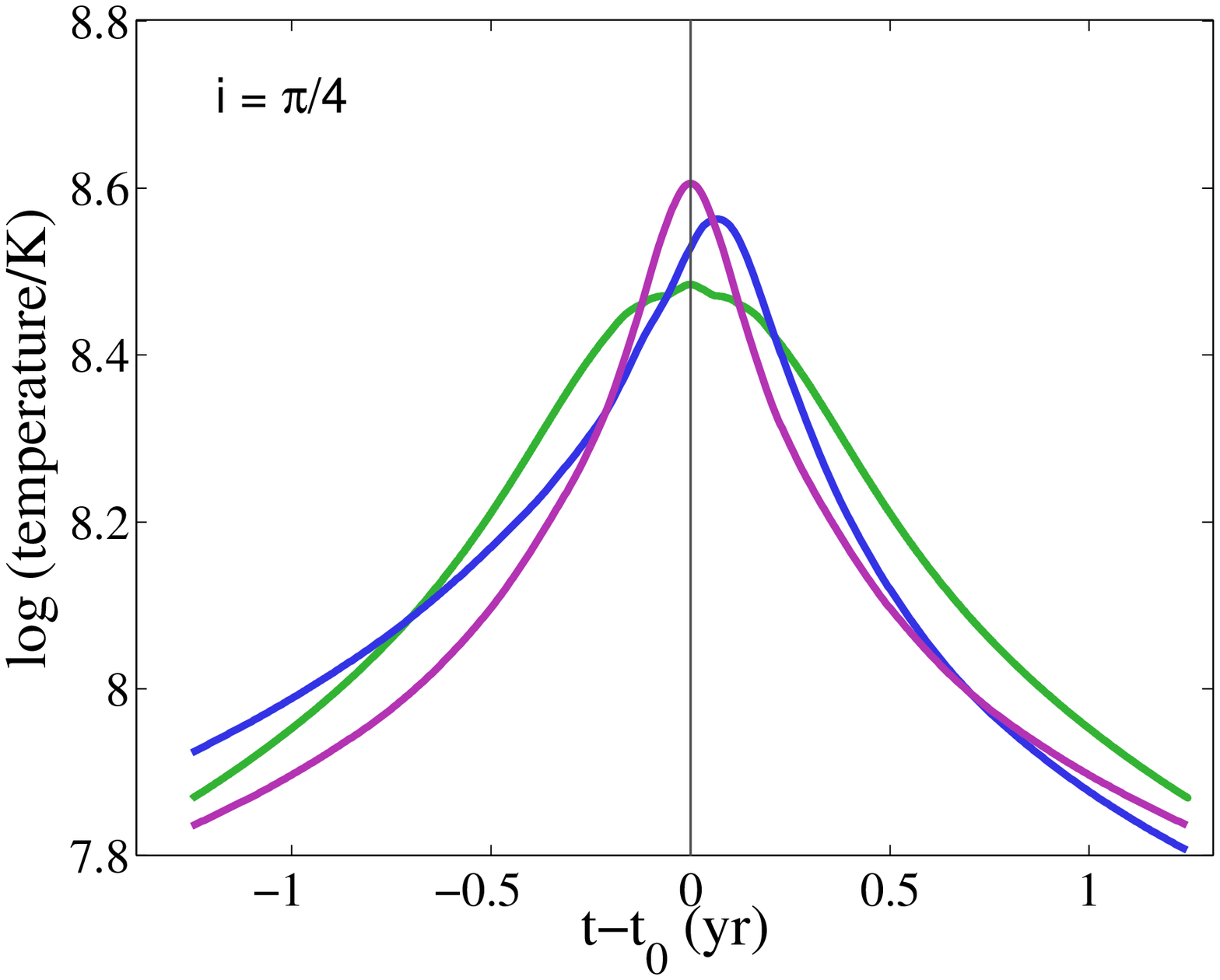}}
\subfigure{\includegraphics[width=.33\textwidth]{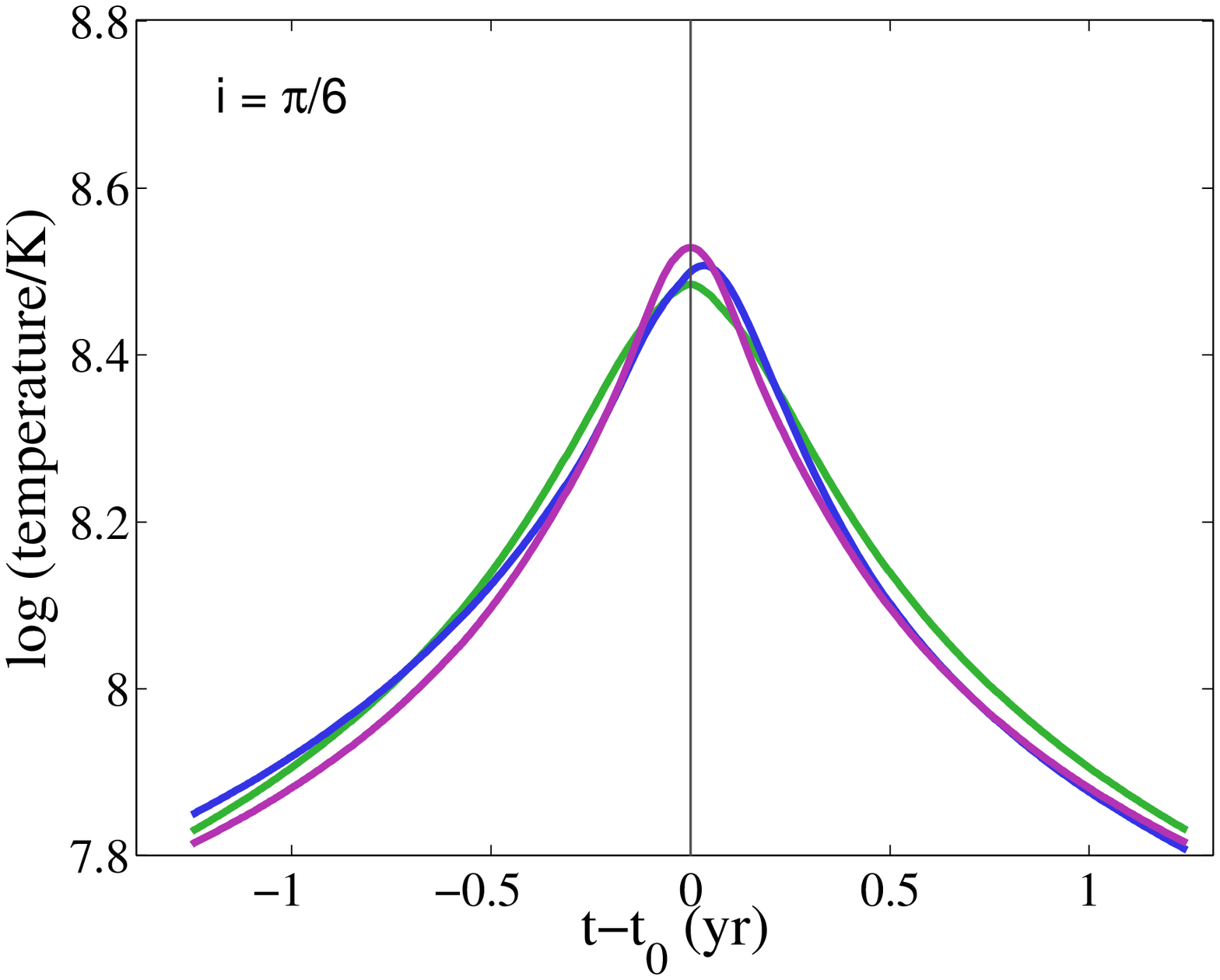}}\\\vspace{-.4cm}
\subfigure{\includegraphics[width=.33\textwidth]{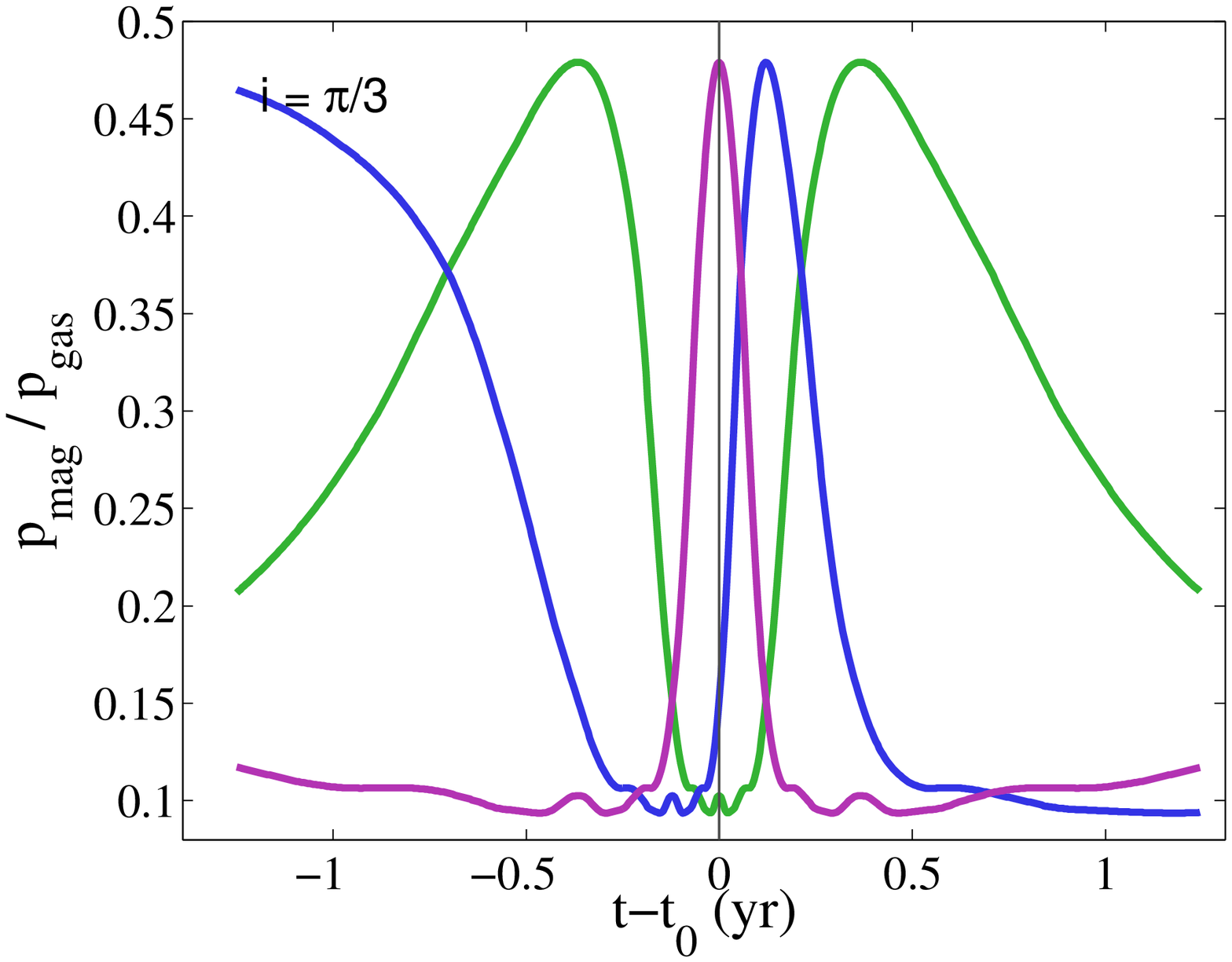}}
\subfigure{\includegraphics[width=.33\textwidth]{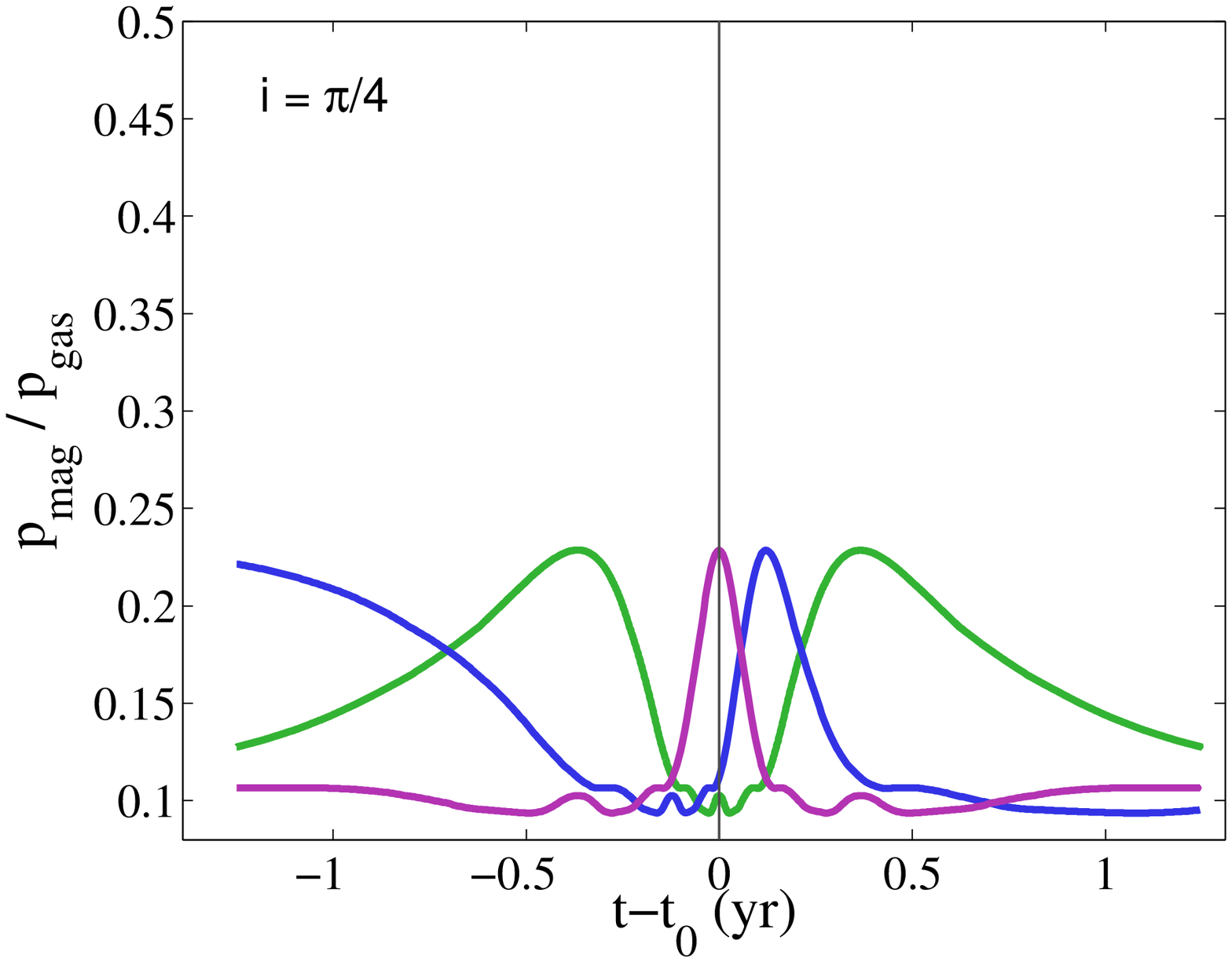}}
\subfigure{\includegraphics[width=.33\textwidth]{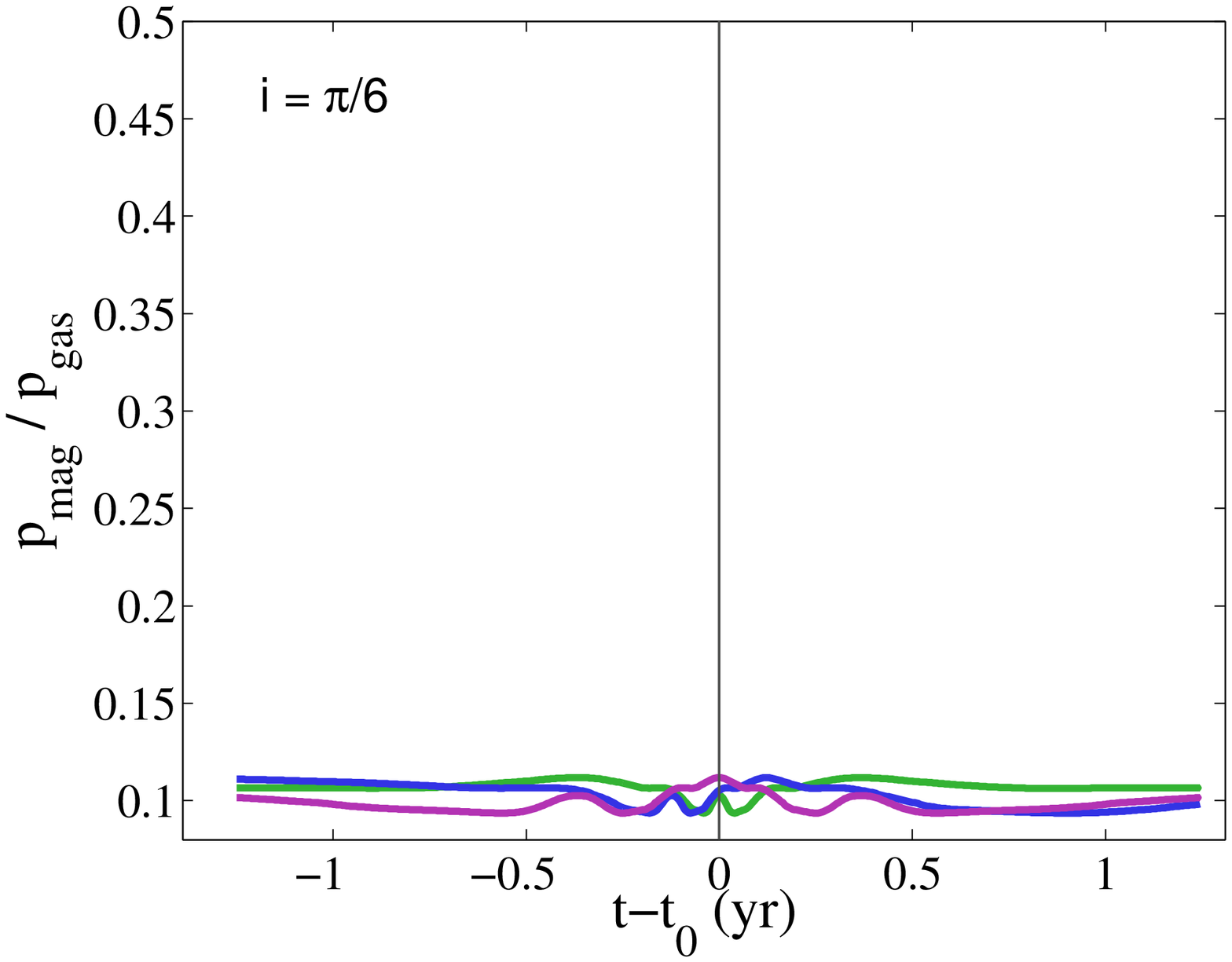}}\\\vspace{-.4cm}
\subfigure{\includegraphics[width=.33\textwidth]{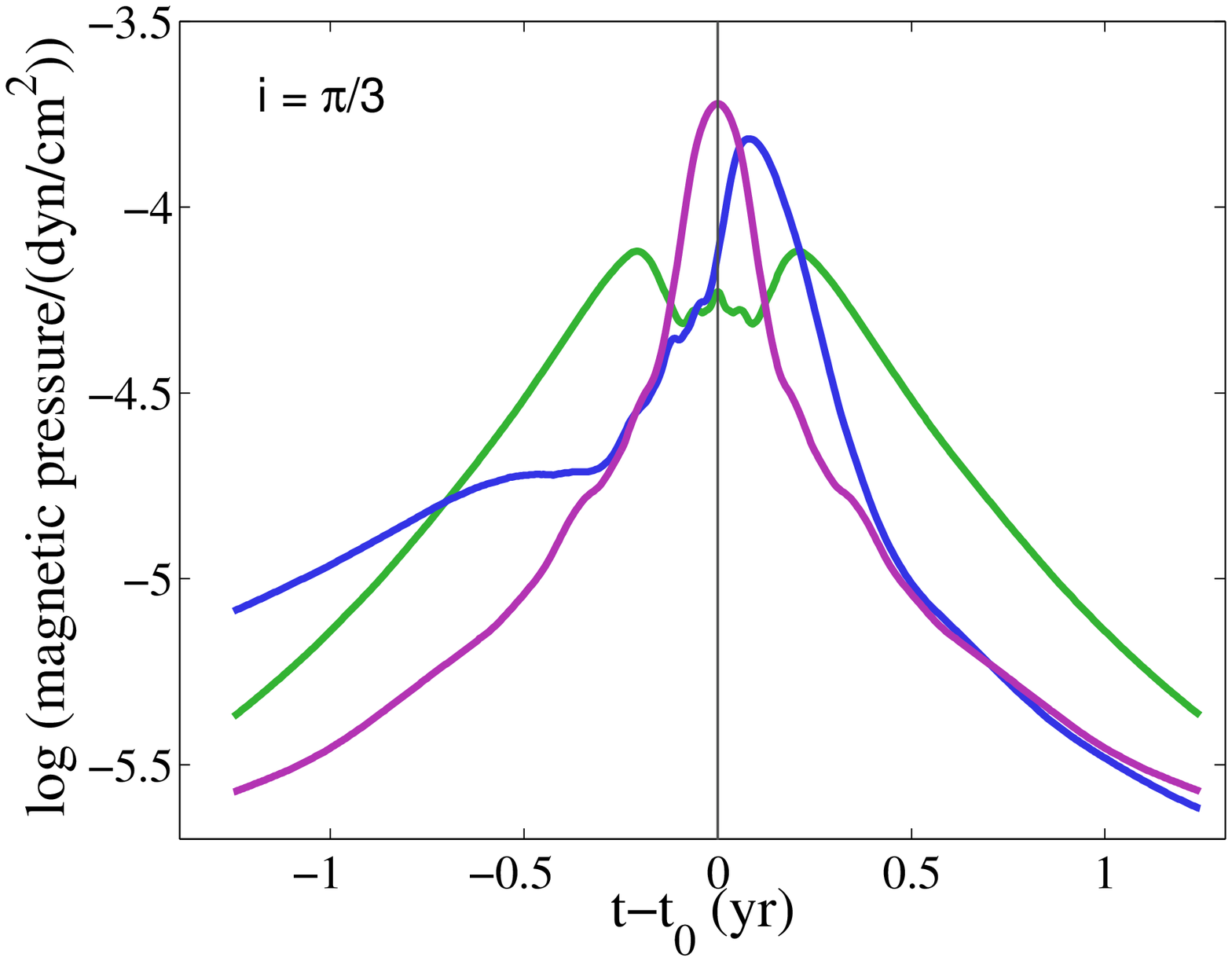}}
\subfigure{\includegraphics[width=.33\textwidth]{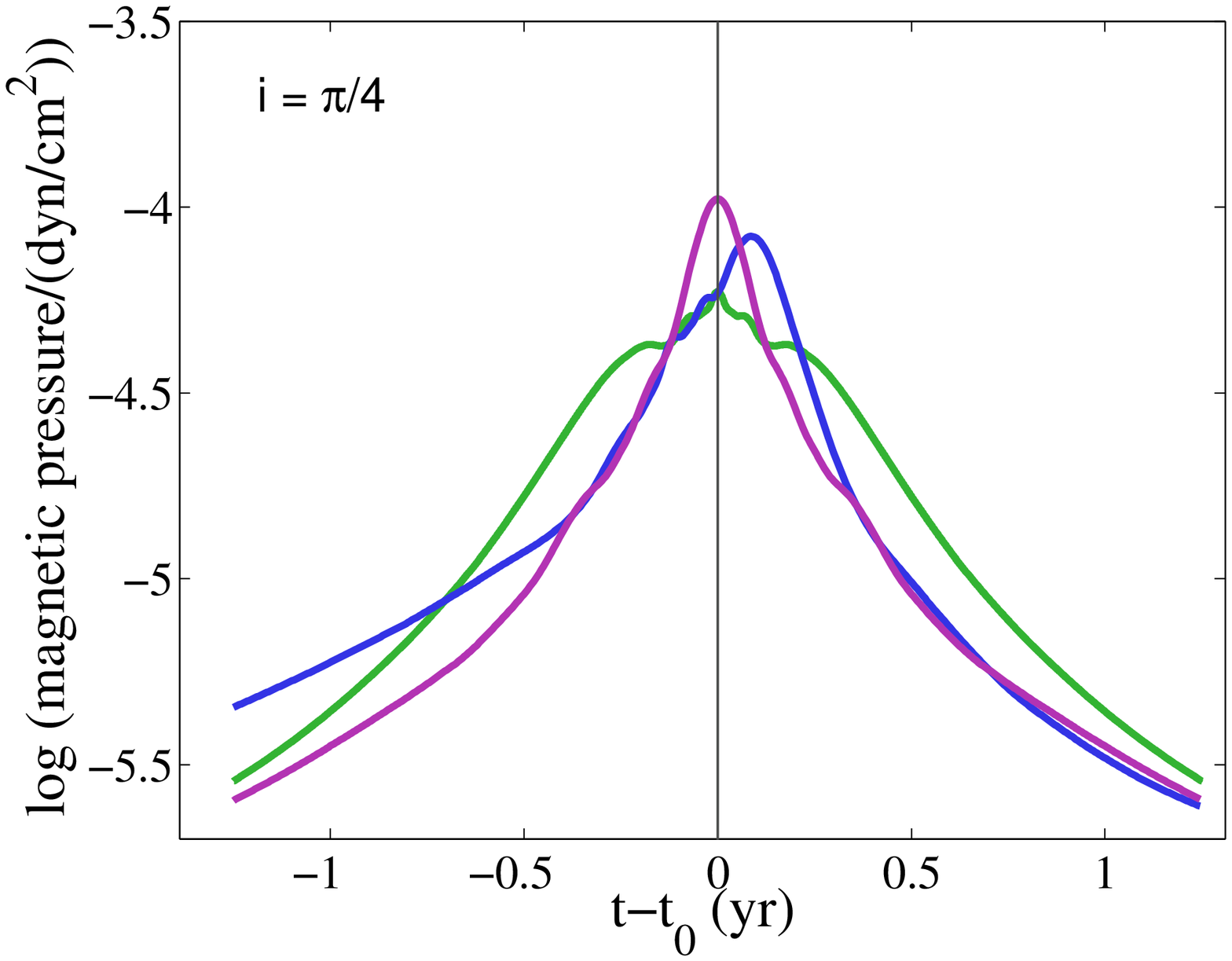}}
\subfigure{\includegraphics[width=.33\textwidth]{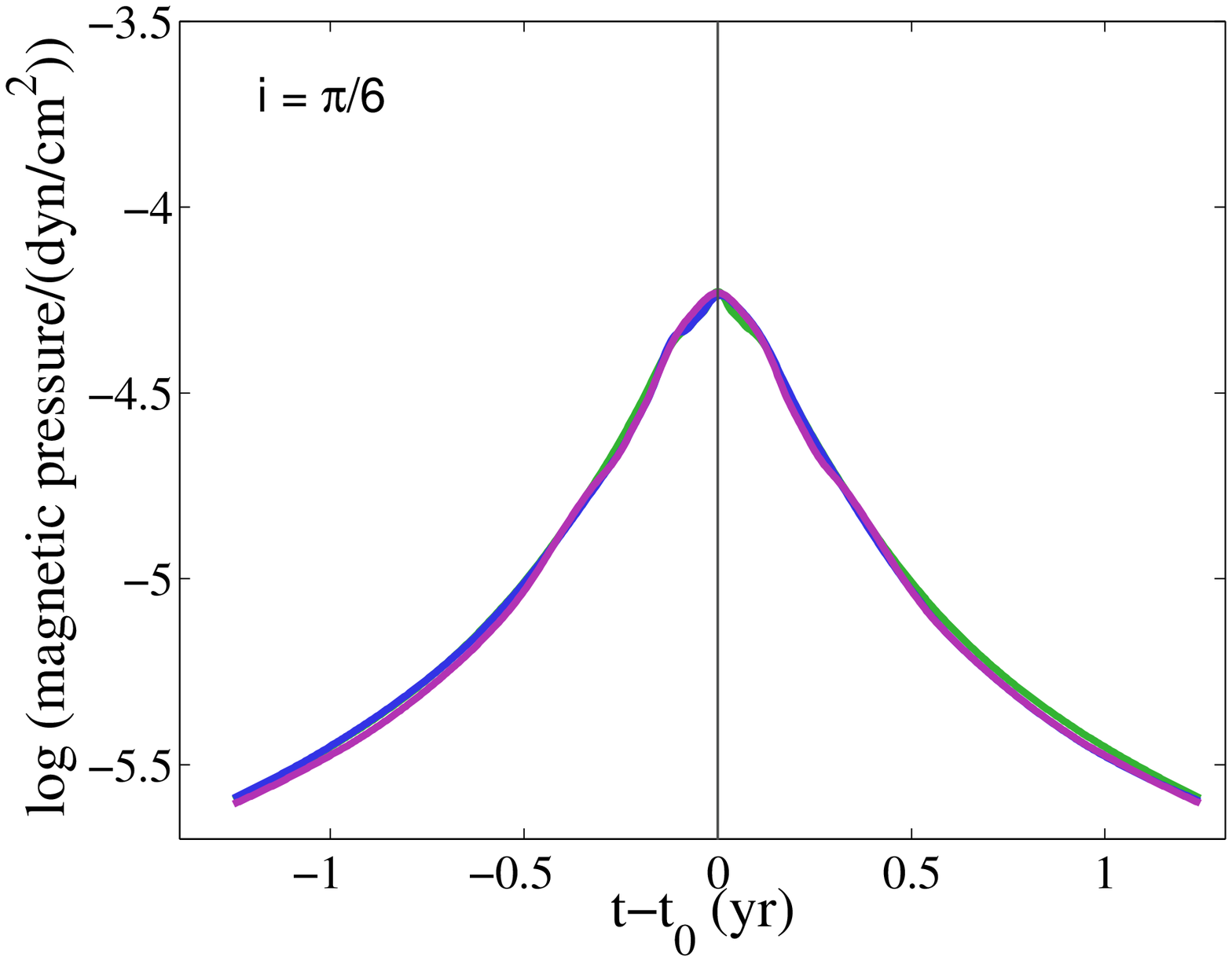}}\\\vspace{-.4cm}
\subfigure{\includegraphics[width=.33\textwidth]{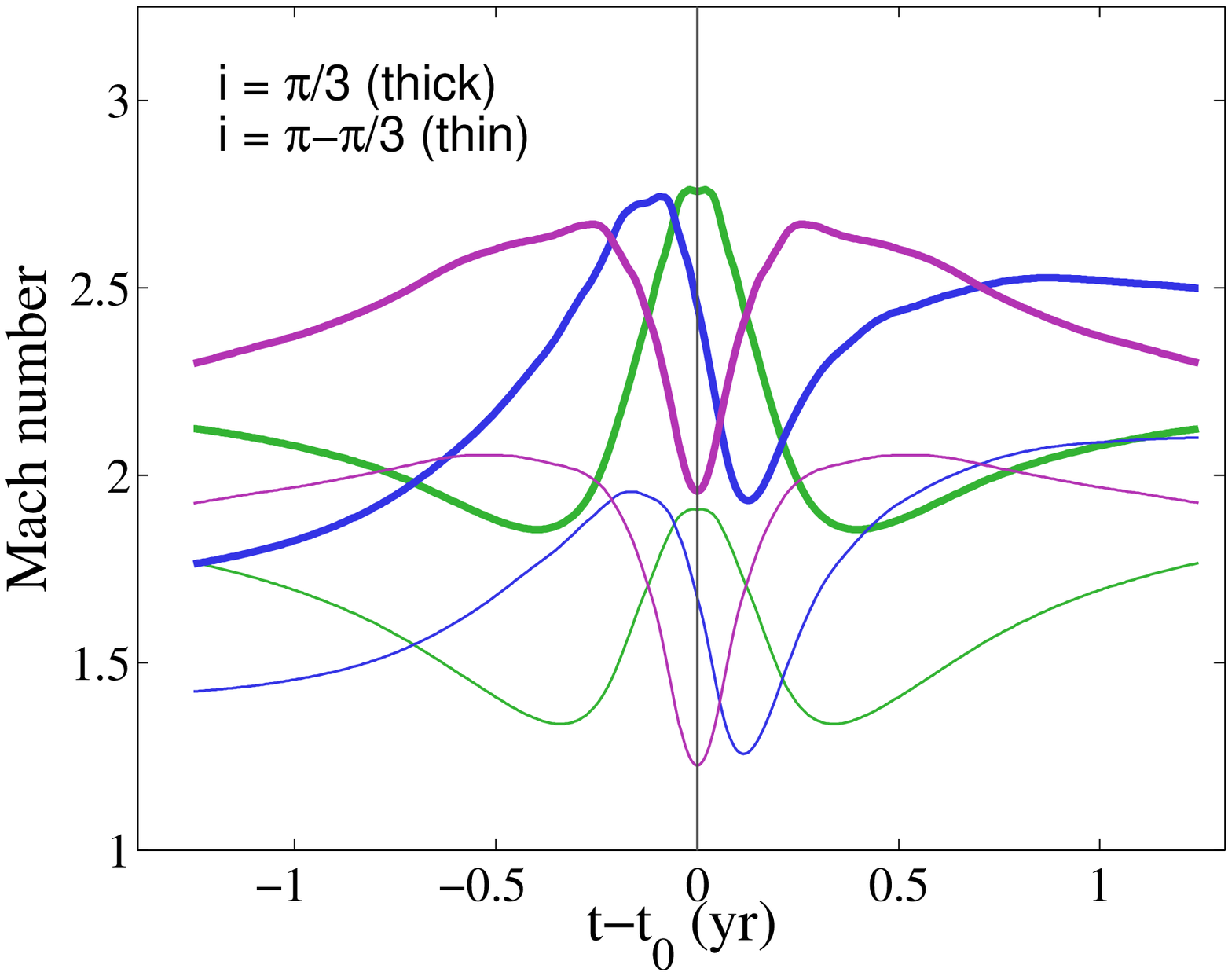}}
\subfigure{\includegraphics[width=.33\textwidth]{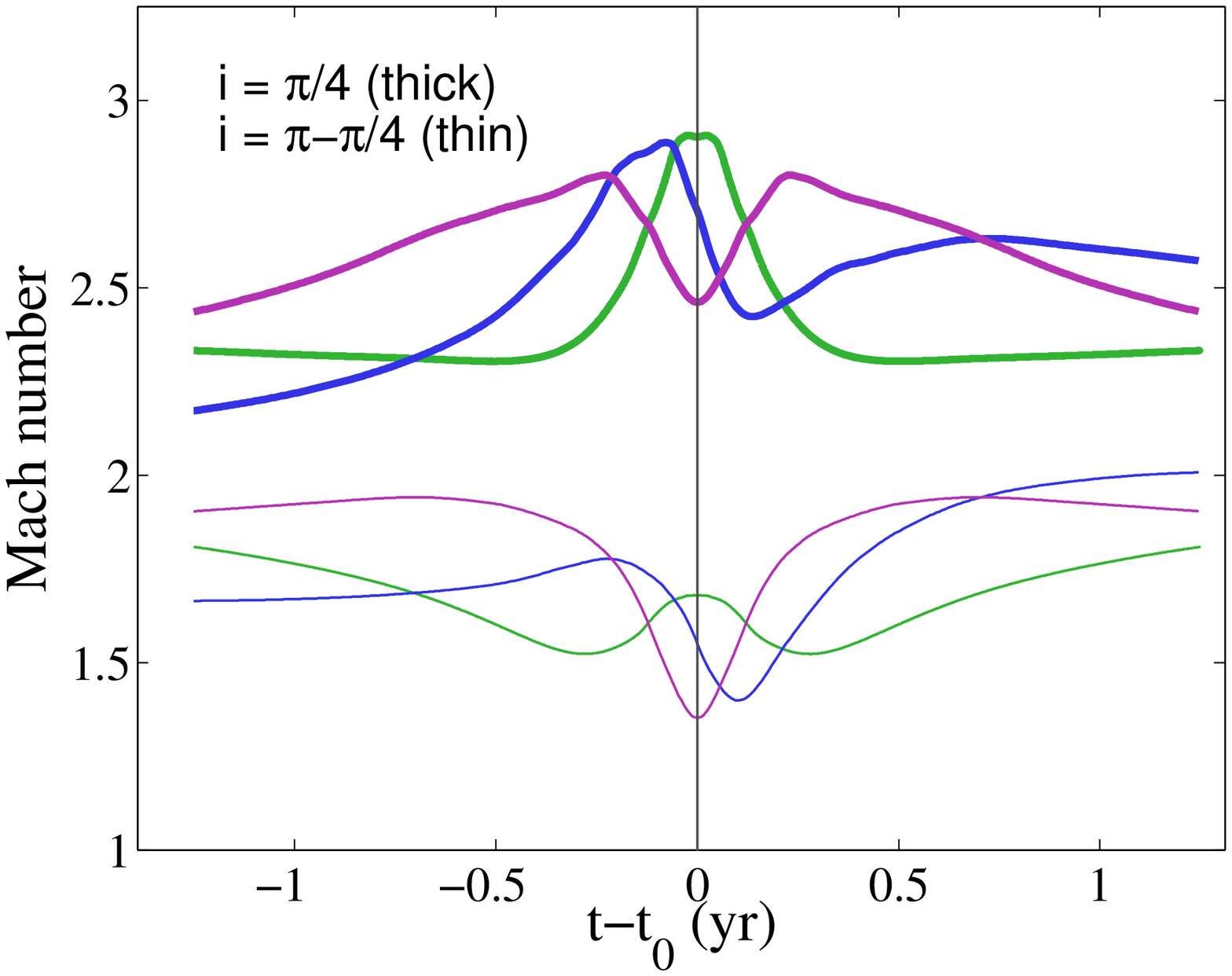}}
\subfigure{\includegraphics[width=.33\textwidth]{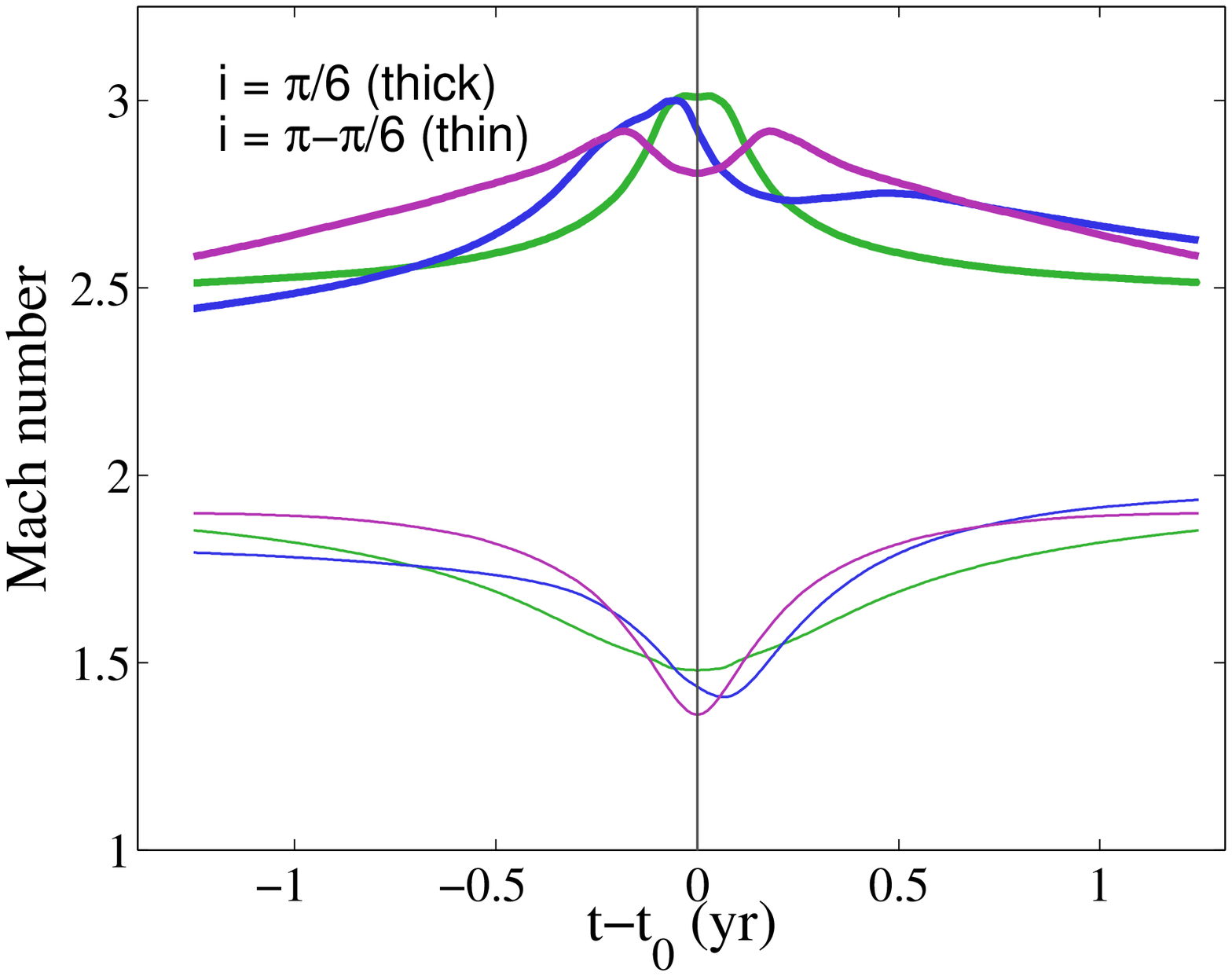}}
\caption{Profiles of (from top to bottom) density, temperature, magnetic 
to gas pressure ratio, magnetic pressure and Mach number along the cloud orbit for
inclinations: $i=\pi/3$ (left), $\pi/4$ (middle) and $\pi/6$ (right
column) as a function of time ($t_0$ denotes the epoch of pericenter of the shock
front). Thick and thin lines in the bottom-most
set of panels correspond to counter- and co-rotating orbits,
respectively. Colors denote the value of the argument of periapsis
$\omega$ and correspond to the orbits shown in
Fig.~\ref{f.trajectories}.} \label{f.evol1}
\end{figure*}

The top-most panels in Figure~\ref{f.evol1} show the variation of
density along the orbit. The columns correspond to inclination angles
$i=\pi/3$ (left, high inclination; but also $i=\pi - \pi/3$, since the
scalar fields are identical when reflected), $\pi/4$ (middle), and
$\pi/6$ (right, low inclination). In Figure~\ref{f.evol1}, colors denote
orbits with different values of the argument of periapsis
$\omega$. The horizontal axis shows time relative to the epoch of periastron $t_0$.
The green lines correspond to $\omega=0$, where the cloud
crosses the disk equatorial plane only once, at
pericenter. Consequently, all these density curves have a single
maximum at $t=t_0$. Blue curves correspond to $\omega=\pi/4$, where
the cloud crosses the disk plane twice. For the high inclination case
there are two local maxima. The first one corresponds roughly to the
first crossing of the equatorial plane ($t=t_0-0.12$), while the second
occurs much earlier than the second crossing ($\sim t_0+0.3$ vs
$t_0+1.8$) because of the fact that the decrease of density with
increasing radius dominates over the effect of the latitude
away from the midplane. The second peak is not visible for low 
inclinations because in these cases the cloud never goes very far 
from the mid-plane in between the two crossings. For the same reason, 
all the density curves in the last column (low inclination case) lie 
almost on top of each other. Indeed, in the limit $i\to0$
(which corresponds to the orbital plane coinciding with the disk 
plane), $\omega$ will have no impact on any of the curves.

The second row of panels shows the variation of temperature along the
cloud orbit. As discussed in Section~\ref{s.disk}, at a given radius
the temperature is the lowest at the mid-plane and increases towards
the polar axis. As a result, for the highly inclined orbit shown in
the left panel, the green curve in which the orbit crosses the disk
plane exactly at pericenter has twin peaks, a few months before
and after pericenter. This is when the cloud comes closest to the
polar axis and encounters the hottest gas. The opposite behavior is 
seen for the purple curve with two symmetric disk plane crossings. 
Here, there is a single peak in the temperature exactly at pericenter, 
when the cloud moves away from the mid-plane.  As in the case of the 
density, for orbits with low inclinations, all the curves are similar.

The third row of panels shows the magnetic to gas pressure
ratio. Close to the mid-plane this parameter is approximately equal to
$0.1$, while closer to the poles it takes on larger
values. Qualitatively, the behavior is similar to that of the temperature,
except that, being a ratio of pressures, there is no variation with
radius. The shapes of the various curves in the different panels are
then easy to understand.  The fourth row of panels shows the magnetic
pressure $P_{\rm mag}=B^2/8\pi$. This quantity does vary with radius
and behaves quite a lot like the temperature.  Thus, the curves show
striking similarity to those shown in the second row of panels.

Finally, the bottom row of panels shows the cloud Mach number with
respect to the disk gas. Here two sets of curves are plotted for each
choice of $i$ and $\omega$, thick curves for counter-rotating orbits
(inclination $=i$) and thin curves for co-rotating orbits (inclination
$=\pi-i$). As expected, the Mach number is lower for
co-rotating orbits. Also, the difference between counter- and
co-rotating orbits increases for lower inclinations (the two curves
coincide for an orbit perpendicular to the disk plane, $i=\pi/2$). The
Mach number for counter-rotating orbits lies in the range $1.8-3.0$
while for co-rotating orbits it is in the range $1.3-2.0$. Most
importantly, we find that the Mach number is always greater than unity
in all cases. This means that a bow shock is expected to form for every
orientation of the orbit and thus we always expect some level of
excess synchrotron emission from shock-accelerated electrons.

\subsection{Electron acceleration in the bow shock of G2} \label{s.acceleration}
A variety of astrophysical evidence suggests that particles are
accelerated efficiently via the Fermi and shock-drift acceleration
mechanisms in shocks (e.g., Blandford \& Eichler 1987). Typically,
these processes give rise to a non-thermal power-law tail in the
energy distribution of the particles.  The parameters of the bow shock
of G2 correspond to an interesting regime that has not been well
studied. The shock is non-relativistic, but the upstream electrons are
quasi-relativistic (assuming thermal equilibrium between electrons and
protons), with temperature varying between $T\simeq10^{8.0} K$ and
$T\simeq10^{8.7} K$ along the cloud orbit (see the second row of
panels in Fig.~\ref{f.evol1}). As a result of the relatively hot
temperature in the accretion flow, the shock has a modest Mach number,
ranging between $\Em \approx 1.5$ and $\Em \approx 3$ (see the
bottom-most row of panels in Fig.~\ref{f.evol1}).

We studied the acceleration of electrons in the bow shock of G2 by
means of two-dimensional first-principles numerical simulations, with
the particle-in-cell code TRISTAN-MP (Spitkovsky 2005). The simulation
setup parallels closely the one employed by Riquelme
\& Spitkovsky (2011) and \cite{narayan+12a}, with the magnetic field 
lying initially in the simulation plane, oriented at an oblique angle
with respect to the flow velocity. We fixed the angle
between the upstream field and the shock velocity to be
$\psi=63^\circ$, and the ratio of magnetic to gas pressure in the
accretion flow to be $P_{\rm mag}/P_{\rm gas}=0.1$. Below, we argue
that the same acceleration physics operates across a wide range of
magnetic obliquities, and for ratios of magnetic to gas pressure as
large as $P_{\rm mag}/P_{\rm gas}=0.5$ (see the third row of panels in
Fig.~\ref{f.evol1}). We explored the dependence of our results on the
pre-shock temperature over the range $10^{8.1}K$ up to $10^{8.7}$ K,
as appropriate for the region closest to the cloud pericenter where
most of the emission will be produced, and on the shock Mach number,
which we varied between $2$ and $3.5$.

For computational convenience, we chose a reduced mass ratio $m_{\rm
p}/m_{\rm e}=100$, but we verified that our results remain the same
for larger mass ratios, when all the physical quantities are scaled
appropriately. Specifically, we ran simulations spanning the range
$m_{\rm p}/m_{\rm e}=25-400$, fixing the pre-shock electron
temperature (equal to the proton temperature), the shock Mach number
and the ratio of magnetic to gas pressure, and we measured the time in
units of the inverse proton cyclotron frequency $\omega_{\rm
ci}^{-1}$. We also checked the convergence of our results with respect
to the spatial resolution\footnote{We resolved the electron skin depth
$c/\omega_{\rm pe}$ with 10 computational cells, and the electron
Debye length $\lambda_{\rm D,e}= \sqrt{k T/m_{\rm e}
c^2}\,c/\omega_{\rm pe}$ with a few cells.} and the number of
computational particles per cell (up to 128 per cell).

We find that at the shock, a fraction of the incoming electrons are
reflected backward by the shock-compressed magnetic field (Matsukiyo
et al. 2011) or by scattering off of electron whistler waves excited
in the shock transition layer (Riquelme \& Spitkovsky 2011). For
quasi-relativistic electron temperatures, the reflected electrons are
fast enough to remain ahead of the shock, resisting advection
downstream by the oblique pre-shock field. As the shock-reflected
electrons gyrate around the shock, they get energized by shock-drift
acceleration (e.g., Begelman \& Kirk 1990) and form a local
non-thermal population, just upstream of the shock. The constraint
that the reflected electrons be fast enough to outrun the shock along
the oblique field, thus participating in the shock-drift acceleration,
can be recast as an upper limit on the field obliquity $\psi$. If
the characteristic electron thermal velocity is $v_{\rm th,e}=\sqrt{k
T/m_{\rm e}}$ and the shock velocity is $v_{\rm sh}$, electron
acceleration is possible only if $\psi\lesssim \psi_{\rm crit}$,
where the critical obliquity angle $\psi_{\rm crit}$ satisfies
$v_{\rm th,e}\cos\psi_{\rm crit}= v_{\rm sh}$. So, the criterion for
efficient electron acceleration can be rewritten as
\begin{equation}\label{eq:thetacrit}
\psi\lesssim\psi_{\rm crit}\simeq\arccos
\left(\Em \sqrt{m_{\rm e}/m_{\rm p}}\right)~,
\end{equation}
showing that, even for our largest Mach number ($\Em=3.5$), electron
acceleration will be a common by-product of the shock evolution, being
allowed for all angles $\psi\lesssim85^\circ$, if we take the
realistic mass ratio $m_{\rm p}/m_{\rm e}=1836$. For our reduced mass
ratio $m_{\rm p}/m_{\rm e}=100$, we chose the obliquity angle
$\psi=63^\circ$ so that the constraint $\psi\lesssim\psi_{\rm
crit}$ required for efficient electron acceleration is satisfied for
all the shock Mach numbers that we investigated.

\begin{figure*}
\centering
 \includegraphics[scale=0.535]{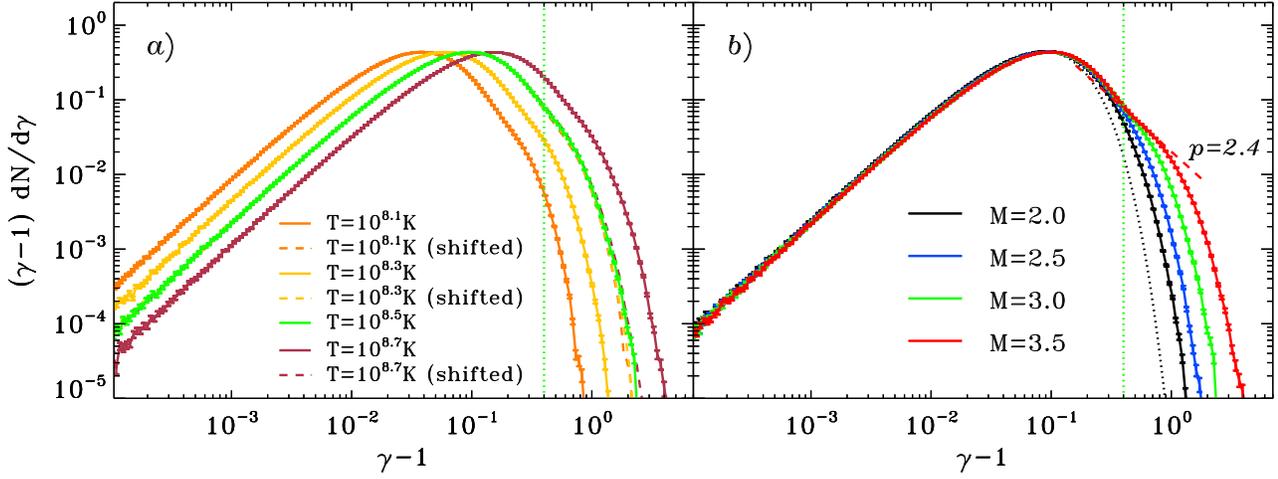}
\caption{Electron energy spectrum just upstream of the shock, 
normalized to the pre-shock electron density, for parameters relevant
to the orbit of G2. (Left) Electron energy spectrum at $\omega_{\rm
ci}t=10.8$ for different upstream temperatures, but fixed Mach number
$\Em=3$. The spectra plotted with dashed lines have been shifted along
the $x$-axis by $(T/10^{8.5} K)^{-1}$, to normalize the energy scale.
(Right) Electron energy spectrum at $\omega_{\rm ci}t=10.8$ for
different Mach numbers, but fixed electron temperature
$T=10^{8.5}K$. In both panels, the vertical green dotted line marks the
location of the low-energy end of the electron power-law tail, whose
slope $p\simeq2.4$ is indicated as a red dashed line in the right panel. 
In the right panel, the black dotted curve shows a Maxwellian distribution 
with $T=10^{8.5}K$.}  \mbox{}
\label{xinyi} 
\end{figure*}

As demonstrated by \cite{narayan+12a} for a pre-shock flow with
temperature $T=10^9K$, the electrons participating in the shock-drift
acceleration process are accelerated ahead of the shock into a power-law 
tail with a slope $p\simeq2.2$ and containing roughly $5\%$ of the incoming 
electrons (i.e., the so-called acceleration efficiency is
$\eta\simeq0.05$).\footnote{With time, a power-law tail of similar
normalization and slope as the pre-shock spectrum will appear in the
post-shock electron distribution.} The upper energy cutoff of the
pre-shock electron spectrum steadily increases and the slope becomes
flatter over time, suggesting that the distribution will asymptote at
late times to a power-law tail with $p\lesssim 2.2$ extending to very
large electron energies (we refer to the left panel of Fig.~2 in
\cite{narayan+12a} for details on the temporal evolution of the
electron spectrum). The electron energization at early times is
governed by the shock-drift mechanism, but the diffusive Fermi process
is expected to dominate at late times. We find that the counter-streaming 
between the incoming flow and the shock-reflected electrons
triggers the Weibel filamentation instability ahead of the shock
(Weibel 1959, Medvedev \& Loeb 1999), with the wavevector
perpendicular to the pre-shock field.\footnote{We point out that the
Weibel mode can only be captured by means of multi-dimensional
simulations and was, therefore, absent in the one-dimensional
experiments of Matsukiyo et al. (2011).} By scattering off of the
turbulence generated by the Weibel instability, the electrons
pre-accelerated by the shock-drift mechanism will be injected into the
Fermi process.

If the electron acceleration in our simulations is still primarily
controlled by the shock-drift mechanism, the electron energy gain
$\Delta\gamma\, m_{\rm e} c^2$ during a time $\Delta t$ will be
\begin{equation}
\Delta\gamma\, m_{\rm e} c^2\sim q \,E\, v_{\rm e/\!/} \,\Delta t~,
\end{equation}
where $E$ is the strength of the background motional electric field
(originating from the drift of the upstream frozen-in magnetic field
toward the shock) and $v_{\rm e/\!/}$ is the component of the electron
velocity along the electric field. By taking $v_{\rm e/\!/}\sim v_{\rm
th,e}$ (or a few times larger than $v_{\rm th,e}$), we can rewrite the
expression above as
\begin{equation}\label{eq:dgamma}
\frac{\Delta\gamma\, m_{\rm e} c^2}{k T}\sim \Em \sin\psi 
\sqrt{\frac{m_{\rm p}}{m_{\rm e}}}\; \omega_{\rm  ci}\,\Delta t~,
\end{equation}
which can be used to interpret the dependence of our results on the
pre-shock conditions, as we now describe.

In Figure~\ref{xinyi}, we show how the electron spectrum ahead of the
shock depends on the upstream temperature (left panel, with fixed Mach
number $\Em=3$) and on the shock Mach number (right panel, with fixed
temperature $T=10^{8.5}K$). In the left panel, we present the electron
spectrum ahead of the shock for four values of the flow temperature:
$T=10^{8.1} K$ (orange solid curve), $T=10^{8.3} K$ (yellow solid curve), 
$T=10^{8.5} K$ (green solid curve) and $T=10^{8.7} K$ (purple solid curve).  
The location of the thermal peak in the electron spectrum scales linearly 
with the pre-shock temperature. When we rescale our spectra to account for 
this effect (i.e., we shift them along the $x$-axis by $(T/10^{8.5} K)^{-1}$), 
the resulting distributions nearly overlap (compare the green solid line
with the three dashed curves). The electron power-law tail
starts at $\gamma-1\simeq\xi\, kT/m_{\rm e}c^2$, where $\xi\simeq7.5$
is nearly insensitive to the pre-shock temperature (vertical green
dotted line), and the tail contains about $5\%$ of the incoming electrons. 
The fact that electron spectra for different upstream temperatures are 
identical, modulo an overall shift in the energy scale, can be easily 
understood from Eqs.~(\ref{eq:thetacrit}) and (\ref{eq:dgamma}). The 
former demonstrates that the condition for efficient injection into the
acceleration process does not depend explicitly on the flow
temperature, at fixed Mach number. In addition, Eq.~(\ref{eq:dgamma})
shows that the temporal evolution of the high-energy spectral cutoff,
normalized to the thermal peak, is insensitive to the upstream
temperature.

The dependence of the electron spectrum on the shock Mach number is
illustrated in the right panel of Figure~\ref{xinyi}, showing that the
non-thermal tail is most pronounced in high-$\Em$ shocks. Yet, some
evidence for electron acceleration to non-thermal energies is present
even in the spectrum of $\Em=2$ shocks (black solid line), whose
high-energy end clearly departs from a Maxwellian distribution (black
dotted line). We point out that the main difference among shocks with
different Mach number is not in the normalization of the non-thermal
tail or in the location of its low-energy end, but in the extent of
the power-law component. Regardless of the Mach number, the tail
contains a fraction $\sim5\%$ of the incoming electrons, and its
low-energy end is at $\gamma-1\simeq\xi\, kT/m_{\rm e}c^2$, where
$\xi\simeq7.5$ is nearly insensitive to the shock Mach number
(vertical green dotted line).  The fact that the acceleration efficiency
is independent of the Mach number follows from Eq.~(\ref{eq:thetacrit}), 
since $\psi\lesssim\psi_{\rm crit}$ for all of our choices of 
$\Em$.\footnote{In shocks with higher Mach
number (and fixed $\psi$), the condition $\psi\lesssim\psi_{\rm
crit}$ will break down, fewer electrons will be reflected back at the
shock (Matsukiyo et al. 2011), so a smaller fraction of the incoming
electrons will be injected into the shock-drift acceleration
process. This will approach the limit of cold upstream plasmas studied
by Riquelme \& Spitkovsky (2011).} Also, the increase in
the high-energy spectral cutoff with Mach number is in agreement with
Eq.~(\ref{eq:dgamma}), stating that the acceleration rate scales
linearly with $\Em$. It follows that, at a given time (in units of
$\omega_{\rm ci}^{-1}$), the spectrum of shocks with higher Mach
number will have stretched to more extreme energies. 

The non-thermal tail in the electron spectrum of $\Em=3.5$ shocks (red 
solid line) can be fitted as a power law of slope $p\simeq 2.4$ (red dashed
line). We expect the power-law tail to become flatter with time, as its 
high-energy cutoff increases, and the spectral slope $p$ should approach the 
value $p\simeq2.2$ found by \citet{narayan+12a} in a flow with $T=10^9K$. 
In shocks  with $\Em\lesssim3$ (black, blue and green curves), given the 
limited extent of the power-law tail, it is still premature to draw firm 
conclusions regarding the appropriate value of the power-law slope at late 
times. Below, we assume that, regardless of $\Em$, the power-law tail in 
steady state will have a slope $p=2.4$. As discussed above, this is likely 
to be a conservative upper limit, since the power-law index at late times 
is expected to approach the value $p\simeq2.2$ obtained by \citet{narayan+12a}.

We have also tested that neither the acceleration efficiency nor the
spectral slope depend significantly on the ratio of magnetic to gas
pressure (we have tried $P_{\rm mag}/P_{\rm gas}=0.1$, 0.3 and 0.5) or
the obliquity angle $\psi$, provided that $\psi\lesssim\psi_{\rm
crit}$. In fact, from Eqs.~(\ref{eq:thetacrit}) and (\ref{eq:dgamma}),
we expect that the ratio $P_{\rm mag}/P_{\rm gas}$ should not
appreciably change the spectral shape, whereas the field obliquity (in
the range $\psi\lesssim\psi_{\rm crit}$) should only affect the
evolution of the high-energy spectral cutoff, as illustrated in
Eq.~(\ref{eq:dgamma}).

\subsection{Radio lightcurves}
\label{s.lightcurves}

As we showed in the previous sections, the bow shock that will form 
ahead of the cloud G2 as it moves supersonically through the ambient disk gas
is likely to accelerate, via the mechanism described in the previous subsection, 
the already hot electrons to relativistic energies. The original thermal 
distribution of electrons will thus be modified and a high energy power law tail 
will develop. When these high energy electrons gyrate magnetic field present 
in the disk, they emit synchrotron radiation. As we will show below, under 
favorable circumstances, the emission from the shocked electrons can overwhelm 
the steady-state radio emission from Sgr~A$^*$.

In our calculations we assume the following parameters to calculate
the total number of accelerated electrons. We take the cloud cross
section (the transverse area of the shock perpendicular to the
relative velocity between the cloud and the accretion flow) to be
$A=\pi R^2$, with $R=10^{15} {\rm cm}$,\footnote{$R=10^{15} {\rm cm}$
is the value assumed by \cite{narayan+12a} and corresponds to $2/3$ of
the measured half-width at half maximum of the cloud
\citep{gillessen+12a}.  The choice of the effective cloud cross
section and the impact of changes in cross section along the the orbit
are discussed in Section~\ref{s.discussion}.}  and use an electron
acceleration efficiency equal to $\eta=0.05$. We take the low-energy
end of the power-law tail at $\gamma-1=\xi(kT/m_{\rm e}c^2)$, where
$\xi=7.5$ and $T$ is the electron temperature, and the slope of the
high-energy tail as $p=2.4$.

Whether or not the shocked electrons will remain close to the shock is
unclear.  We, therefore, consider two distinct limiting cases. In one
scenario, which we call the plowing model, we assume that all
relativistic electrons remain in the shocked gas and radiate in the
instantaneous shock-compressed magnetic field.  In the second
scenario, referred to as the local model, we assume that the shocked
electrons are left behind at their original position in the accretion
flow while the cloud and the shock move on, and that the electrons
radiate in the original uncompressed field.  We believe that the real
situation will be somewhere between these two extremes.

\subsubsection{Results from the plowing model}
\label{s.total}

In the plowing model, all the accelerated electrons remain in the
vicinity of the shock. Because of the long synchrotron cooling
timescale ($>100\,\rm yrs$, see Narayan et al. 2012a), there is no
loss of accelerated particles and hence the number of radiating
relativistic electrons at each energy increases monotonically with
time. However, this does not imply a monotonically increasing
luminosity since the synchrotron emissivity varies with the magnetic
field strength. The field in the shocked gas is proportional to the
field in the ambient accretion flow, which varies as a function of
position along the orbit. As discussed in the previous section (see
Fig.~\ref{f.evol1}), after an initial increase, the magnetic field
strength drops as the cloud moves away from the central black
hole. Therefore, while we expect an initial rapid increase in the
synchrotron luminosity (since both the number of relativistic
electrons and the magnetic field strength increase as the cloud
approaches periapsis), this will be followed by a slower decay (the
number of shocked electrons will continue to increase but the magnetic
field will drop with time).

To calculate the lightcurve, we pre-compute the positions and
velocities of the cloud as a function of time for given orbit angles
$i$ and $\omega$. For each differential time interval between $t$ and
$t+\Delta t$, we calculate the energy distribution of electrons
accelerated in that interval\footnote{The underlying assumption is
that the acceleration time is shorter than $\Delta t$, and that
electrons accelerated at earlier times do not modify the shock
structure at later times (the shock structure is only determined by
the properties of the pre-shock flow at that time).},
\begin{equation}
\label{e.dNdgamma1}
\frac{{\rm d}N}{{\rm d}\gamma}_{\rm local}=
A v \Delta t n_e \eta
\frac{(p-1)(\xi kT/m_{\rm  e}c^2)^{p-1}}{(\gamma-1)^{p}}\quad\text{for}\,\gamma-1>\xi
(kT/m_{\rm  e}c^2), 
\end{equation}
where $n_e$ is the local disk number density and $v$ is the velocity
of the cloud relative to the gas. The differential distributions are
summed to obtain the total distribution of electrons at any given
time according to
\begin{equation}
\frac{{\rm d}N}{{\rm d}\gamma}(t,\gamma)=\sum_{n=1}^{N(t)}
\frac{{\rm d}N}{{\rm d}\gamma}_{\rm local},
\end{equation}
where the sum goes over all the time steps up to the given time $t$.

The spectral synchrotron power is calculated using the standard
formula for a power-law distribution of electrons
\citep{rybicki-book},
\begin{equation}
P_{\nu}(t)=\frac{\sqrt{3}q^3CB}{m_{\rm e}c^2(p+1)}
\Gamma\left(\frac p4+\frac{19}{12}\right)\Gamma
\left(\frac p4-\frac1{12}\right)\left(\frac{2\pi m_{\rm e}c\nu}
{3qB}\right)^{-\frac{p-1}2},
\label{e.Pintr1}
\end{equation}
where 
\be
C=\frac{{\rm d}N}{{\rm d}\gamma}(t,\gamma=2).
\ee 

The strength of the magnetic field $B$ is given by 
\be 
B=\frac{(\hat\gamma+1){\cal M}^2}{(\hat\gamma-1){\cal
M}^2+2} B_{\rm disk}, 
\ee 
which accounts for the amplification of the ambient disk magnetic
field $B_{\rm disk}$ by a compression factor that depends
on the shock Mach number ${\cal M}$ and the adiabatic
index $\hat\gamma=5/3$. Note that only the component 
of the field perpendicular to the shock normal is amplified and 
that, for the sake of simplicity, we ignore the longitudinal component
here. Taking into account the distance to \sgra as $D=8.3\rm kpc$, 
we obtain the observed synchrotron flux as
\be 
F_\nu(t)=\frac{P_{\nu}(t)}{4\pi D^2}.  
\ee

\begin{figure*}
  \centering
\begin{tabular}{MMMM}
&$\omega=0$&$\omega=\pi/4$&$\omega=\pi/2$\\
\begin{sideways}$i=\pi/3$\end{sideways}&\hspace{-.3cm}
\subfigure{\includegraphics[width=.33\textwidth]{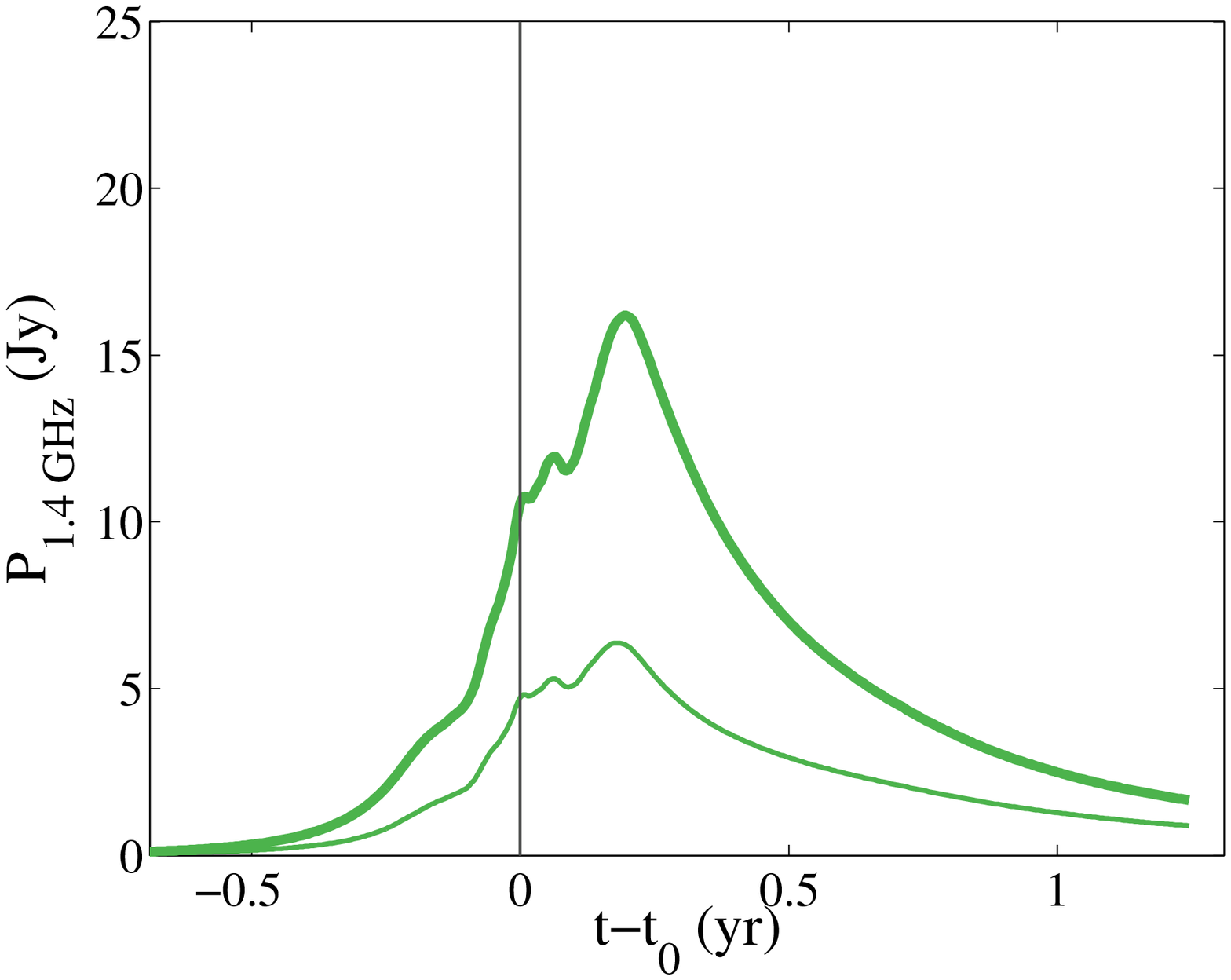}}&\hspace{-.36cm}
\subfigure{\includegraphics[width=.30\textwidth]{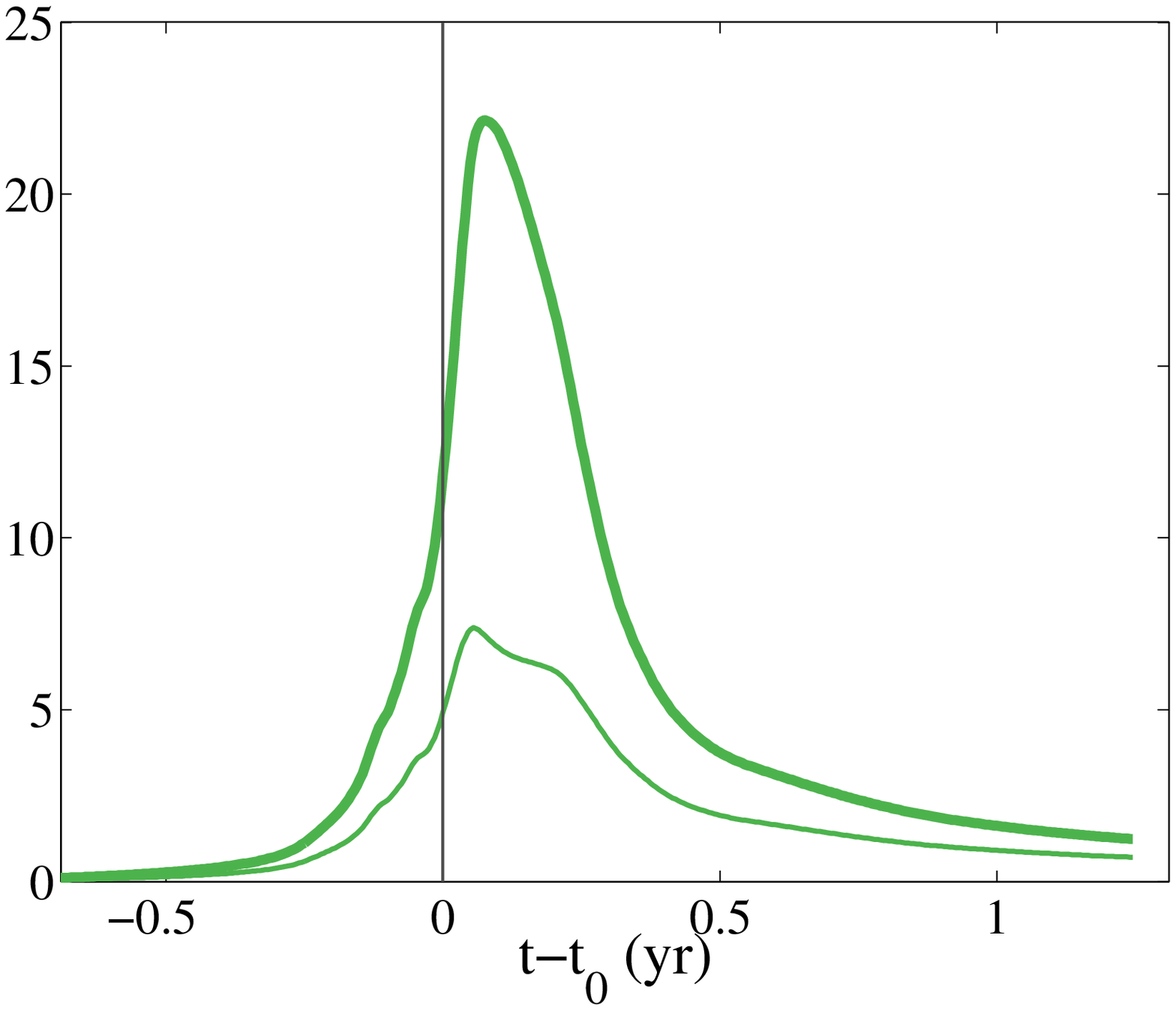}}&\hspace{-.36cm}
\subfigure{\includegraphics[width=.30\textwidth]{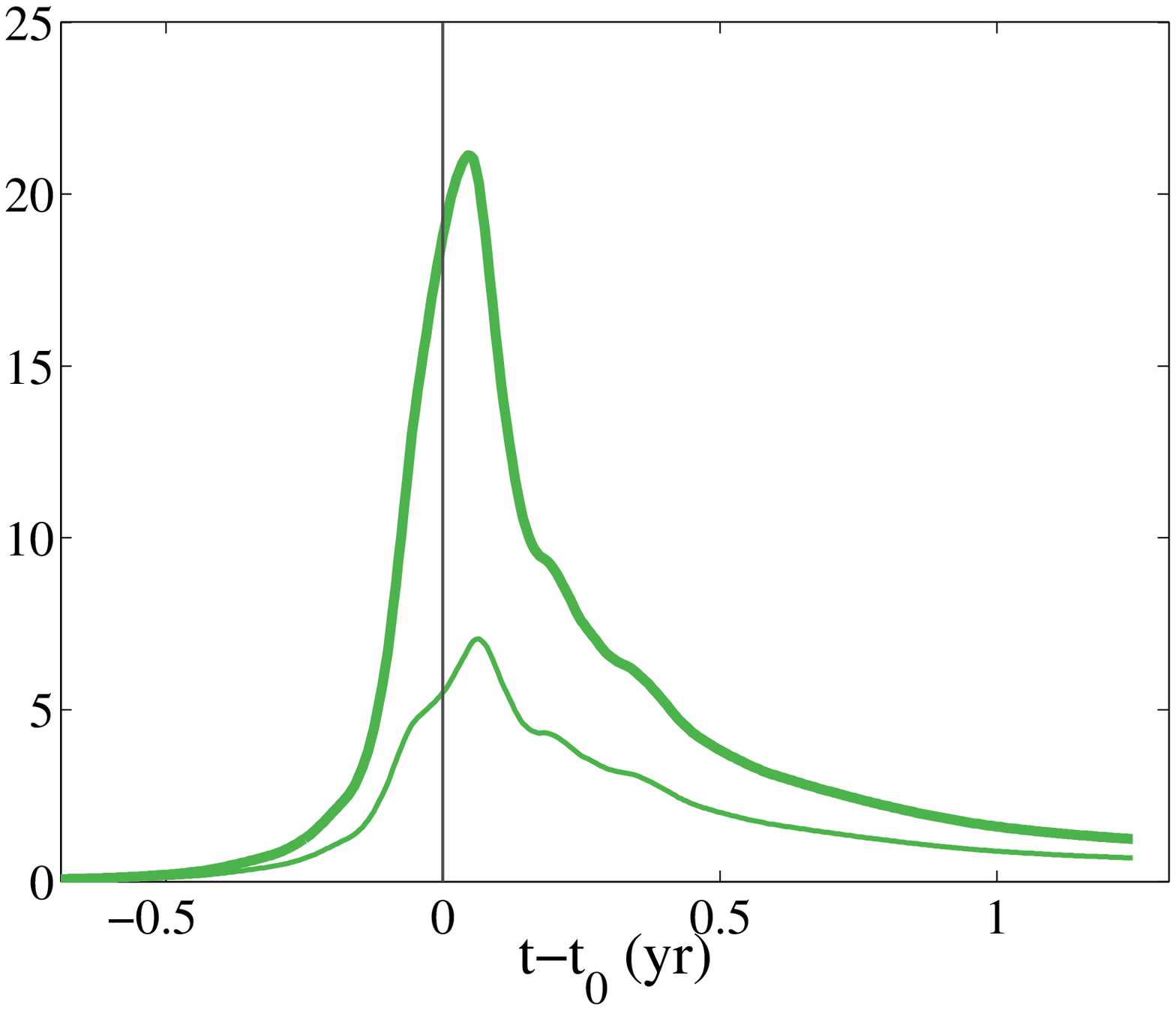}}\\
\begin{sideways}$i=\pi/4$\end{sideways}&\hspace{-.3cm}
\subfigure{\includegraphics[width=.33\textwidth]{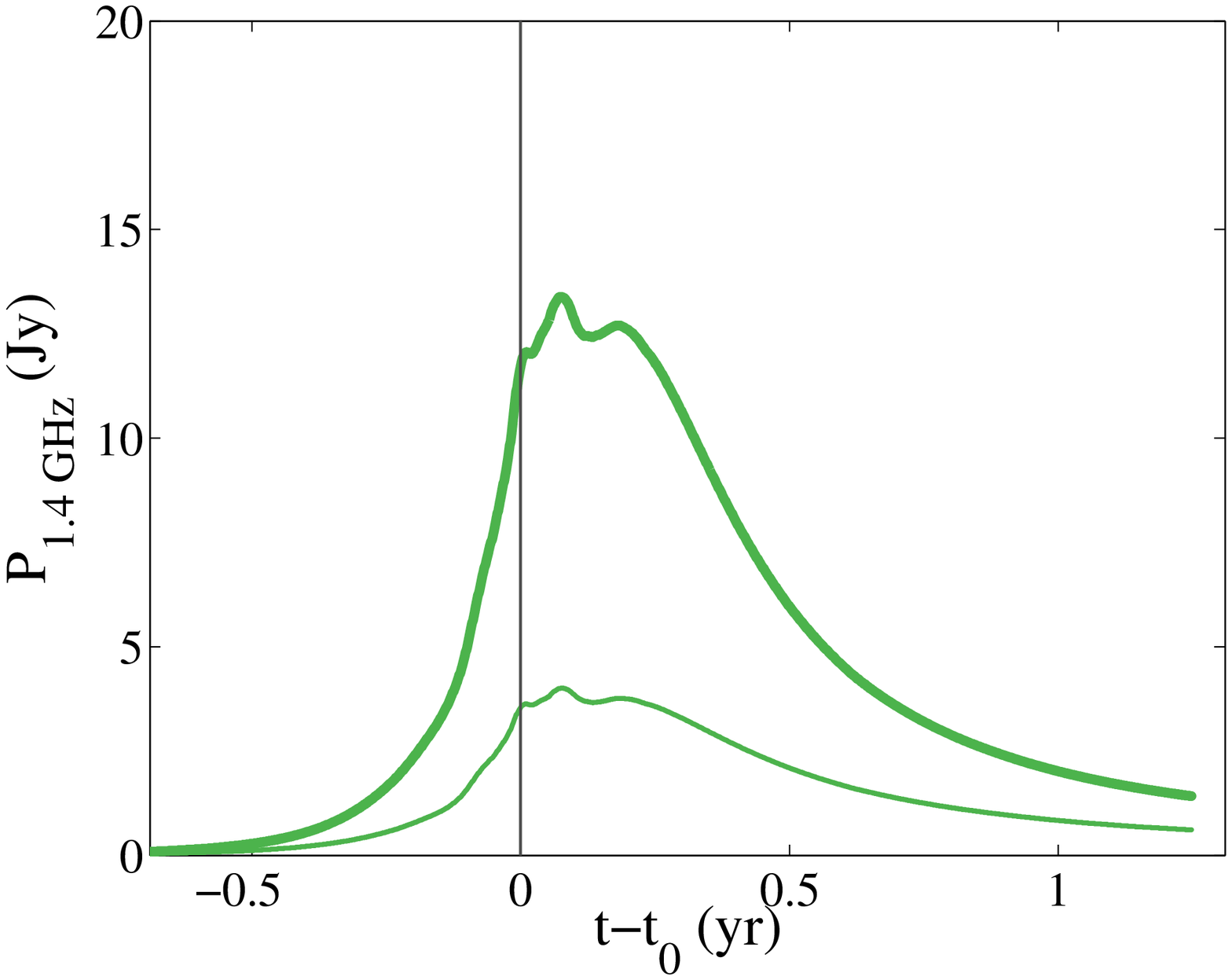}}&\hspace{-.36cm}
\subfigure{\includegraphics[width=.30\textwidth]{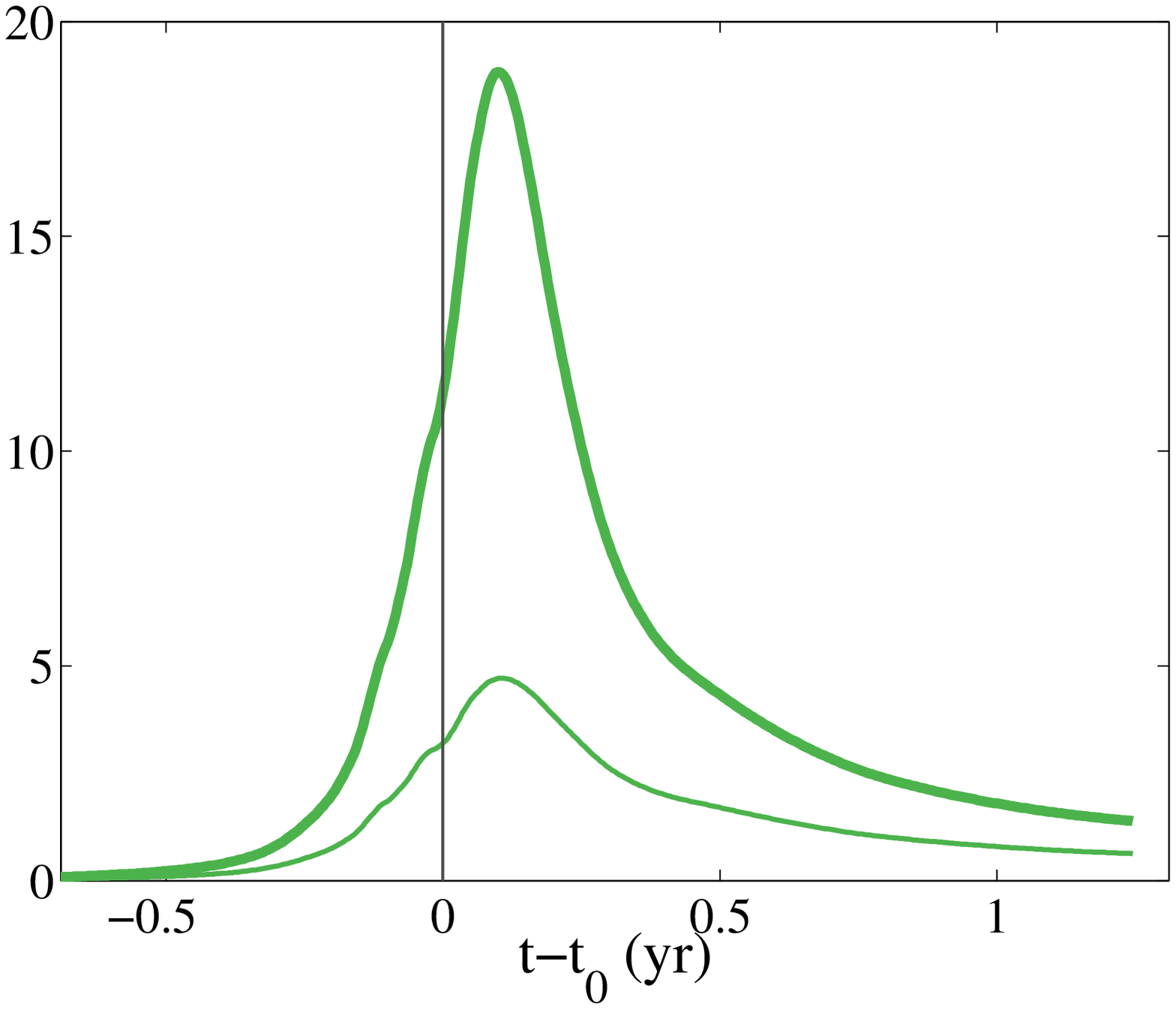}}&\hspace{-.36cm}
\subfigure{\includegraphics[width=.30\textwidth]{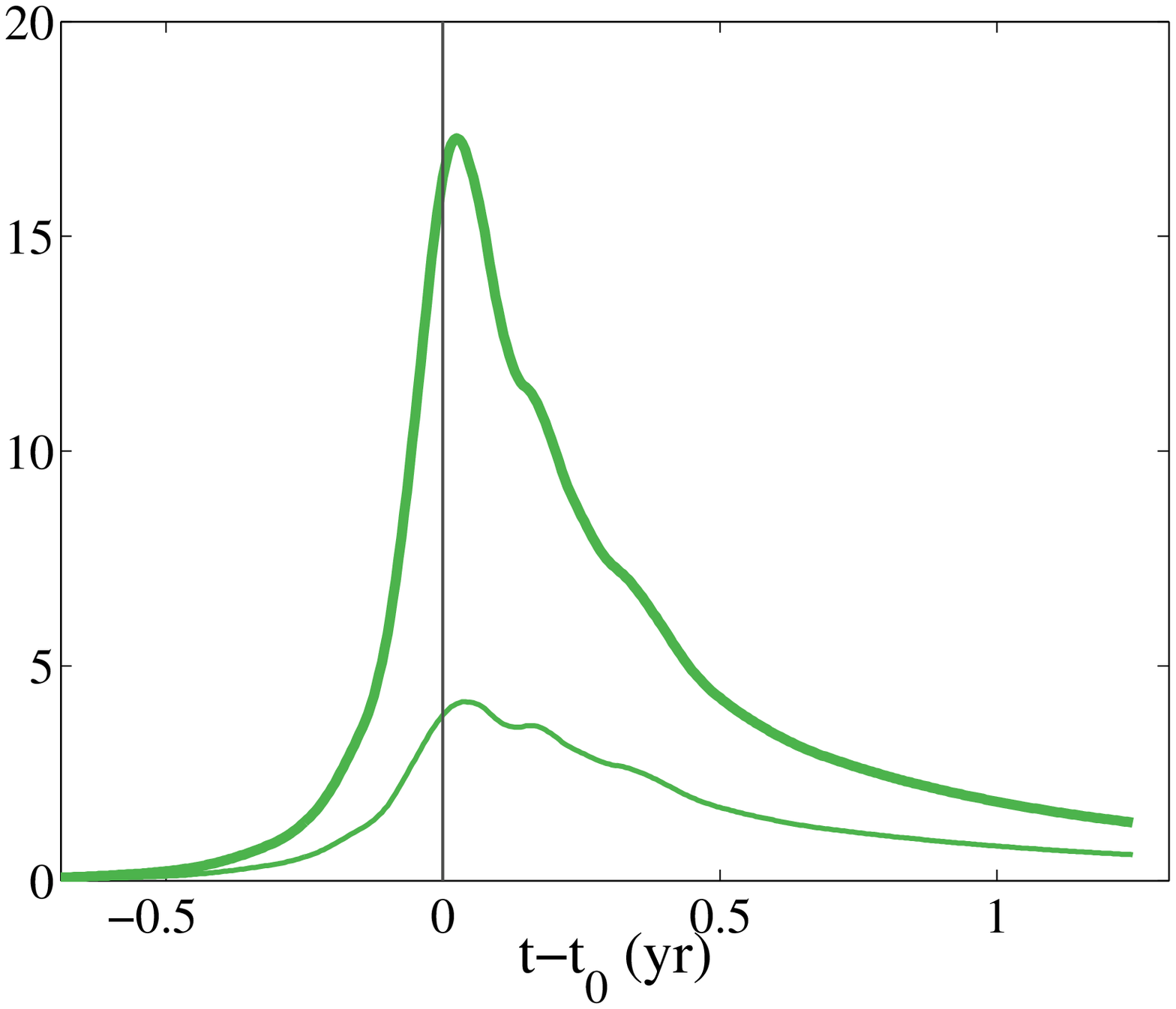}}\\
\begin{sideways}$i=\pi/6$\end{sideways}&\hspace{-.3cm}
\subfigure{\includegraphics[width=.33\textwidth]{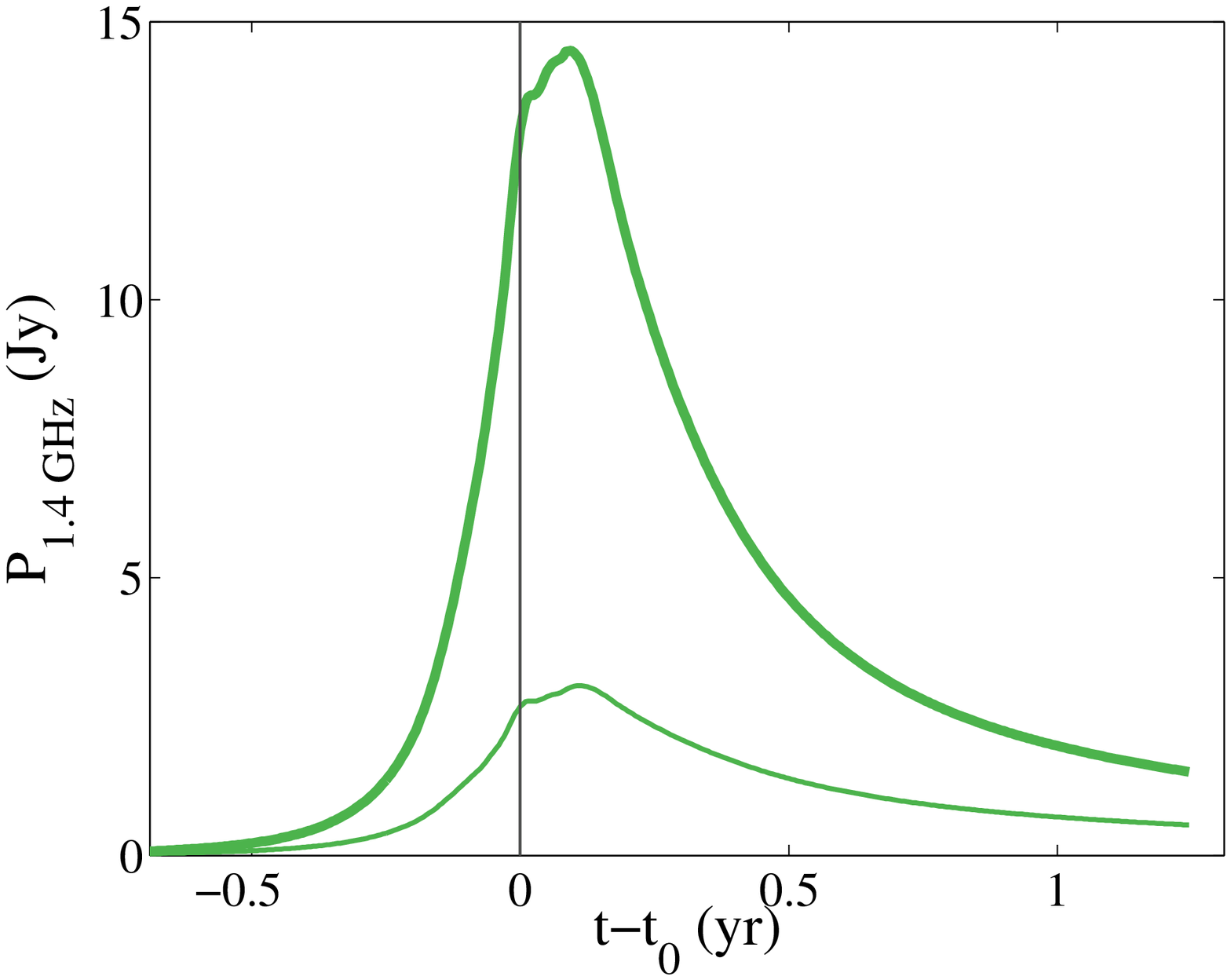}}&\hspace{-.36cm}
\subfigure{\includegraphics[width=.30\textwidth]{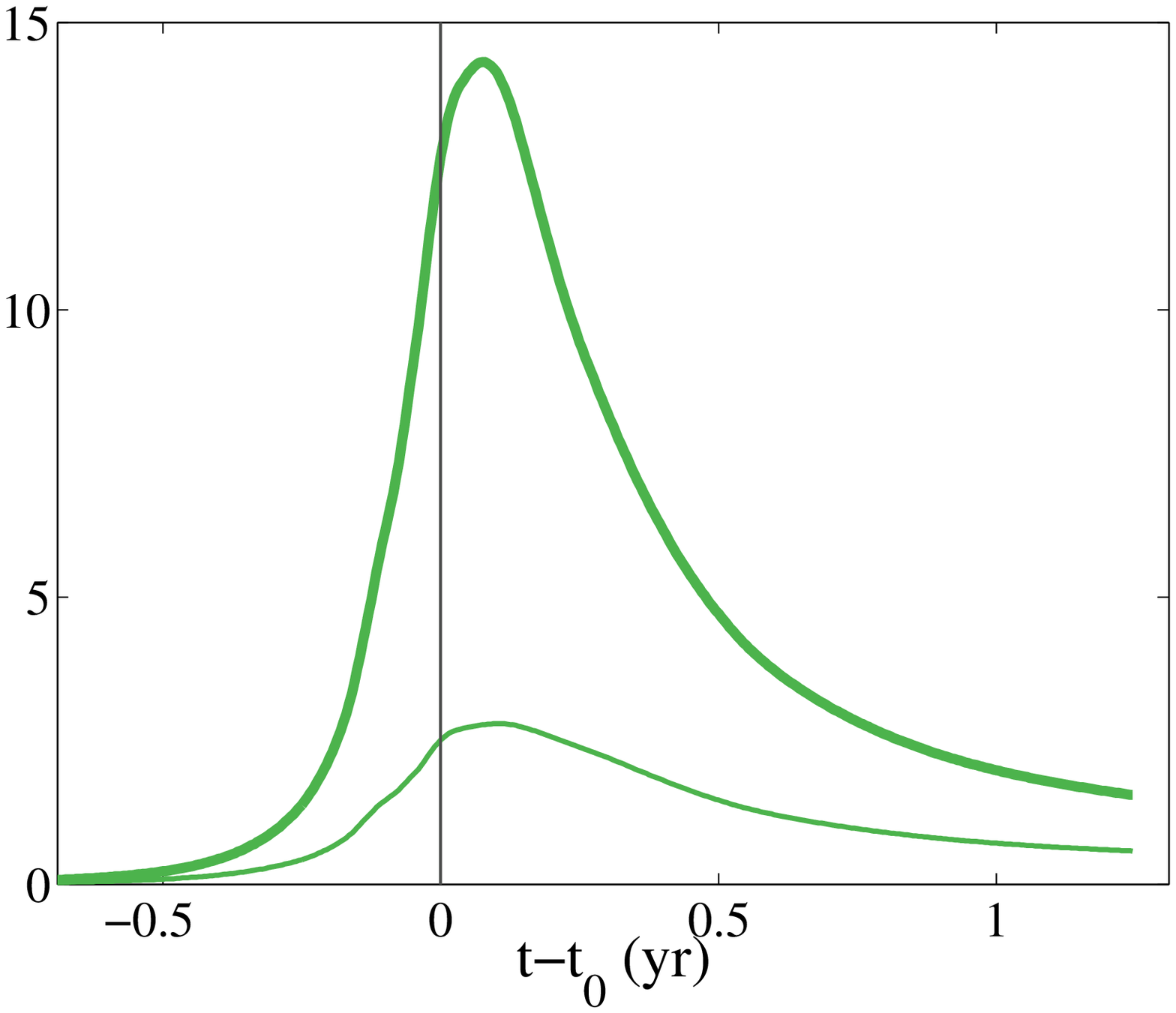}}&\hspace{-.36cm}
\subfigure{\includegraphics[width=.30\textwidth]{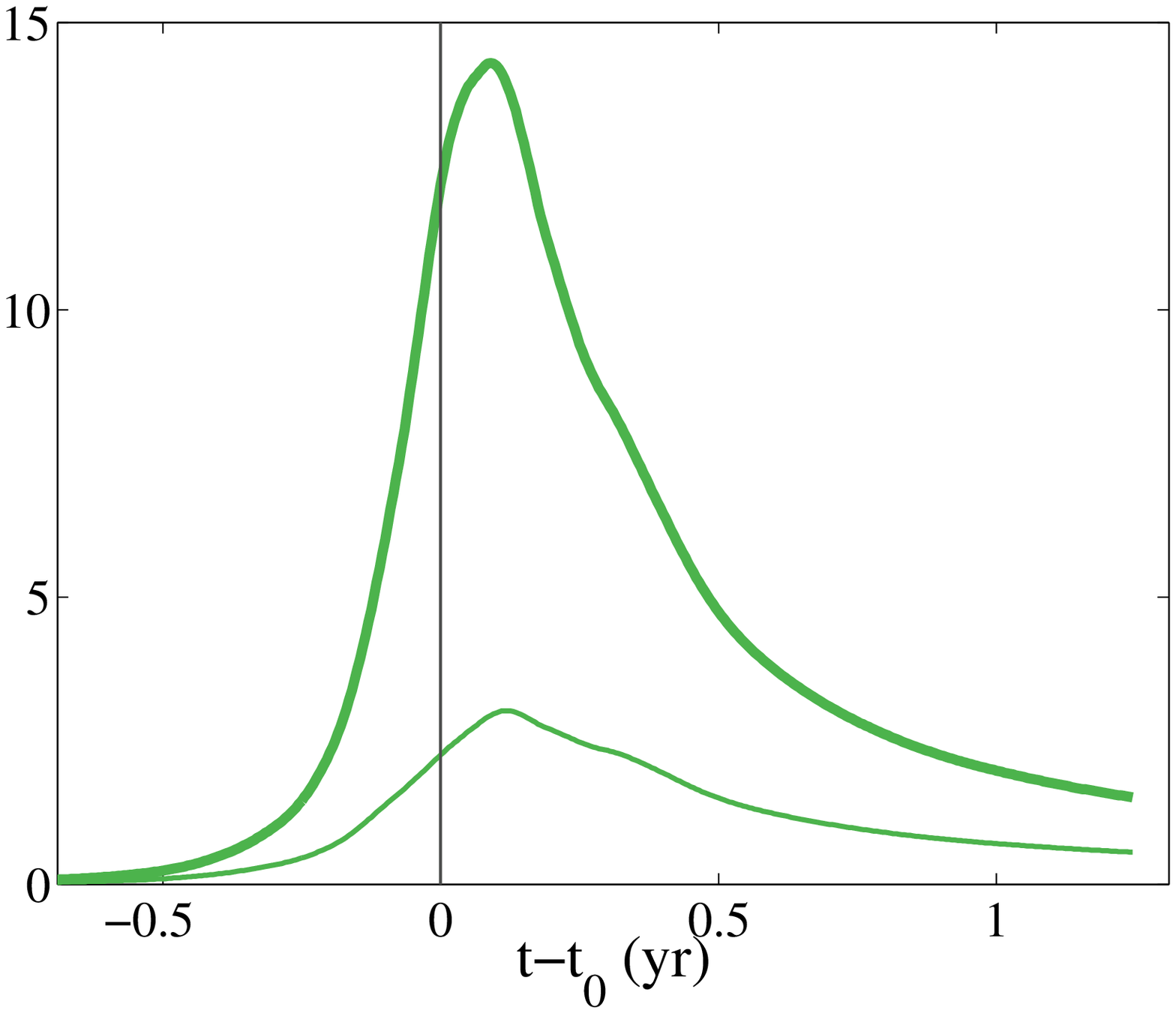}}\\
\end{tabular}

\caption{Radio lightcurves at 1.4 GHz for different orbit orientations 
assuming the plowing model for acceleration
(Section~\ref{s.total}). Rows correspond to inclination angles
$i=\pi/3$, $\pi/4$ and $\pi/6$ (top to bottom). Columns correspond to
the argument of periapsis $\omega=0$, $\pi/4$ and $\pi/2$ (left to
right). Thick and thin lines are for counter- and co-rotating orbits,
respectively. The vertical line corresponds to the bow shock epoch
of periastron $t_0$.}  \label{f.lc1}
\end{figure*}

Figure~\ref{f.lc1} shows a set of lightcurves calculated with the
above procedure for $\nu=1.4\,\rm GHz$ (where self-absorption is
negligible, see Section~\ref{s.spectra}). The rows correspond to
different inclination angles, starting from $i=\pi/3$ (top), through
$i=\pi/4$ (middle), to $i=\pi/6$ (bottom), while the columns
correspond, from left to right, to different arguments of periapsis,
$\omega=0$, $\pi/4$, and $\pi/2$.  The thick and thin lines refer to
counter- and co-rotating orbits, respectively. All the curves show the
expected behavior, i.e., a relatively sharp increase when the cloud
approaches pericenter, followed by a slower decay when the cloud and
the shocked electrons move to regions with a weaker magnetic
field. For all orbit inclinations, the radio emission is predicted
to start rising a few months before the shock reaches pericenter.
Co-rotating orbits produce significantly weaker emission due to the
lower relative velocity of the cloud with respect to the gas, which
results in a smaller number of accelerated electrons as well as a
weaker shock-compressed field (since the shock Mach number is
smaller). The maximal flux for counter-rotating orbits is in the range
$13-22 \,\rm Jy$, with the exact value depending on the orbit
orientation.  For co-rotating orbits, the range is $3-7 \,\rm
Jy$. In most cases, the maximal emission is occurs roughly a month
after the shock reaches periapsis. For some orbit orientations two
peaks are predicted in the light curve.

\subsubsection{Results from the local model}
\label{s.local}

In the local model, electrons are assumed to move out of the shocked
region and to remain in their original location in the accretion flow, 
radiating in the original uncompressed magnetic field. 
We neglect adiabatic expansion losses of the accelerated electrons, 
since the expansion time 
is typically larger than the evolution timescale 
of the cloud along its orbit. In this approximation, each accelerated
electron radiates under the same conditions throughout the entire
period of the cloud encounter; hence, the procedure described in the
previous section has to be modified. Equation~\ref{e.dNdgamma1}, which
describes the energy distribution of electrons accelerated in a given
time step, still holds, but instead of summing up the number of
electrons and calculating the emission, here we sum up the radiation
from electrons at each location that are accelerated in all the time
steps up to a given time $t$, i.e.,
\be\label{e.pnulocal}
P_{\nu}(t)=\sum_{n=1}^{N(t)}P_{\nu,\,\rm local}(t), 
\ee where
$P_{\nu,\, \rm local}(t)$ is calculated via Equation~(\ref{e.Pintr1}) but
with 
\be
C=\frac{{\rm d}N}{{\rm d}\gamma}_{\rm local}(t,\gamma=2).
\ee 
This last quantity corresponds to the number of accelerated electrons
in a given time step at time $t$ and $B=B_{\rm disk}$ is the original
uncompressed disk magnetic field.

\begin{figure*}
  \centering
\begin{tabular}{MMMM}
&$\omega=0$&$\omega=\pi/4$&$\omega=\pi/2$\\
\begin{sideways}$i=\pi/3$\end{sideways}&\hspace{-.3cm}
\subfigure{\includegraphics[width=.33\textwidth]{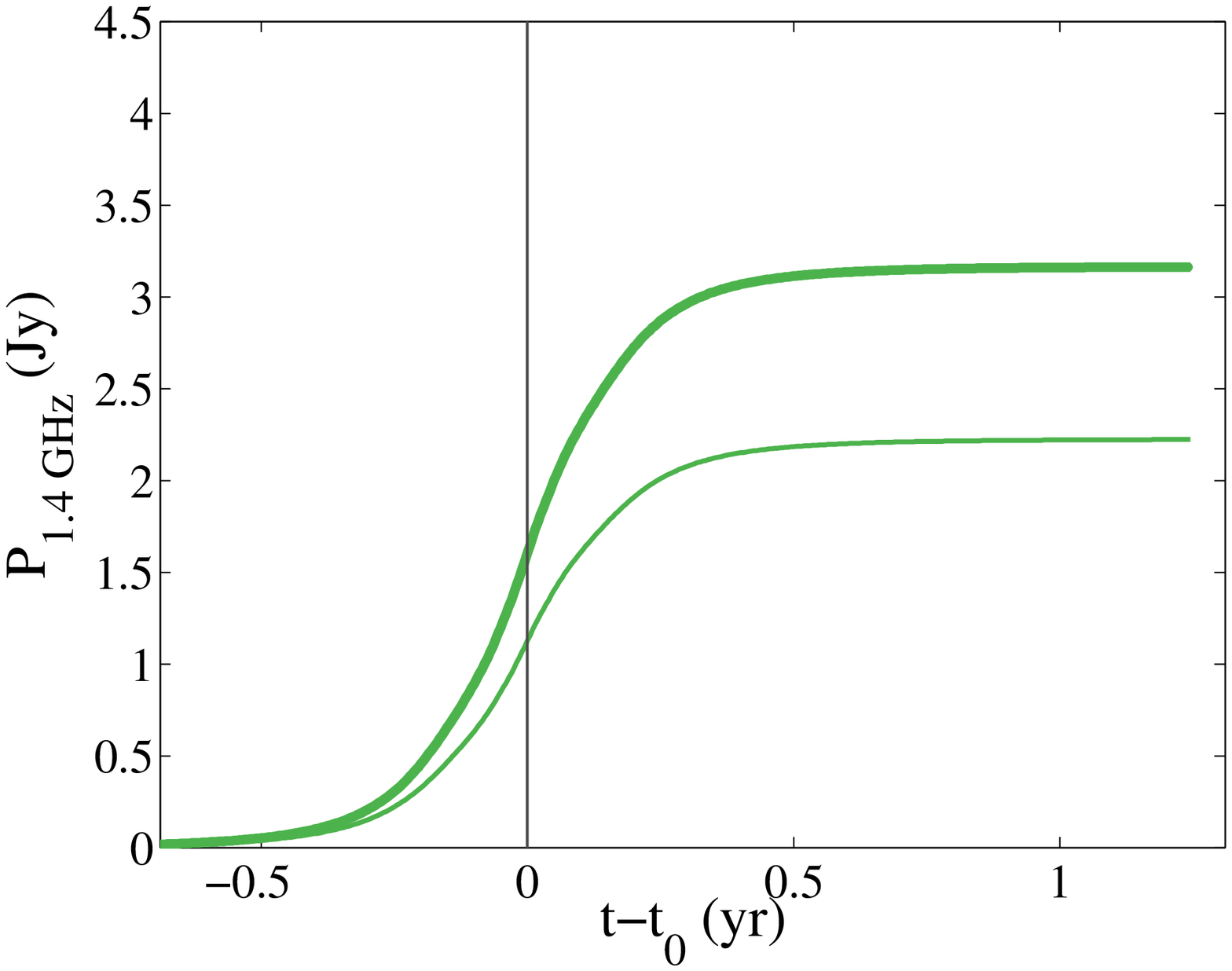}}&\hspace{-.36cm}
\subfigure{\includegraphics[width=.3\textwidth]{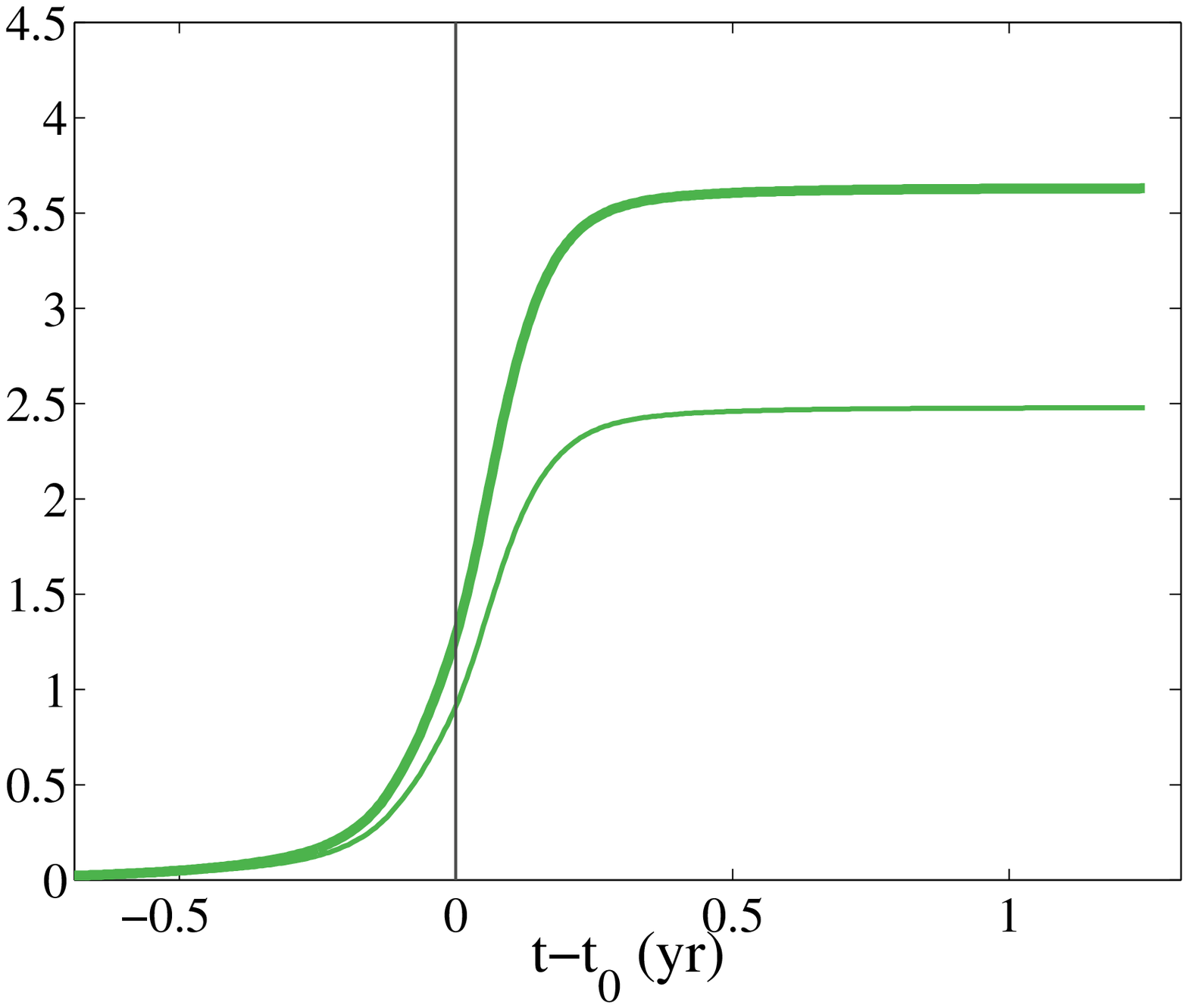}}&\hspace{-.36cm}
\subfigure{\includegraphics[width=.3\textwidth]{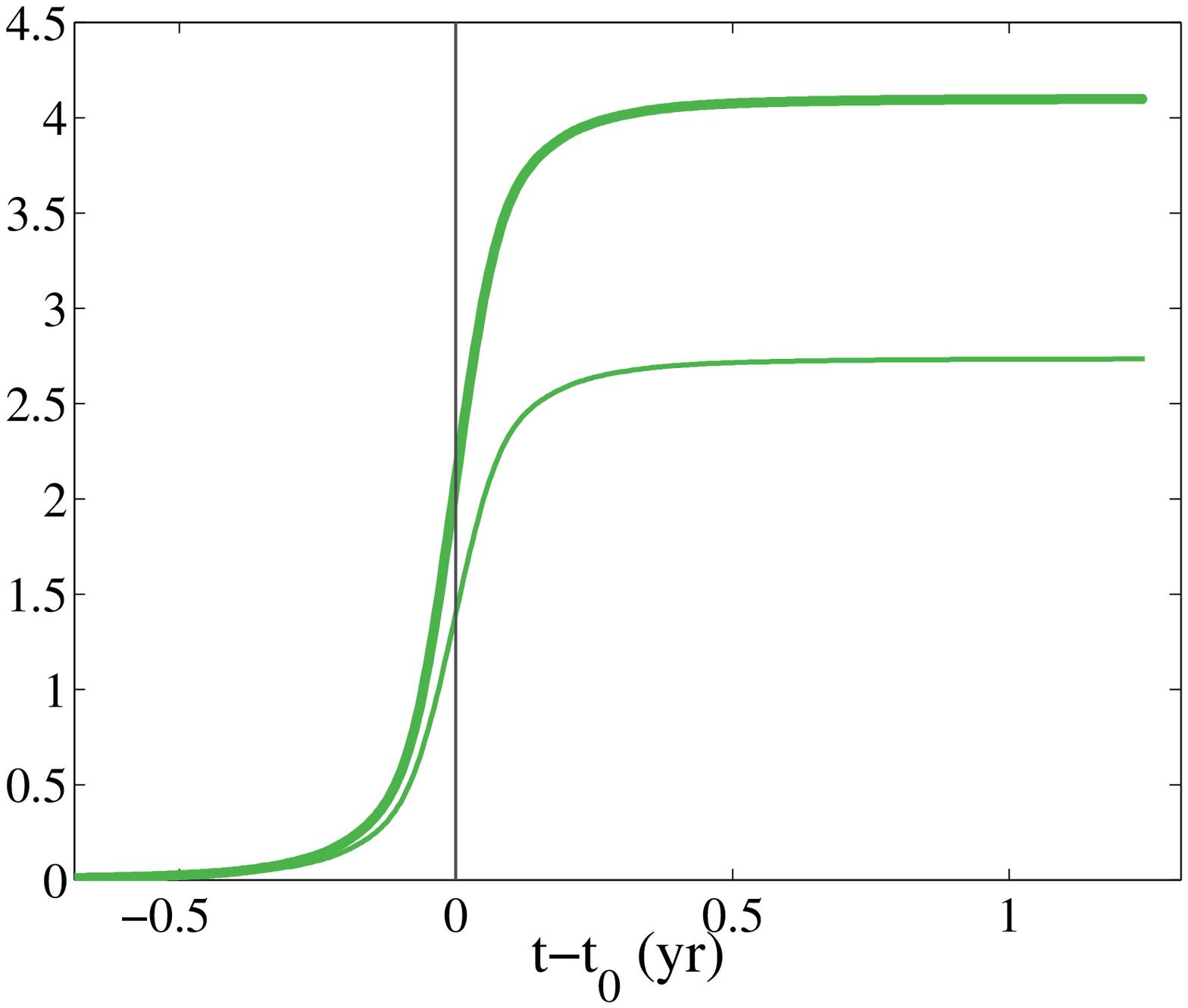}}\\
\begin{sideways}$i=\pi/4$\end{sideways}&\hspace{-.3cm}
\subfigure{\includegraphics[width=.33\textwidth]{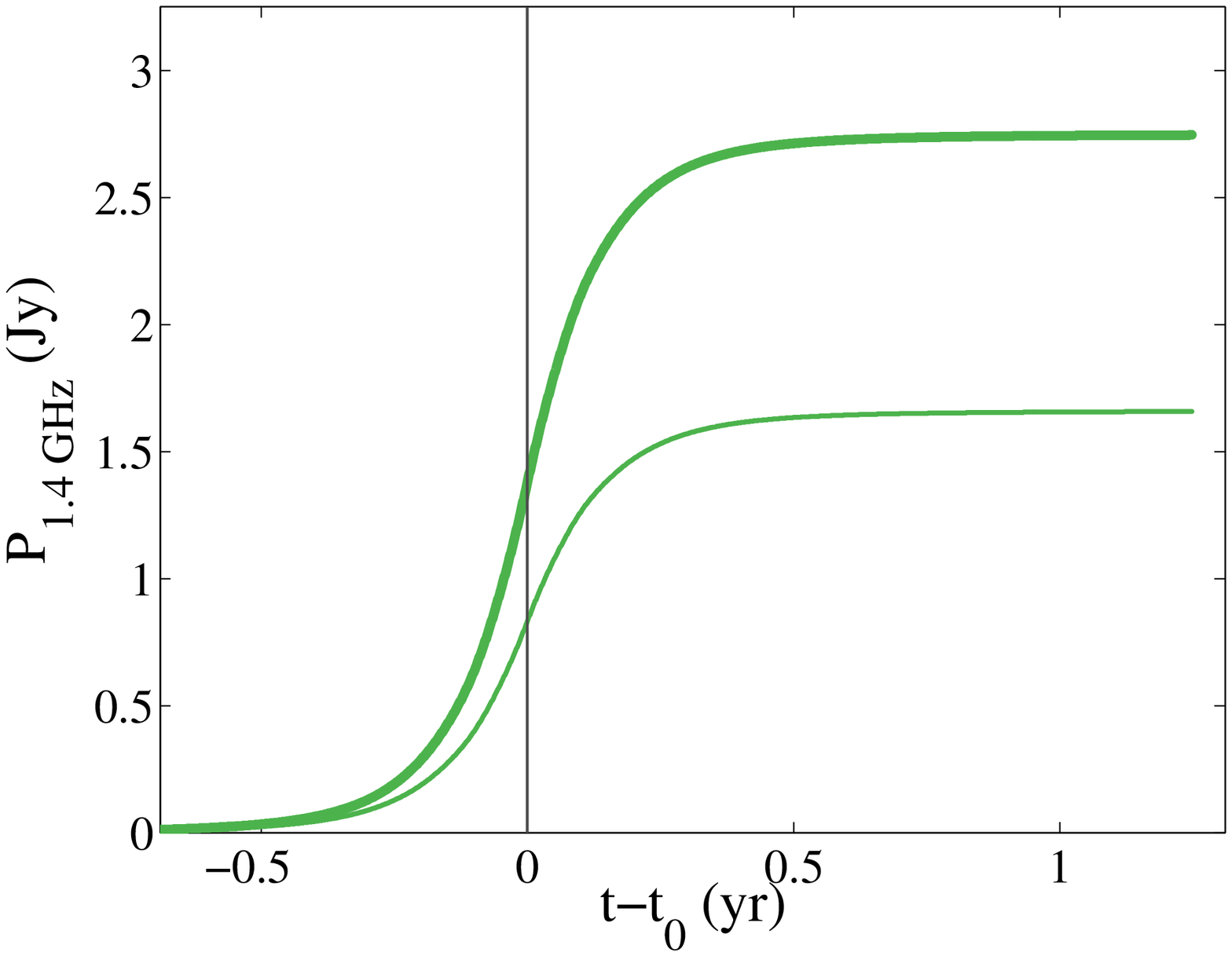}}&\hspace{-.36cm}
\subfigure{\includegraphics[width=.3\textwidth]{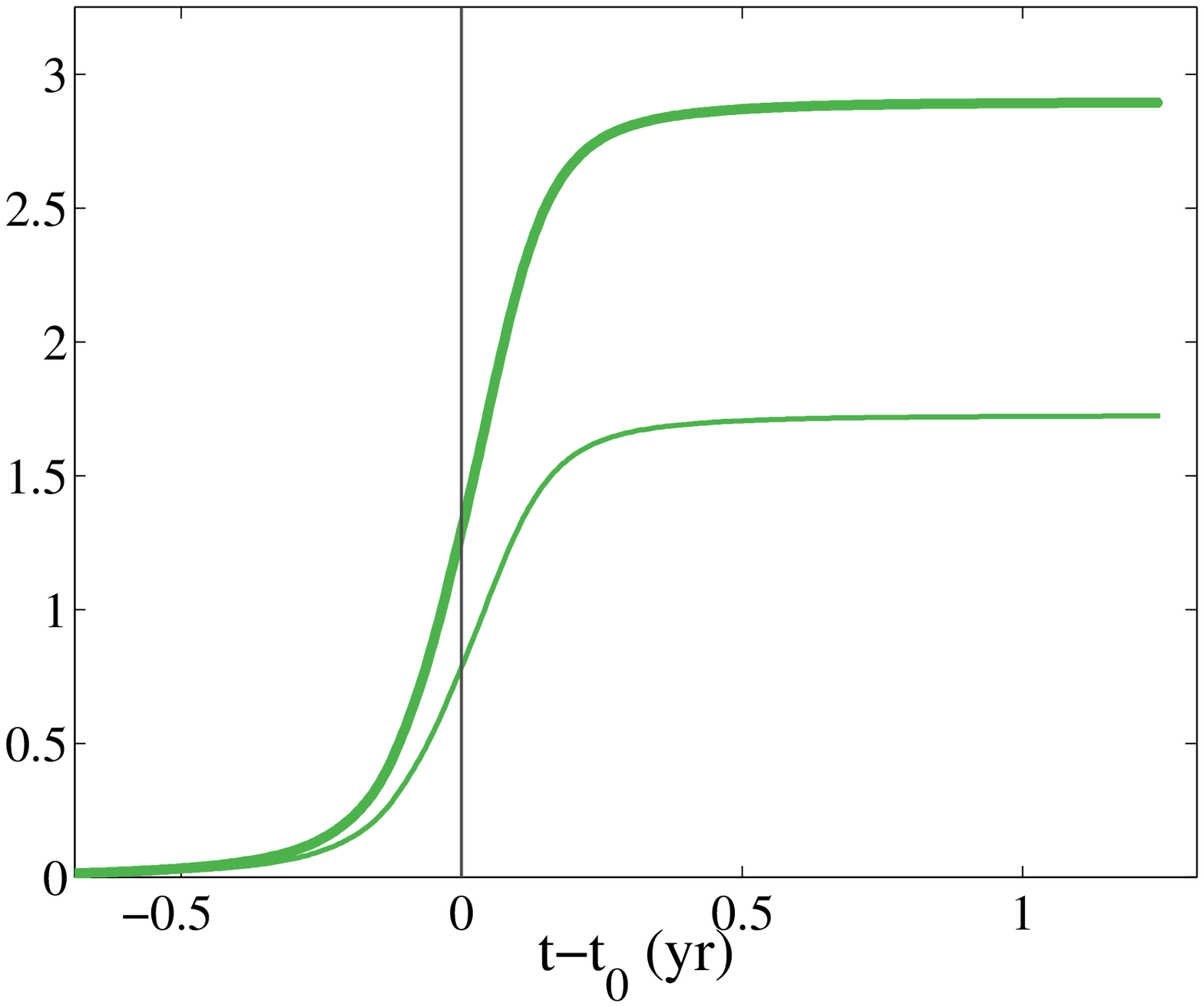}}&\hspace{-.36cm}
\subfigure{\includegraphics[width=.3\textwidth]{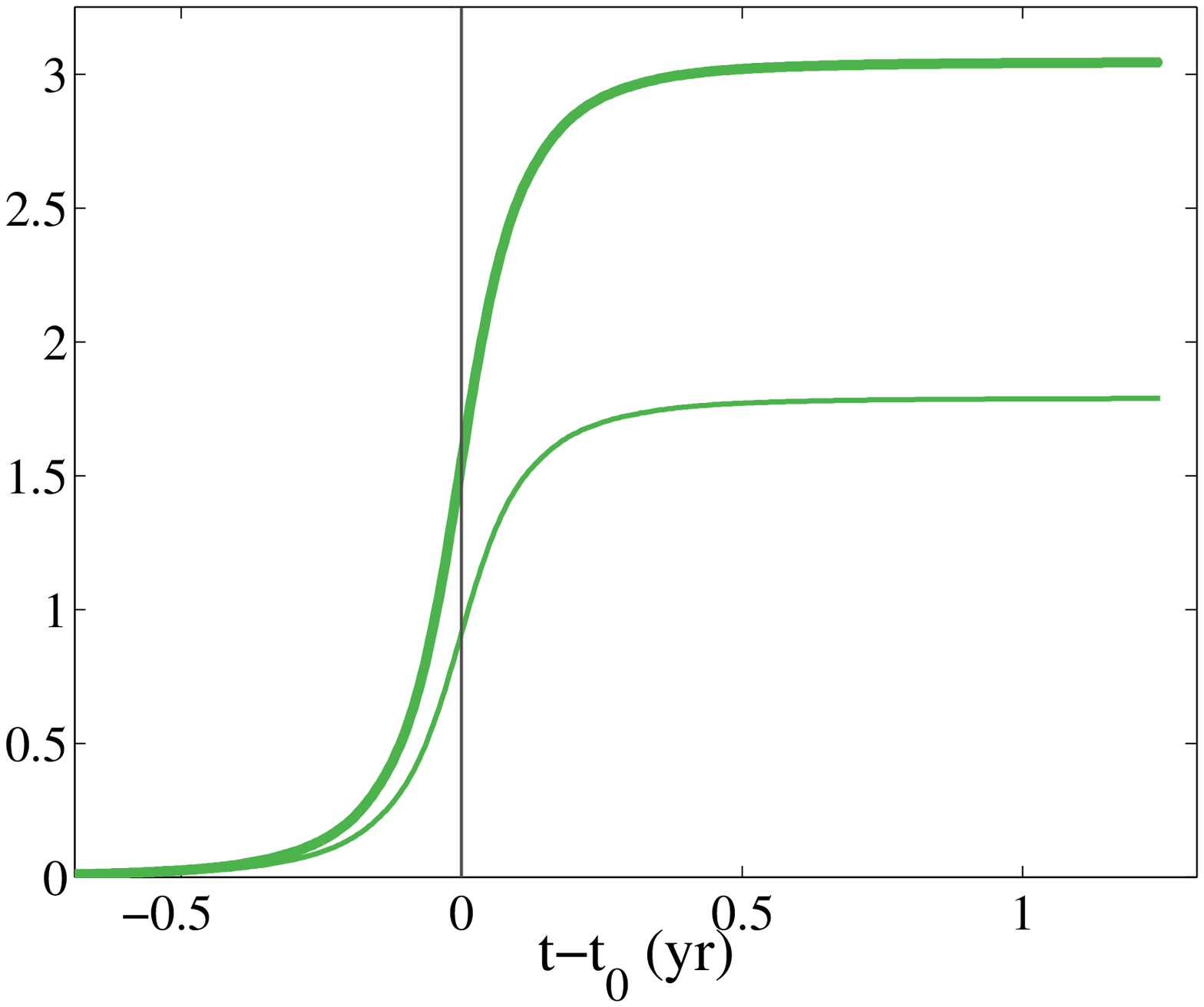}}\\
\begin{sideways}$i=\pi/6$\end{sideways}&\hspace{-.3cm}
\subfigure{\includegraphics[width=.33\textwidth]{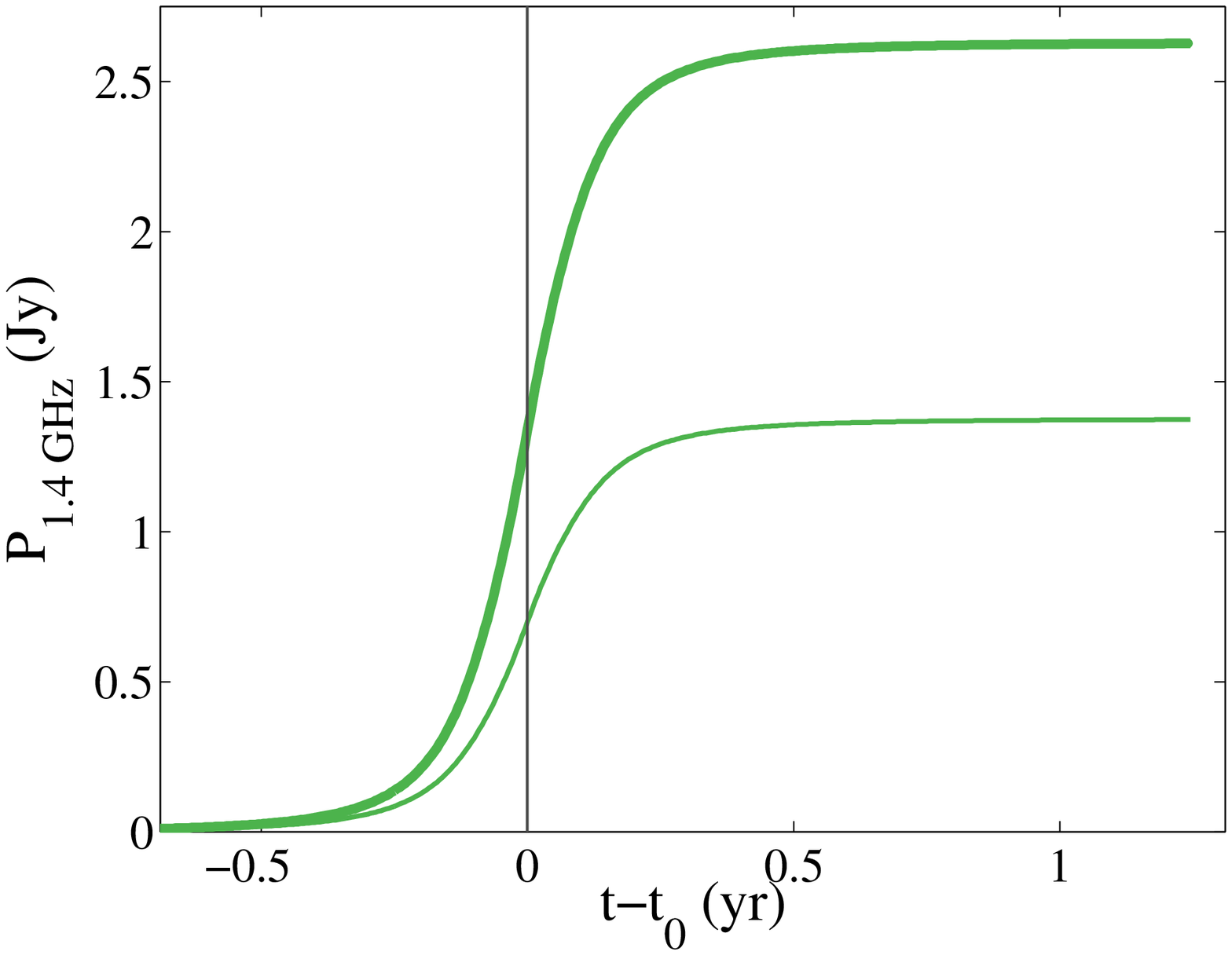}}&\hspace{-.36cm}
\subfigure{\includegraphics[width=.3\textwidth]{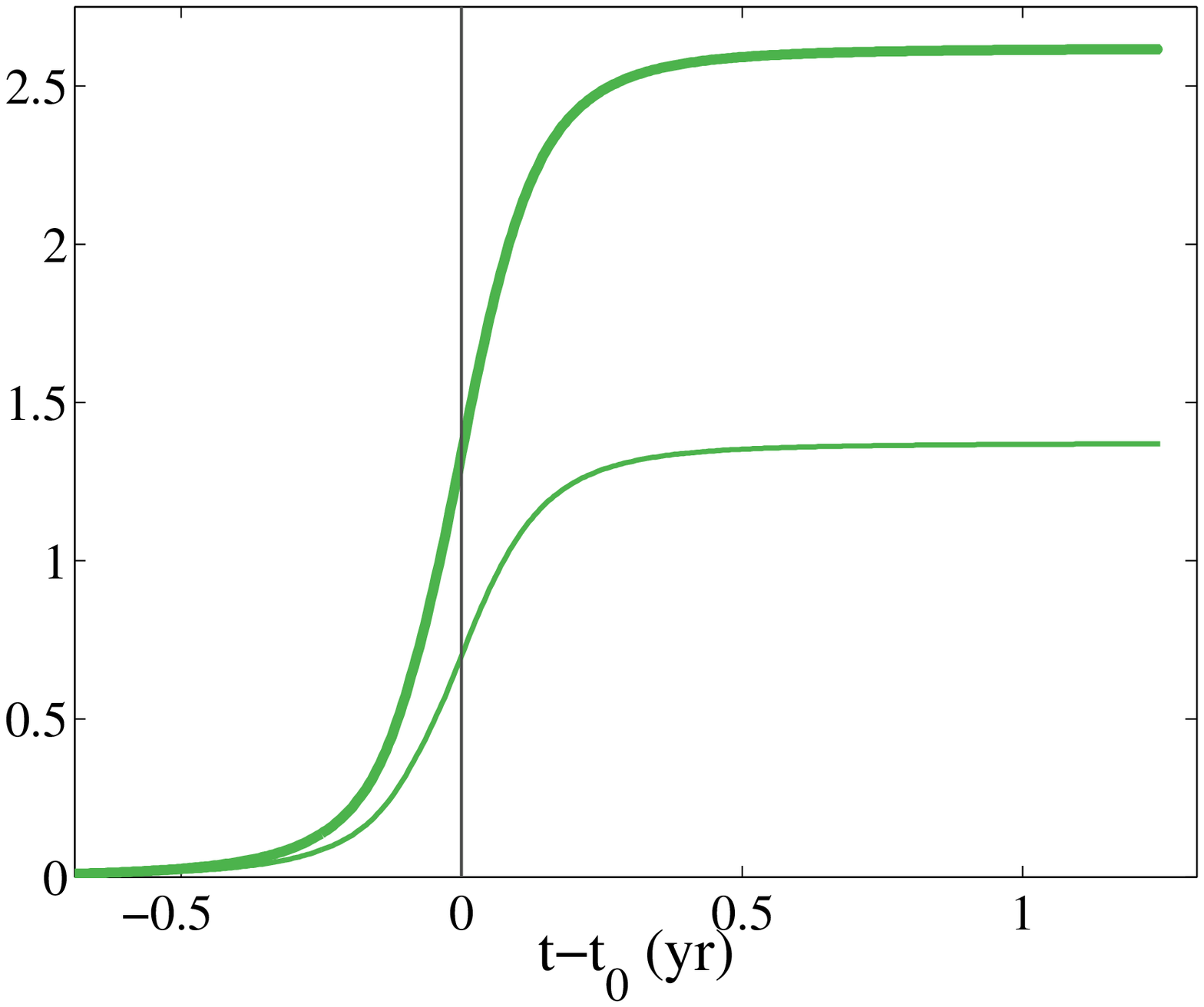}}&\hspace{-.36cm}
\subfigure{\includegraphics[width=.3\textwidth]{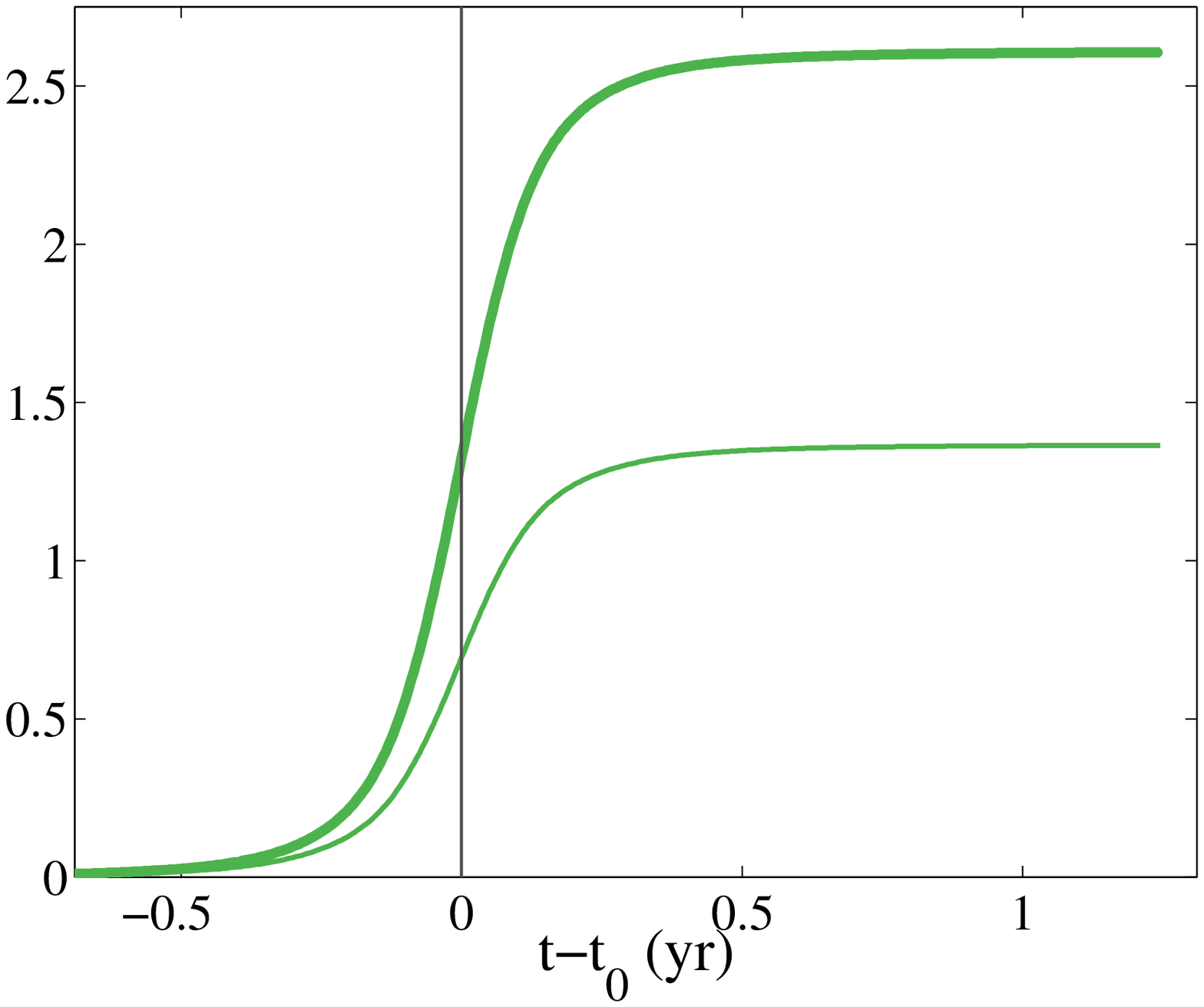}}\\
\end{tabular}

\caption{Similar to Figure~\ref{f.lc1} but for the local model (Section~\ref{s.local}).}
  \label{f.lc2}
\end{figure*}

By construction, in the local model, the emission increases
monotonically with time. However, because the magnetic field is not
amplified, the total emission is lower than in the plowing model.
Figure~\ref{f.lc2} shows a set of light curves that exhibit exactly
such behavior. All of the curves start to increase a few months
before the shock reaches pericenter but none of them reaches a flux
that significantly exceeds $4\,\rm Jy$. The difference in the
magnitude of the maximum flux between the plowing and local models,
which is roughly a factor of 5, is caused by two factors. First, the
magnetic field is weaker in the local model than in the plowing model,
leading to a corresponding decrease in the synchrotron emissivity.
Second, because the accelerated electrons are not accumulated in the
local model, fewer electrons reach the most favorable radius for
emission. Depending on the orbit orientation, the maximal flux at
$1.4\,\rm GHz$ lies in the $1.4 - 4.1\,\rm Jy$ range.

\subsection{Spectra}
\label{s.spectra}

In the preceding calculations, we assumed that the energy distribution of the 
shocked electrons is described by a single power law with slope $p=2.4$. This 
results in a power-law synchrotron spectrum ($F_\nu\propto \nu^\alpha$) with
spectral index $\alpha=-(p-1)/2$ (Eq.~\ref{e.Pintr1}), which implies that
the flux is highest at low frequencies. However, at very low
frequencies, the emission is limited by synchrotron self-absorption, 
which was estimated by Narayan et al. (2012a) to affect frequencies 
$\nu\lesssim 1\rm GHz$. In this section, we calculate synchrotron 
self-absorption for the parameters adopted in this work and plot the 
expected spectral energy distribution.

Synchrotron emission at a given frequency is self-absorbed when its
total intensity, calculated according to Equation~(\ref{e.Pintr1}), 
exceeds the intensity of the local blackbody emission at the characteristic
temperature corresponding to the energy of electrons responsible for
emission at that frequency.  The emission is then limited to the 
blackbody value.

Synchrotron emission peaks at frequency $\nu$ for electrons with
Lorentz factor $\gamma$ given by \citep{rybicki-book}, 
\be
\gamma^2\approx\frac{4\pi m_{\rm e} c\nu}{3qB}, 
\ee 
which corresponds to a temperature $T$ equal to
\be 
T\approx\frac{\gamma m_{\rm e}c^2}{3k_{\rm B}}.
\ee 
The spectral power emitted by a black body with this temperature
from a surface area $A$ is 
\be 
P_{\nu,\,{\rm BB}}=A\pi B_\nu(T), 
\ee
where $B_\nu(T)$ is the intensity of blackbody radiation, 
\be
B_\nu(T) =\frac{ 2 h\nu^{3}}{c^2}\frac{1}{ e^{h\nu/kT}-1}.  
\ee 
The critical frequency below which self-absorption limits the synchrotron
emission is given by the condition
\be 
P_{\nu,\,\rm BB}=P_\nu, 
\ee
where $P_\nu$ is given by Equation~(\ref{e.Pintr1}) or 
Equation~(\ref{e.pnulocal}). For simplicity, we will not treat the 
transition regime in detail. Instead, we will assume that the total 
spectrum can be represented as a broken power law, with a slope equal 
to $-(p-1)/2$ in the high-frequency range and equal to $5/2$ in 
the low-frequency segment that is self-absorbed below the 
critical frequency. 

In the case of the plowing model, we assume that the radiating
electrons are contained within a sphere of radius
$R=10^{15}\rm cm$ and $A=4\pi R^2$. For the local model, we
correct the emission for self-absorption independently at each cloud
location assuming that the differential emitting area is $\Delta
A=2\pi R \rm \Delta r$, where $\rm \Delta r$ corresponds to the
distance covered by the cloud between time $t$ and $t+\Delta t$. Thus,
the emitting area is effectively the surface of a tube of radius $R$
traced by the cloud moving along its orbit.

\begin{figure*}
  \centering
\includegraphics[width=.45\textwidth]{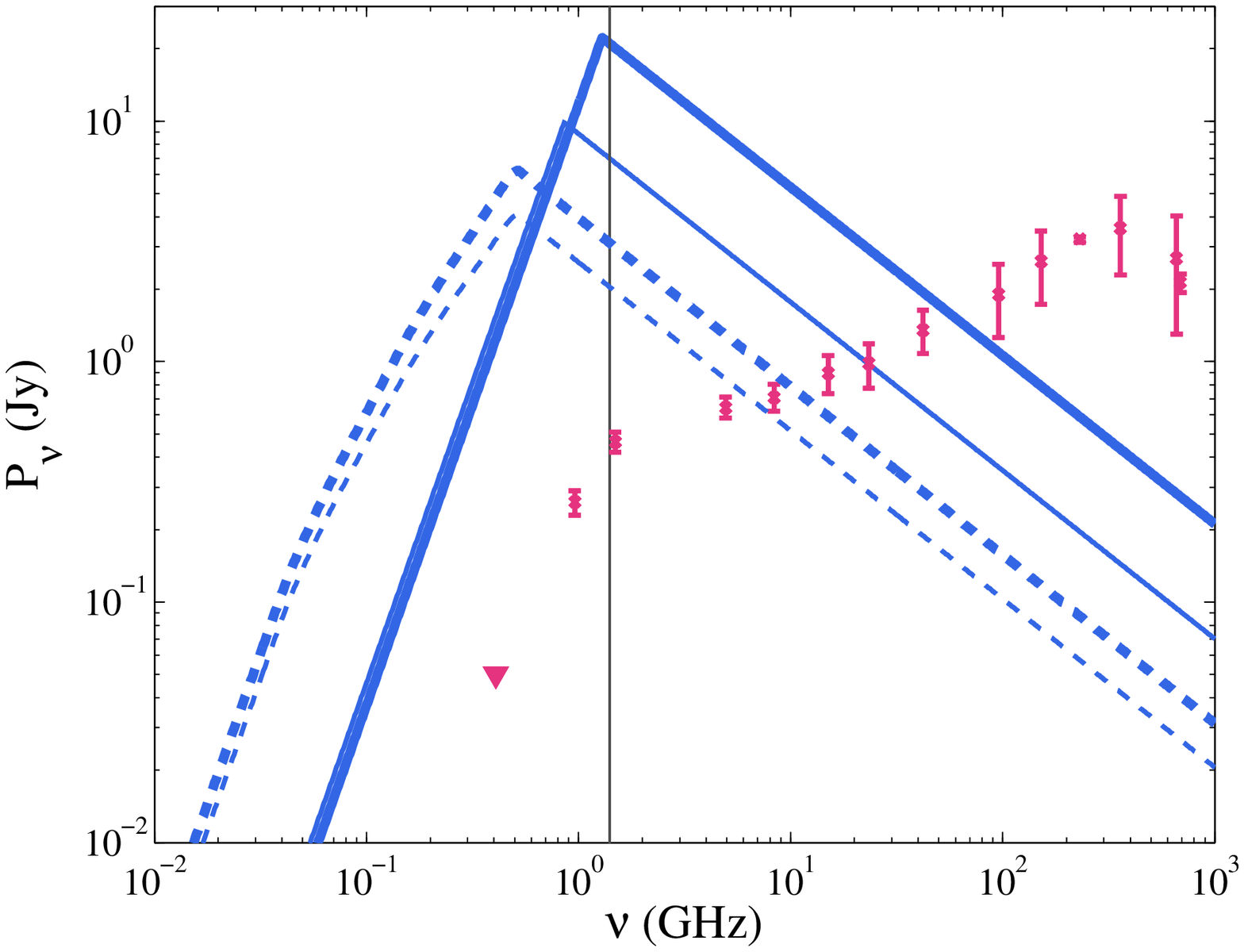}
\includegraphics[width=.45\textwidth]{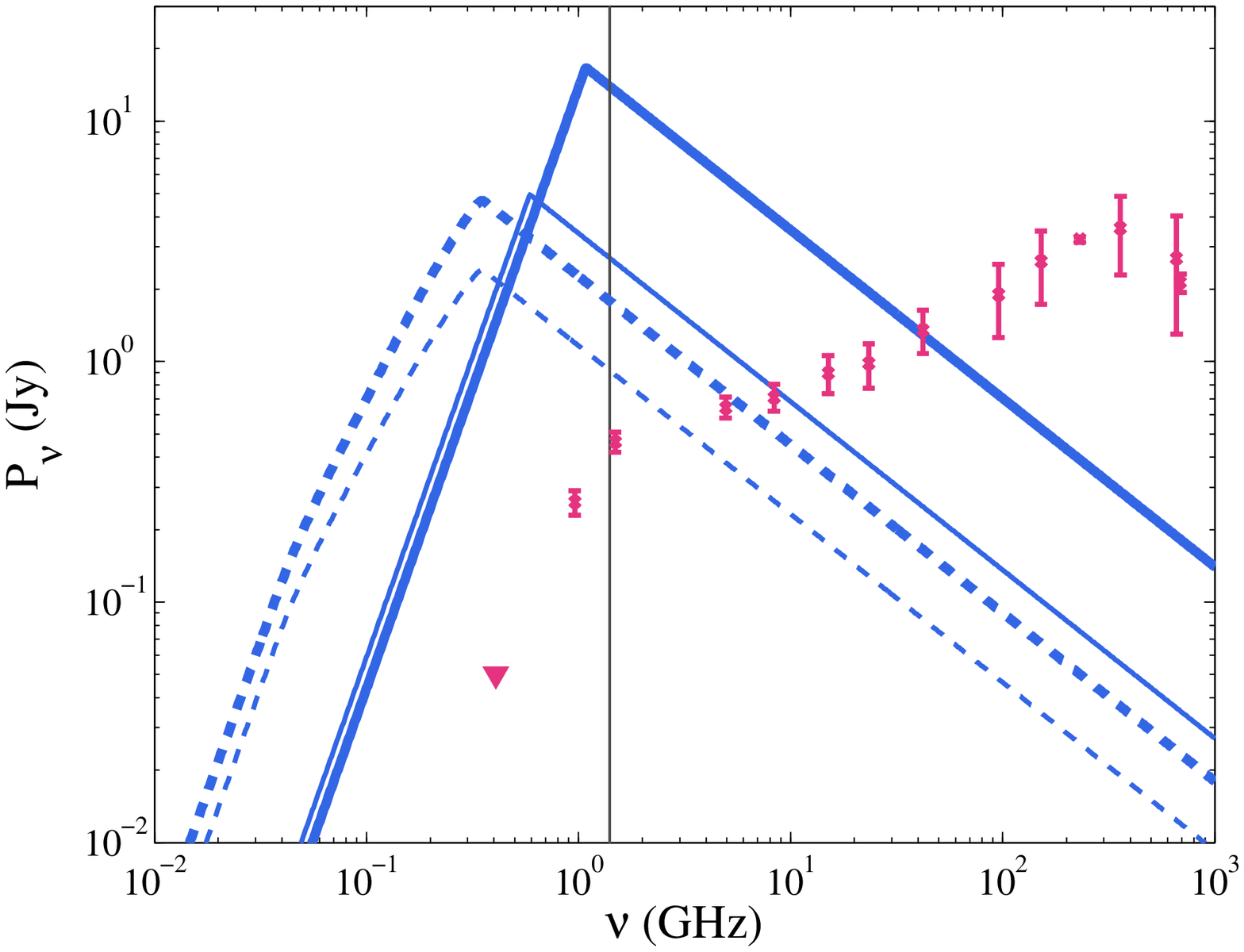}\\
\caption{Spectra of bow shock synchrotron emission for most ($i=\pi/3$, 
$\omega=\pi/2$, left) and least ($i=\pi/6$, $\omega=\pi/2$, right panel)
favorable orbit orientation (out of the orientations considered)
calculated at $t=t_0+0.05$. Solid and dashed lines correspond to the
plowing and local acceleration models, respectively. Thick lines show
counter-rotating orbits while thin lines denote co-rotating
orbits. Magenta points show the quiescent radio emission from \sgra. The vertical line shows the frequency $\nu=1.4\,\rm GHz$ for which the lightcurves have been calculated.}
\label{f.spectra}
\end{figure*}

Figure~\ref{f.spectra} shows our calculated spectra at $t=t_0+0.05$ for
the most and least favorable orbit orientations considered in this
work. The left and right panels correspond to the top-right and
bottom-left panels in Figures~\ref{f.lc1} and \ref{f.lc2}, respectively,
i.e., to orbit orientations producing the most and the least
amount of emission. The solid lines in Figure~\ref{f.spectra} correspond 
to the plowing model and the dashed lines to the local model. As before, 
thick and thin lines represent the counter- and co-rotating orbits, 
respectively. The flux levels obtained with the various models extend 
over a wide range, reflecting our lack of knowledge of the orbit orientation 
as well as the bow shock structure.

The magenta data points in Figure~\ref{f.spectra} show the intrinsic
radio emission of \sgra compiled from the literature 
\citep{davies+76,falcke00,zhao+03,marrone+08}. If
bow shock-related synchrotron radiation is to be unequivocally
detected, the emission must exceed the quiescent emission of 
Sgr~A$^*$. It is clear that, at frequencies above $100\,\rm GHz$, none of
the models predicts a detectable level of emission. The maximum
emission in the plowing model occurs at $\sim 1\rm GHz$ and the flux
here is well above the intrinsic emission of \sgra. However, at the
same frequency, the local model predicts the bow shock emission to be
roughly comparable to the intrinsic flux. All our models predict
detectable emission at levels above the intrinsic emission of Sgr~A$^*$ 
at frequencies in the range $0.1<\nu<1\,\rm GHz$.

\subsection{When is $t_0$?}
\label{s.t0}

All of the light curves and spectra we calculated so far are
expressed as a function of relative time with respect to the time
$t_0$ when the bow shock reaches pericenter.  The exact time
corresponding to $t_0$ is unfortunately uncertain. Since the bow shock
is expected to form in front of the cloud, $t_0$ is certainly earlier
than the time $t_{0,\rm CM}$ when the cloud CM reaches
pericenter. However, the exact location of the bow shock is poorly
constrained. As a rough estimate, we can assume that the shock forms
at the location corresponding to the ``surface'' of the original
spherical cloud at $t=2000.0$, i.e., at a distance $R=1.5\times 10^{15}\rm cm$
\citep{gillessen+12a} ahead of the cloud CM along the orbit. This
particular location precedes CM by $\sim 5$ months. Therefore, we
estimate the time at which the bow shock reaches pericenter to be
$t_0\approx 2013.2$, i.e., the bow shock should reach pericenter in
March 2013. Naturally, this estimate is rather approximate since the
internal structure of the cloud is unknown.  

\section{Discussion}
\label{s.discussion}

\subsection{Uncertainties}

Apart from the uncertainty in $t_0$ discussed above, the model adopted
here for calculating radio light curves relies on a number of other
simplifications and assumptions. The most crucial of these is the
adopted cross section of the shock region, which determines the number
of shocked electrons and, therefore, the intensity of emission. We
assumed that the effective area is constant in time, $A=\pi R^2$, with
$R=10^{15}\,\rm cm$ corresponding to $2/3$ of the half-width at
half-maximum as measured by \cite{gillessen+12a}. However, given that
the cloud density is much larger than the density of the surrounding
gas, it is possible that it will effectively plow gas with even larger
area, thus increasing the magnitude of the flux enhancement.

A related uncertainty arises from the assumed shape of the cloud.  G2
is not expected to keep its spherical shape throughout the passage:
simulations show that the tidal gravitational field will first elongate
it along the orbit, then squeeze it in the orbital plane at pericenter
\citep{gillessen+12a,saitoh+12}, and finally let it decompress. However, 
the spatial extent of the cloud is large and the whole cloud will not
become compressed at the same time. In other words, when the center of
mass reaches the pericenter, only the middle part of the cloud is
squeezed while its front has already decompressed and the tail lags
behind.

\begin{figure*}
  \centering
\includegraphics[width=.255\textwidth]{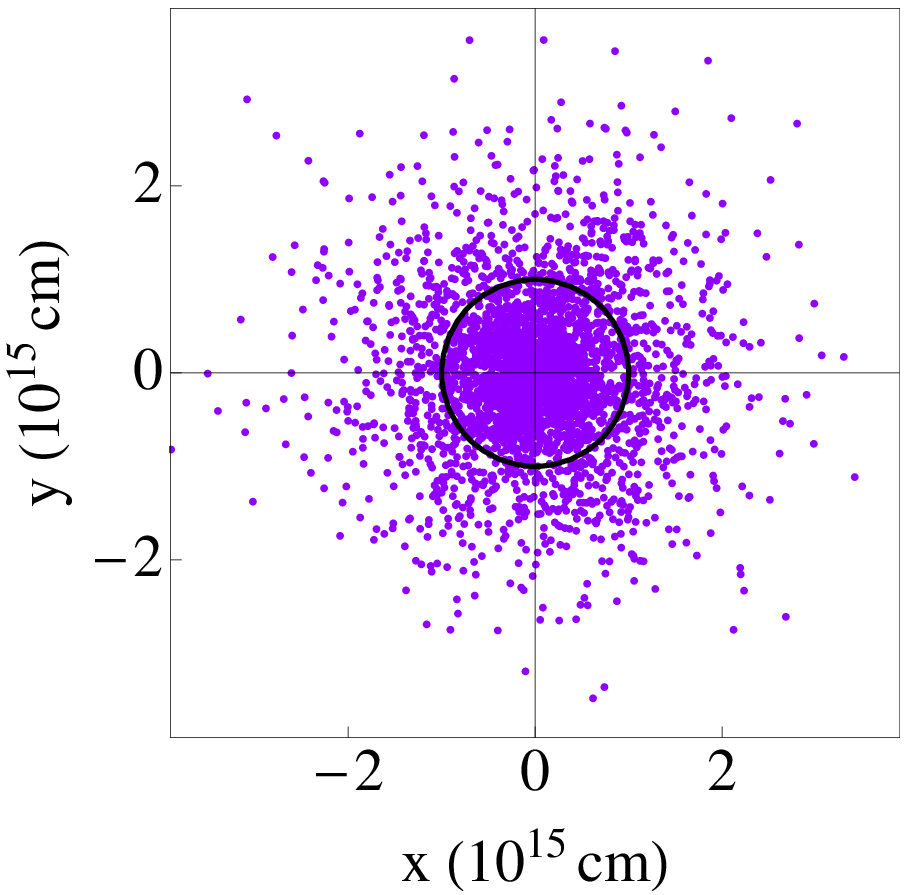}
\includegraphics[width=.23\textwidth]{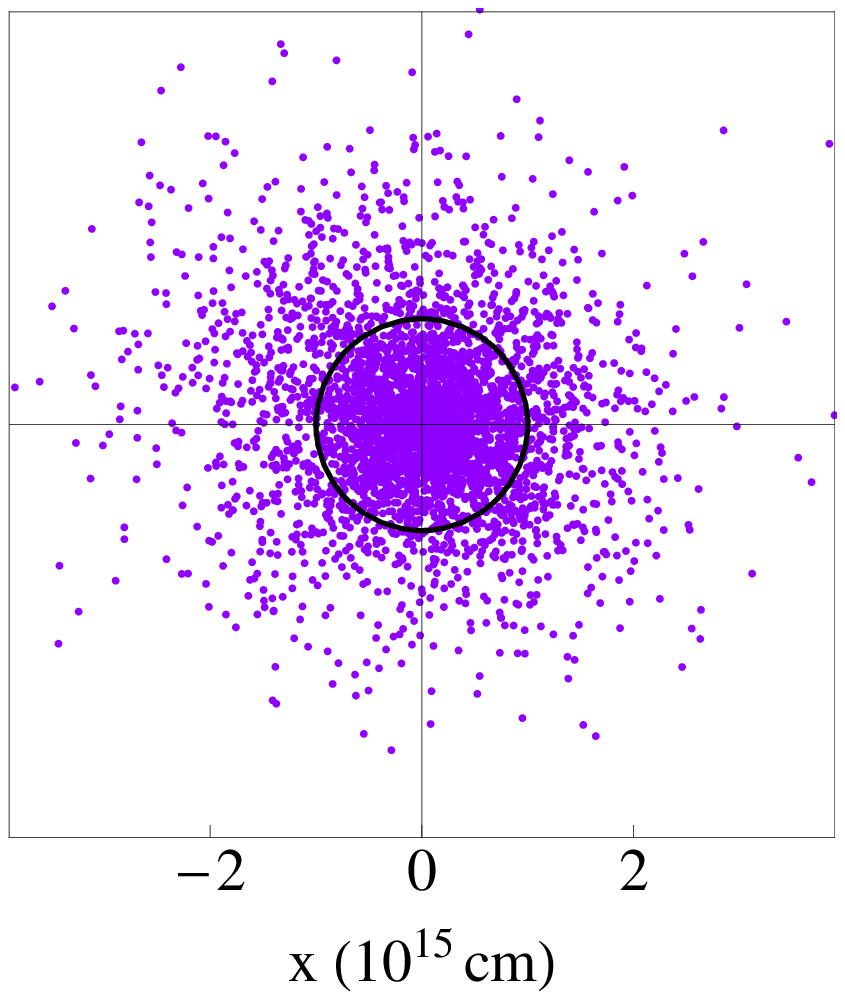}
\includegraphics[width=.23\textwidth]{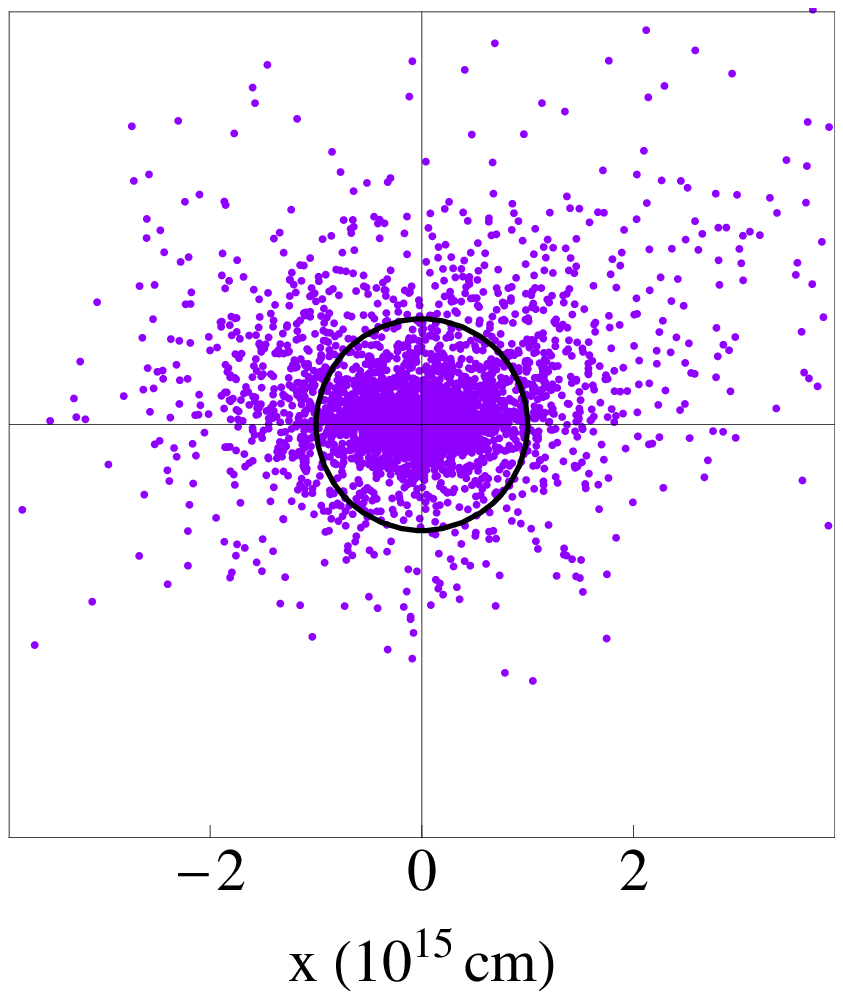}
\includegraphics[width=.23\textwidth]{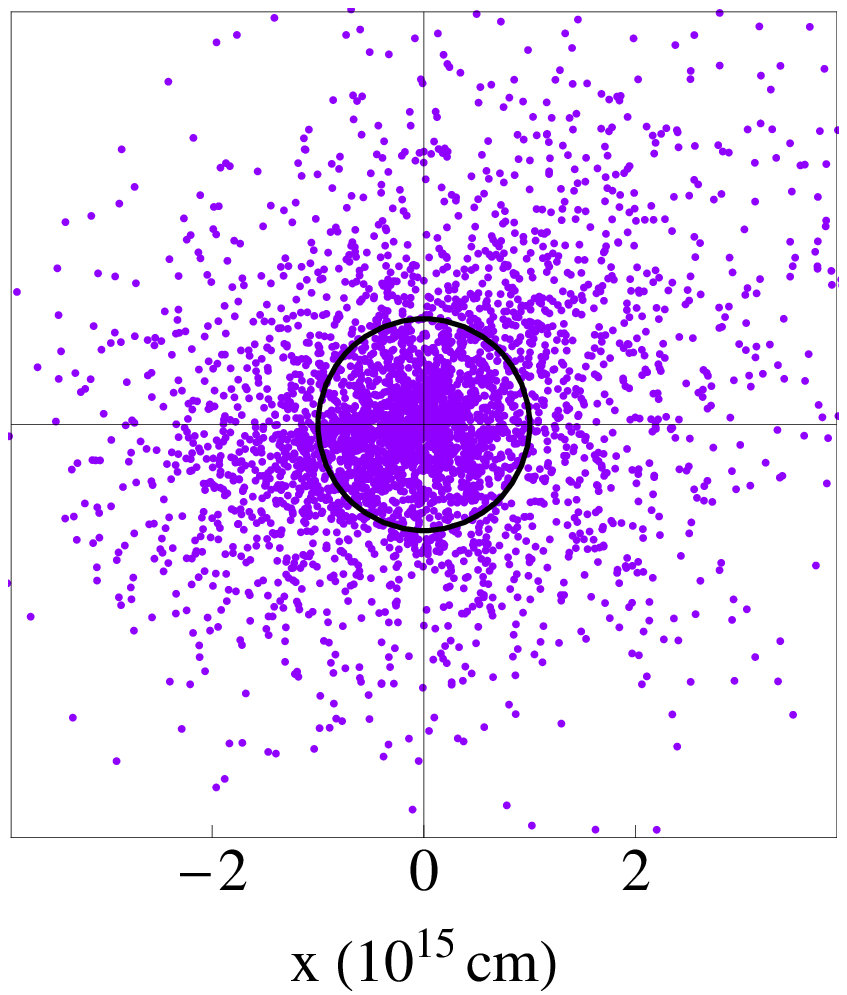}
\caption{Cloud profile on a plane perpendicular to its orbit 
calculated with $5000$ test particles for (left to right) $t=2000.0$,
$2013.0$, $2013.69$ (epoch of periastron) and $2014.5$. The black
circle shows the region within $R=10^{15}\,\rm cm$
corresponding to the assumed cloud crossection.}
\label{f.cross}
\end{figure*}

To quantify the effects of tidal compression, we calculated
the trajectories of $5000$ test particles following the approach of
\cite{gillessen+12a}. We adopted the set of parameters given therein: 
a spherical cloud at $t=2000.0$ with Gaussian distribution of density
with full-width at half-maximum (FWHM) equal $3\times 10^{15}\,\rm cm$
and a velocity dispersion $\sigma_v=130\,\rm km/s$. At four epochs
($t=2000.0$, $2013.0$, $2013.69$ and $2014.5$), we projected each
particle onto the plane perpendicular to its nearest section of the
cloud orbit and plotted their distribution in
Figure~\ref{f.cross}. Due to the large density of the cloud when
compared to the disk, the whole region covered by test particles may
be considered likely to plow the gas. The black circle corresponding
to $R=10^{15}\,\rm cm$ is always well within the projected cloud cross
section. Our value is, therefore, rather conservative and the actual
radio fluxes may be significantly stronger than those given in
Section~\ref{s.lightcurves} if the effective area is indeed larger.
In such a case, the critical frequency for the self-absorption will
decrease, leading to higher flux at low frequencies, which may make
the bow-shock emission easier to observe in that frequency range.

In Section~\ref{s.acceleration} we discussed the mechanism of
electron acceleration. Given the long computational time required to
obtain the power-law index of the accelerated electrons, we used only
a coarse sampling of physical parameters throughout each potential
orbit of the cloud. However, the impact of this coarse gridding is not
overwhelming. For example, taking a power-law index $p=2.6$ instead
of $p=2.4$ drives the calculated fluxes down only by a factor of
$\sim 2$ and does not alter our conclusions regarding detectability. 
On the other hand, the value $p=2.4$ adopted here is likely to be a 
conservative upper limit. As described in section 4.2, the electron 
non-thermal tail should evolve toward flatter slopes at later times, 
approaching the value $p=2.2$ adopted by Narayan et al (2012a).

Finally, we note that there are uncertainties arising from 
the model for the accretion disk. While the gas temperature should
always be close to virial, the disk density is poorly constrained by
direct observations between the known values at the Bondi radius and
the innermost regions where there is an estimate of the accretion rate
(see also Psaltis 2012). We adopted a density profile $\rho\propto
R^{-1}$ in between these two limits. In principle, the density may be
a more complicated function of radius, which could lead to a different
density along the orbit than predicted here. Given that the radio
fluxes computed here are directly proportional to the local density of
the accretion flow, they can easily be scaled for other models of the
accretion flow.

\subsection{Suggested Radio Monitoring Observations}
\label{s.proposed}

The calculations presented in this paper indicate that the observed
radio emission from \sgra\ should undergo a significant brightening
around or shortly after the pericentric passage of
the bow shock, i.e., in spring 2013, assuming $t_0$ corresponds
to March 2013 (see Section \ref{s.t0}).
 In
fact, given the $\sim 22$~mas displacement of the expected bow shock
from the center of the accretion flow, where the long-term radio
emission originates, the bow shock emission should appear as a
new radio source that is, in principle, resolvable from the existing
emission (Narayan et al.\ 2012a). In addition, we showed here that,
under both the plowing and the local acceleration model assumptions,
the spectrum of the bow shock emission should be different from that
of quiescent emission in a number of ways. First, the emission is
expected to peak at lower frequencies, around $\sim 0.3-1$~GHz.
Second, the bow shock should produce detectable emission at $\nu
\lesssim 1$~GHz, where the quiescent radio emission is undetectable.

The Karl V. Jansky Very Large Array (JVLA) is ideally positioned to
detect and monitor the radio flux from the bow shock of G2 at
frequencies around 1\;GHz and above. Our ``local'' case with the least
favorable orbit orientation produces the lowest level of flux
enhancement. It nevertheless predicts a flux of 1.4\;Jy at 1.4\;GHz,
which is comparable to the current quiescent flux from \sgra\ at that
frequency (see Figure~8). The plowing case, on the other hand,
produces a radio signal that can be an order of magnitude or more
higher than the current flux at 1.4\;GHz but is shorter lived, with the
flux enhancement decaying on a timescale of the order of a
year. Monitoring the Galactic Center with the JVLA in 2013,
therefore, should be able to not only detect the new radio emission
from the bow shock but also to clearly distinguish the plowing and
local models we considered here.

Given that the spectrum of the bow shock emission may also peak below
1\;GHz for certain orbit orientations and acceleration models, as we
discussed in Section 4, monitoring the Galactic Center at frequencies
ranging from $\lesssim 100$\;MHz to 1\;GHz provides another optimal
opportunity to probe the shock physics and the conditions of the
accretion flow around pericenter. Furthermore, because \sgra\ does not
normally produce emission at these low frequencies, the detection of a
bright radio source will deliver a definitive signature of the
emission associated with G2. To this end, the Giant Metrewave Radio
Telescope (GMRT), which operates at 5 frequencies between 50\;MHz and
610\;MHz, as well as at 1.4\;GHz, is ideally placed and provides a
strong complement to the JVLA. Given its latitude of 19$^\circ$N, it is
at a highly favorable location to observe the Galactic Center. In
addition, GMRT has a distinct advantage in collecting area owing to 30
dishes with 45\;m diameter each.

We encourage frequent monitoring at available bands between 50 MHz and
100 GHz during 2013 in order to capture the increase in the 
flux from \sgra, to measure its rise time, and to detect the possible 
decay a year after pericentre. 

\section{Summary}
\label{s.summary}

The G2 cloud is expected to penetrate the accretion flow near \sgra
with a supersonic velocity and shock the gas in front of it. We showed 
that the shock parameters and the pre-shock electron temperatures are 
conducive to accelerating electrons to relativistic energies. 
These electrons can emit synchrotron radiation in the magnetic
field of the accreting gas. 

In this paper, we presented detailed predictions for radio light 
curves during the passage of the cloud.  We adopted a
state-of-the-art global numerical solution of a radiatively inefficient
accretion flow as a base for our disk model and obtained physical disk
parameters at large distances by extrapolating the numerical solution
from within its converged, steady-state region. The parameters describing 
the electron acceleration process were determined by a set of
particle-in-cell numerical simulations set up for the expected
conditions as a function of the flow temperature and the shock Mach
number. We considered a set of possible orientations of the orbit with
respect to the accretion disk and two models for the particle
acceleration: the plowing case, where all the electrons are kept in the
shocked region and radiate in the shock-amplified magnetic field, and
the local model, where the accelerated electrons are left behind the cloud
front and radiate in their local, unshocked magnetic fields. 

In both models, assuming the bow shock leads the cloud CM by half a year, 
we found that the radio luminosity is expected to reach peak values in
spring 2013. For the plowing scenario this maximum will be followed
by a decay, while for the local scenario the luminosity will stay
constant due to the long cooling time of electrons. Calculated maximum
luminosities at $\nu=1.4\,\rm GHz$ span the range $1.4 - 22\,\rm Jy$,
depending on the adopted orbit orientation and acceleration mode, and
are likely to exceed the intrinsic luminosity of \sgra.  At lower
frequencies, the predicted emission will overwhelm the intrinsic
luminosity for all the models considered by roughly an order of
magnitude.  We, therefore, recommend an observational campaign in 2013
at frequencies in the range $100\,{\rm MHz} - 1\,{\rm GHz}$.

\section{Acknowledgements}

We thank Dimitrios Psaltis for useful comments. A.S. and R.N.
were supported in part by NASA grant NNX11AE16G. L.S. is supported by
NASA through Einstein Postdoctoral Fellowship grant number PF1-120090
awarded by the Chandra X-ray Center, which is operated by the
Smithsonian Astrophysical Observatory for NASA under contract
NAS8-03060. F.\"O. acknowledges support from NSF grant AST-1108753 and
from the Radcliffe Institute for Advanced Study at Harvard University.
The particle-in-cell simulations were performed on the Odyssey cluster
at Harvard University, on XSEDE resources under contract
No. TG-AST120010, and on NASA High-End Computing (HEC) resources
through the NASA Advanced Supercomputing (NAS) Division at Ames
Research Center.
 
\bibliographystyle{mn2e}

{\small
\begin{thebibliography}{}

\bibitem[Anninos et al.(2005)]{anninosetal05} Anninos, P., Fragile, 
P.~C., \& Salmonson, J.~D.\ 2005, \apj, 635, 723 

\bibitem[Baganoff et al.(2001)]{2001Natur.413...45B} Baganoff, F.~K., 
Bautz, M.~W., Brandt, W.~N., et al.\ 2001, \nat, 413, 45 

\bibitem[Baganoff et al.(2003)]{2003ApJ...591..891B} Baganoff, F.~K., 
Maeda, Y., Morris, M., et al.\ 2003, \apj, 591, 891 

\bibitem[Bartko et al.(2009)]{bartko+09} Bartko, H., Martins, F., 
Fritz, T.~K., et al.\ 2009, \apj, 697, 1741 

\bibitem[{{Begelman} \& {Kirk}(1990)}]{begelman_kirk_90}
{Begelman}, M.~C. \& {Kirk}, J.~G. 1990, \apj, 353, 66

\bibitem[{{Blandford} \& {Eichler}(1987)}]{blandford_eichler_87}
{Blandford}, R. \& {Eichler}, D. 1987, Physical Reports, 154, 1

\bibitem[Broderick et al.(2011)]{broderick+11} Broderick, A.~E., 
Fish, V.~L., Doeleman, S.~S., \& Loeb, A.\ 2011, \apj, 735, 110 

\bibitem[Chan et al.(2009)]{2009ApJ...701..521C} Chan, C.-k., Liu, S., 
Fryer, C.~L., et al.\ 2009, \apj, 701, 521 

\bibitem[Davies et al.(1976)]{davies+76} Davies, R.~D., Walsh, 
D., \& Booth, R.~S.\ 1976, \mnras, 177, 319 

\bibitem[Del Zanna et 
al.(2007)]{delzannaetal07} Del Zanna, L., Zanotti, O., Bucciantini, N., 
\& Londrillo, P.\ 2007, \aap, 473, 11 

\bibitem[De Villiers et al.(2003)]{devilliersetal03} De Villiers, J.-P., 
Hawley, J.~F., \& Krolik, J.~H.\ 2003, \apj, 599, 1238 

\bibitem[Falcke \& Markoff(2000)]{falcke00} Falcke, H., \& Markoff, 
S.\ 2000, \aap, 362, 113 

\bibitem[Gaburov et al.(2012)]{gaburov+12} Gaburov, E., Johansen, 
A., \& Levin, Y.\ 2012, \apj, 758, 103 

\bibitem[Gammie et al.(2003)]{gammieetal03} Gammie, C.~F., McKinney, 
J.~C., \& T{\'o}th, G.\ 2003, \apj, 589, 444 

\bibitem[Genzel et al.(2010)]{genzel+10} Genzel, R., Eisenhauer, 
F., \& Gillessen, S.\ 2010, Reviews of Modern Physics, 82, 3121 

\bibitem[Gillessen et al.(2012a)]{gillessen+12a} Gillessen, S., 
Genzel, R., Fritz, T.~K., et al.\ 2012a, \nat, 481, 51 

\bibitem[Gillessen et al.(2012b)]{gillessen+12b} Gillessen, S., 
Genzel, R., Fritz, T.~K., et al.\ 2012b, arXiv:1209.2272 

\bibitem[Levin 
\& Beloborodov(2003)]{2003ApJ...590L..33L} Levin, Y., \& Beloborodov, A.~M.\ 2003, \apjl, 590, L33 

\bibitem[Marrone et al.(2008)]{marrone+08} Marrone, D.~P., 
Baganoff, F.~K., Morris, M.~R., et al.\ 2008, \apj, 682, 373 

\bibitem[Marrone et al.(2007)]{marrone07} Marrone, D.~P., Moran, 
J.~M., Zhao, J.-H., \& Rao, R.\ 2007, \apjl, 654, L57 

\bibitem[Matsukiyo et al.(2011)]{matsu11} Matsukiyo, S., Ohira, Y., 
Yamazaki, R., \& Umeda, T.  2011, ApJ, 742, 47

\bibitem[Matsumoto et al.(2012)]{matsu12}{Matsumoto}, Y., {Amano}, T., 
\& {Hoshino}, M. \ 2012, ApJ, 755, 109

\bibitem[McKinney et al.(2012)]{mtb12} McKinney, J.~C., 
Tchekhovskoy, A., \& Blandford, R.~D.\ 2012, \mnras, 423, 3083 

\bibitem[Mo{\'s}cibrodzka et al.(2009)]{moscibrodzka+09} 
Mo{\'s}cibrodzka, M., Gammie, C.~F., Dolence, J.~C., Shiokawa, H., 
\& Leung, P.~K.\ 2009, \apj, 706, 497 

\bibitem[Narayan et al.(2003)]{2003PASJ...55L..69N} Narayan, R., 
Igumenshchev, I.~V., \& Abramowicz, M.~A.\ 2003, \pasj, 55, L69 

\bibitem[Narayan \& McClintock(2008)]{narayanmcclintock08} Narayan, 
R., \& McClintock, J.~E.\ 2008, \nar, 51, 733 

\bibitem[Narayan et al.(2012a)]{narayan+12a} Narayan, R., {\"O}zel, 
F., \& Sironi, L.\ 2012a, \apjl, 757, L20 

\bibitem[Narayan et al.(2012b)]{narayan+12b} Narayan, R., Sadowski, 
A., Penna, R.~F., \& Kulkarni, A.~K.\ 2012b, MNRAS, accepted

\bibitem[Narayan \& Yi(1994)]{narayanyi94} Narayan, R., \& Yi, I.\ 1994, 
\apjl, 428, L13 

\bibitem[Narayan 
\& Yi(1995)]{narayanyi95} Narayan, R., \& Yi, I.\ 1995, \apj, 452, 710

\bibitem[Narayan et al.(1995)]{narayan+95}
Narayan, R., Yi, I., \& Mahadevan, R.\ 1995, \nat, 374, 623

\bibitem[\protect\citeauthoryear{{Tchekhovskoy} \& {McKinney}}
{{Tchekhovskoy} \& {McKinney}}{2012}]{tm12a}
{Tchekhovskoy} A.,  {McKinney} J.~C.,  2012, \mnras, 423, L55

\bibitem[Quataert et al.(1999)]{1999ApJ...517L.101Q} Quataert, E., Narayan, 
R., \& Reid, M.~J.\ 1999, \apjl, 517, L101

\bibitem[Pang et al.(2011)]{pang+11} Pang, B., Pen, U.-L., Matzner, C.~D.,
Green, S.~R., \& Liebend\"orfer, M.\ 2011, \mnras, 415, 1228

\bibitem[Pen et al.(2003)]{pen+03} Pen, U.-L., Matzner, C.~D., 
\& Wong, S.\ 2003, \apjl, 596, L207 

\bibitem[Psaltis(2012)]{2012ApJ...759..130P} Psaltis, D.\ 2012, \apj, 759, 
130 


\bibitem[Riquelme \& Spitkovsky(2011)]{riquelme11} Riquelme, M.~A., \& 
Spitkovsky, A.\ 2011, ApJ, 733, 63

\bibitem[Rybicki \& Lightman(1979)]{rybicki-book} Rybicki, G.~B., \& 
Lightman, A.~P.\ 1979, New York, Wiley-Interscience, 1979.~393 p.,  

\bibitem[Saitoh et al.(2012)]{saitoh+12} Saitoh, T.~R., Makino, 
J., Asaki, Y., et al.\ 2012, arXiv:1212.0349 

\bibitem[{{Spitkovsky}(2005)}]{spitkovsky_05}
{Spitkovsky}, A. 2005, {in AIP Conf. Ser., Vol.~801, 345}

\bibitem[Vallado(2007)]{vallado-book} Vallado, D.~A.\ 2007, 
Fundamentals of Astrodynamics and Applications, by D.A.~Vallado.~Berlin: 
Springer, 2007.~ ISBN: 978-0-387-71831-6,  

\bibitem[{{Weibel}(1959)}]{weibel_59} {Weibel}, E.~S. 1959, Physical 
Review Letters, 2, 83

\bibitem[Yuan et al.(2012b)]{2012ApJ...761..130Y} Yuan, F., Bu, D., 
\& Wu, M.\ 2012, \apj, 761, 130 

\bibitem[Yuan et al.(2003)]{yuan+03} Yuan, F., Quataert, E., 
\& Narayan, R.\ 2003, \apj, 598, 301 

\bibitem[Yuan et al.(2012a)]{2012ApJ...761..129Y} Yuan, F., Wu, M., 
\& Bu, D.\ 2012, \apj, 761, 129 

\bibitem[Zhao et al.(2003)]{zhao+03} Zhao, J.-H., Young, K.~H., 
Herrnstein, R.~M., et al.\ 2003, \apjl, 586, L29 


\end{thebibliography}
}

\appendix
\section{Approximated model for Sgr A$^*$ accretion flow}
\label{ap.1}

\begin{figure}
  \centering
\subfigure{\includegraphics[height=.95\columnwidth,angle=270]{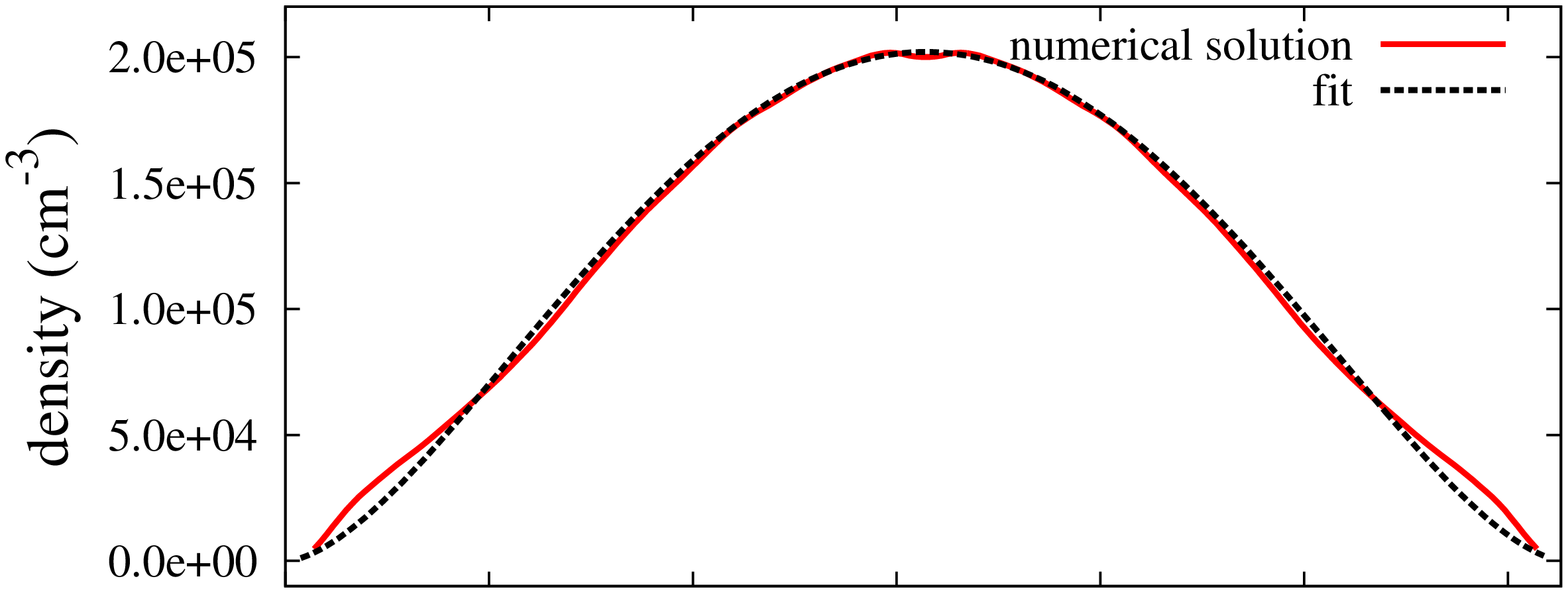}}\vspace{-1.cm}
\subfigure{\includegraphics[height=.95\columnwidth,angle=270]{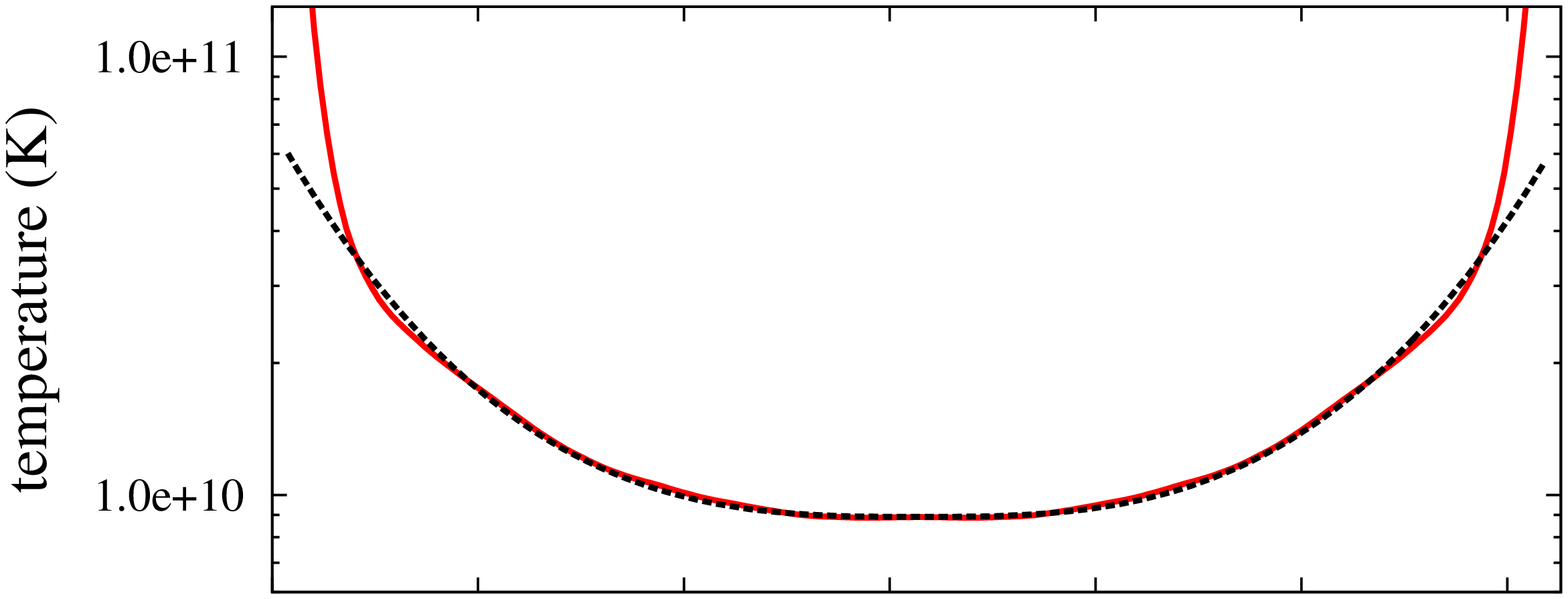}}\vspace{-1.cm}
\subfigure{\includegraphics[height=.95\columnwidth,angle=270]{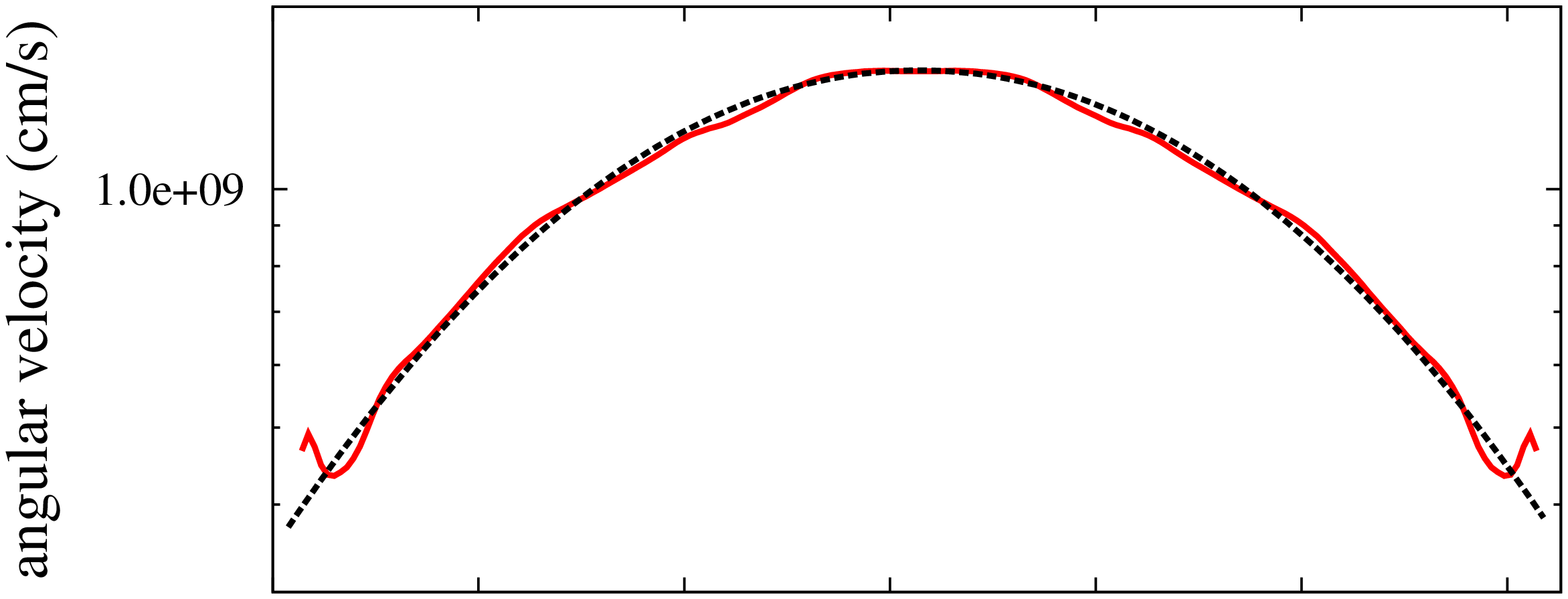}}\vspace{-1.cm}
\subfigure{\includegraphics[height=.95\columnwidth,angle=270]{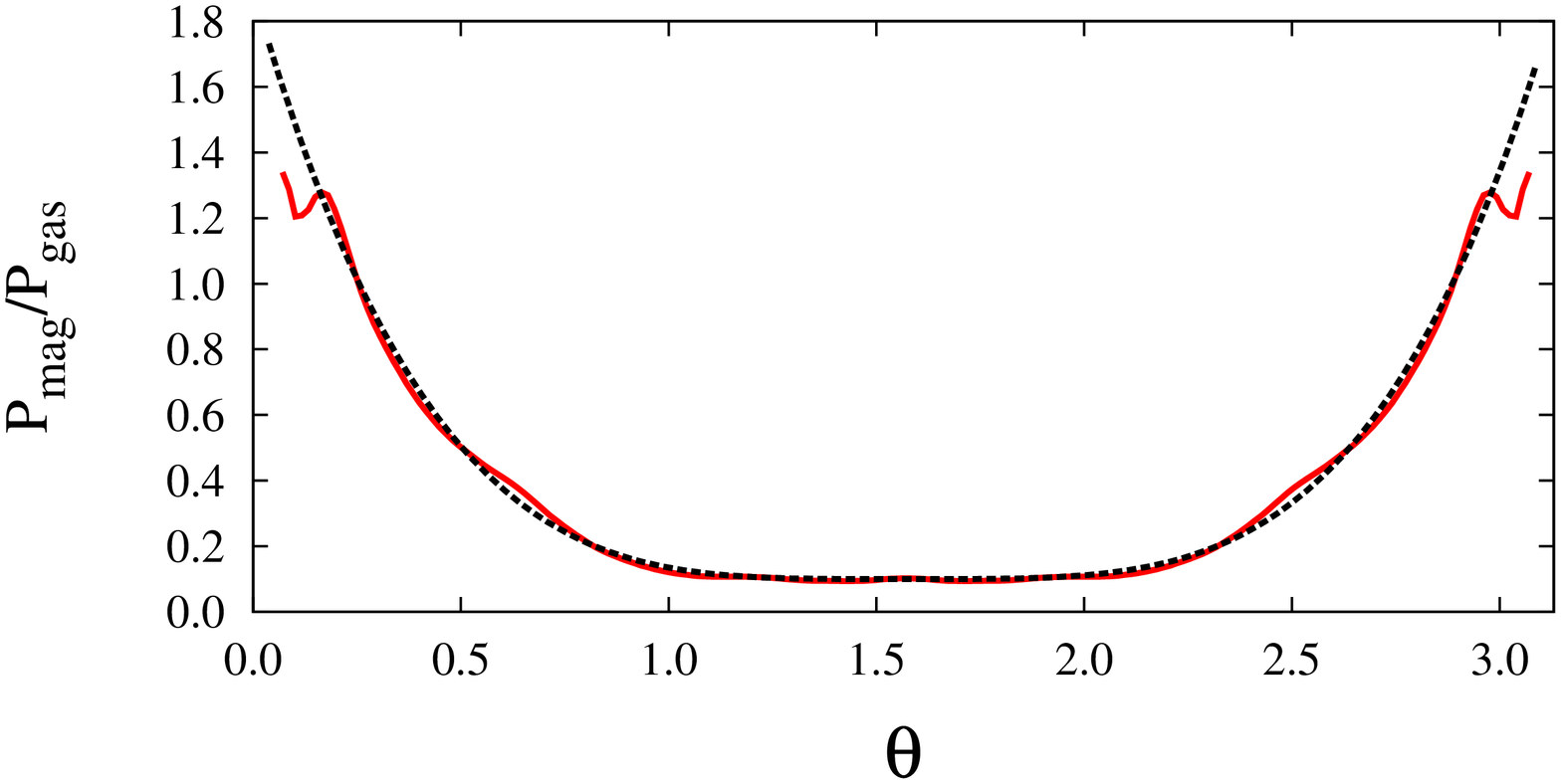}}
\caption{Vertical profiles of (top to bottom) density, gas temperature, 
azimuthal velocity, and magnetic to gas pressure ratio at $R=150R_{\rm
G}$ for the numerical solution (red solid) and the analytical fit
given in Table~\ref{t.fits} (black dashed line). $\theta$ is the polar
angle and $\theta=\pi/2$ corresponds to the equatorial plane.} \label{f.extrfit}
\end{figure}

For convenience, we give here analytical fits to the vertical
structure of the disk at radius $R=150R_{\rm G}$, based on the GRMHD
simulation discussed in Section~\ref{s.disk}. These profiles may be
used to anchor extrapolations of disk properties to larger radii
$R>150R_{\rm G}$. Table~\ref{t.fits} gives expressions for vertical
profiles of density, temperature, azimuthal velocity and magnetic to
gas pressure ratio together with the proposed slopes for radial
extrapolation. Radial and vertical velocities may be neglected.
Figure~\ref{f.extrfit} compares the numerical solution (red solid)
with the fits (black dashed lines). The fits were obtained by least
squares fitting to functions corresponding to expressions in
Table~\ref{t.fits} in the region $0.15<\theta<\pi-0.15$. The region
closest to the polar axis is least reliable and was not taken into
account.

\begin{table}
\caption{Analytical fits to disk vertical structure at $R=150R_{\rm G}$}
\label{t.fits}
\centering\begin{tabular}{@{}ccc}
\hline
\hline 
quantity & fit & extrapolation\\
\hline
&&\vspace{-.15cm}\\\vspace{.2cm}
density [$\rm cm^{-3}$] & $\rho(\theta)=2.02\times10^5\left(1-\left(\frac{\theta-\pi/2}{\pi/2}\right)^2\right)^{1.69}$ & $\propto R^{-1}$ \\\vspace{.2cm}
temperature [$\rm K$] & $\log T(\theta)=9.95+0.24\left(|\theta-\pi/2|\right)^{2.93}$ & $\propto R^{-1}$ \\\vspace{.2cm}
$\phi$ velocity [$\rm cm/s$] & $\log v_\phi(\theta)=9.15-0.24\left(|\theta-\pi/2|\right)^{2.04}$ & $\propto R^{-1/2}$ \\\vspace{.2cm}
$\chi=P_{\rm mag}/P_{\rm gas}$& $\chi(\theta)=0.1+0.31\left(|\theta-\pi/2|\right)^{3.89} $ & $\propto R^{0}$ \\
\hline
\hline
\end{tabular}
\end{table}

\end{document}